# Ionic Electro active Polymer-Based Soft Actuators and Their Applications in Microfluidic Micropumps, Microvalves, and Micromixers: A Review

Mohsen Annabestani, Mahdi Fardmanesh

***Abstract*****-- This paper provides a detailed review on applications of ionic electroactive polymer (*i*-EAP) soft actuators as active elements of microfluidic micropumps, microvalves, and micromixers. The related works that have so far been presented by various research groups in the field have been collected in this review, which in the presented comparative procedure here shows the progress of this field during the time. Microfluidic technology as a pioneer field in bioengineering needs to have some active components likes pumps, valves, and mixers to obtain efficient functionality. Most of the conventional microfluidic active components are challenging to assemble and control outside the laboratory since they need some special facilities, which is not cost-effective. This is while some of the main targets of the microfluidic devices are for the development of point of care (POC) diagnostic systems, home usability, ubiquitousness, and low-cost ability. To solve these problems, i-EAPs have shown promising features to be the proper candidates as active elements of the microfluidic devices. Ionic polymer-metal composites (IPMCs), Conducting polymer actuators (CPAs), and Ionic carbon nanotube-based actuators (i-CNTAs) are three main types of the *i*-EAPs. In this paper, the working principles, fabrication, and microfluidic-based applications of these three categories are described in details. To have a proper comparison between all the reported i-EAP based microfluidic devices, several important features, i.e., applied voltage/frequency, device materials, electrode material, membrane material, and the main measured index of each device, have been reviewed.**

***Index Terms*****—EAPs, *i*-EAPs, Microfluidic, Micropump, Microvalve, Micromixer, Ionic Polymer-Metal Composites (IPMCs), Nafion, Conductive Polymer Actuators (CPAs), Polypyrrole, ionic CNT Actuators (*i*-CNTAs), Bucky Gel Actuators (BGAs)**

## I. Introduction

Microfluidic systems, sometimes known as micro-total analysis systems (μ-TAS), is a field of science which mostly deals with the fluids that are geometrically limited to small feature sizes. Due to some factors such as energy dissipation, fluidic resistance, and surface tension, fluids behave differently at the small scales and different theory and technology for dealing with these behaviors are inquired. Microfluidics is playing a leading role in these kinds of problems, and it can help researchers and industries to find practical solutions for this question that how we can use and manipulate the fluids in a tiny volumes [1-3]. Microfluidics has many applications in the fields of medicine [4-8], diagnostics especially Point Of Care (POC) tests [5, 9], chemical analysis [10, 11], electronic industries [12, 13], etc. Two of the significant applications of microfluidic chips are the development of Lab-on-a-chip (LOC) and Organ-on-a-chip (OOC) devices. LOC is a technology which is integrated with the laboratory-synthesis operations and also analysis of chemicals in a small size, as well as portable and low-cost chips. LOC has many advantages, for example, it is portable, and we can use it where the samples are acquired needless to have any laboratory facilities. Some of LOC devices are easy to use and operable at home with skill-less persons. It also can help to cover health issues and problems in broad area of developing countries because of being affordable and on time diagnose of the vast prevalent diseases like HIV-Aids [14-17], Ebola [18, 19], Zika [20], etc. The convergence of tissue engineering, cell biology, and LOC devices make OOC technologies. OOC technologies can model the interaction between live organs and for drug screening and understanding of drug effects and it can offer a great alternative to conventional preclinical models[21]. OOC is an emerging multidisciplinary field, and we can find many of chips that they have been developed for a variety of organs like lung, heart, liver, vessel, tumor, etc. [22]

All microfluidic devices, LOCs, and OOCs, are categorized into two main categories as active and passive devices [23]. Active devices use at least one external stimulation in order to do an action, for example, applying dielectrophoresis [24], magnetic [25], optical [26], acoustic [27, 28] or any other stimulation for sorting of the particles. However, in passive devices, the primary approach is trying to do actions by making some interactions between the device components such as micro-channels structure and their geometry, the flow field, etc. [29, 30]. There is a variety of applications that they use passive chips, but sometimes a microfluidic system needs some active operation (e.g., active sorting) or active components such as pumps, valves, and mixers in order to efficient manipulation of the fluids [31]. Between these active components, pumps are more special because most of the pumps need some active elements like diaphragms. But we can also find some passive valves or especially passive mixing chips. Active pumps, valves, and mixers are actuated by pneumatic, thermopneumatic, hydraulic, piezoelectric, electromagnetic, electrostatic displacements or Electroactive-polymer (EAP) actuators[31]. Most of the conventional micropump and microvalve active actuators are challenging to assemble and control, and they need some special facilities and are not affordable enough. These constraints are against some of the main targeted advantages of microfluidic devices such as POC development, use at home, availability, low-cost ability, etc. But between these actuators EAPs are different, and they do not have the mentioned constraints. EAPs are low-cost actuators that we can miniaturize them in small sizes in order to integrate them into MEMSs and μ-TASs. Moreover, EAPs are electrically driven actuators that some of them need relatively very low input voltage (>5V) while this range of power is ubiquitous everywhere by a battery or cell phone charger. Considering these advantages of EAPs ensure us that they are very good candidates for using as active elements of microfluidic devices.

EAP actuators are a group of polymer-based composites that respond mechanically to electrical stimulation. They can imitate the behavior of biological muscles; hence sometimes they are called artificial muscles [32, 33]. EAP types are divided into two main groups: Field-activated EAPs (*f*-EAPs) and ionic EAPs (*i*-EAPs) [32]. *f*-EAPs work by the electrostatic force produced by the electric field applied between the electrodes on the membrane or by the charge on a local scale. Based on the molecular, microscopic, or macroscopic phenomena in response to the applied electric field, a strain is produced and this strain creates the main actuation of *f*-EAP actuators [32]. Ferroelectric Polymers (FPs) [34], Dielectric Elastomers (DEs) [35] and Electrostrictive Graft Polymers (EGPs) [36] are the most essential types of *f*-EAPs that usually use poly (vinylidene fluoride-trifluoroethylene) [P(VDF-TrFE)], Polyurethane, and modified copolymer–PVDF-TrFE as the material of their membranes, respectively [32].

An alternative way of creating actuation in the EAPs is using mobile ions in their polymeric membrane. Applying an electric field leads to move the ions, and consequently, this movement carries the solvent and makes swelling or contraction in the membrane or electrodes finally making the desired actuation of EAP[33]. These groups of EAPs that work by ion transport phenomena are called ionic EAPs or the same *i*-EAPs. In contrary to *f*-EAPs, *i*-EAPs work with the low input voltage (≥5V), but due to small spacing between ions and charges and a large amount of charge, the energy of this low applied voltage is high. As depicted in **Fig.1,** most usable and most popular form of an *i*-EAP actuator is a thin strip in the form of a cantilever beam that has a polymeric membrane coated by two electrodes (Some of the *i*-EAPs may have less or more than three layers). Based on the material of their membranes and electrodes, and also their working principles, we can categorize EAPs in three main groups, Ionic polymer-metal composites (IPMCs) [37, 38], Conducting or sometimes conjugated polymer actuators (CPAs) [39] and Ionic carbon nanotube-based actuators (*i*-CNTAs) [40]. In this paper, we talk about the application of *i*-EAPs in active microfluidic micropump, microvalves, and micromixers.

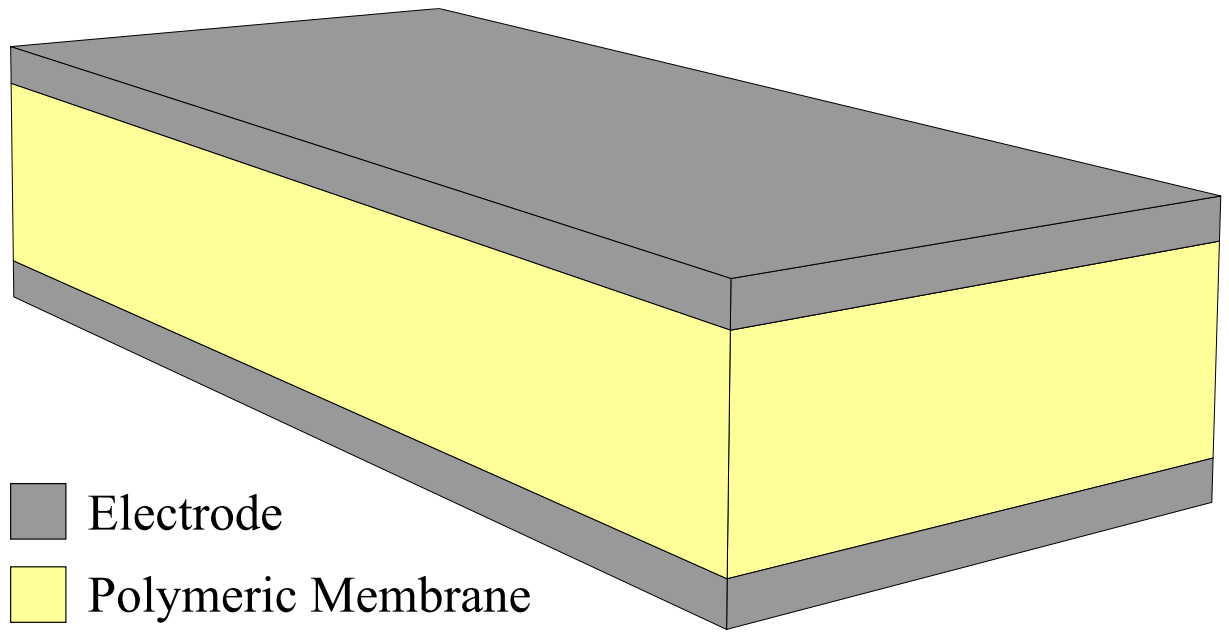

**Fig. 1. The most usable and most popular form of *i*-EAPs.**

In the rest of this paper, we have a detailed review in all *i*-EAP based microfluidic micropumps, microvalves, and micromixers, for which five parts are defined. In the first three parts, we will review all IPMC, CPA, and *i*-CNTA related devices and in fourth part, a disscusion will be presented based on a table consisting of all important information and features about all the reviewed devices. And finaly in the last part a conclusion is provided to figure the points of work out.

## II. Ionic polymer-metal composite Actuators

Ionic polymer-metal composites or in abbreviation IPMCs are a group of soft *i*-EAPs that can be used as actuators and sensors [37, 38, 41, 42]. Due to promising features of IPMCs, it has a wide range of usage, especially in biomedical applications [43-55]. IPMC's density is low, and its toughness is high, having very large stimulus strain. Lightness and the low driving voltage (>5 V) are also other attractive properties of IPMCs. Of course, this smart material has some drawbacks that should be considered, for example, it is not strong enough for some applications, or there are some unwanted effects in its performance like back relaxation effects [56-58]. The general structure of IPMC is a cantilever beam which is a combination of a thin membrane (usually Nafion) coated by two metallic plates (usually Pt) on both sides. In response to the applied voltage through the thickness of an IPMC, its internal ionic content (usually hydrated sodium cations) flow toward the cathode side. This transportation makes the Nafion inflation around the cathode side and leads to a bending response towards the anode side as shown in **Fig.2**. Inversely, if we bend or deform the IPMC, a low voltage is measured between two electrodes across the membrane depending on the applied deformation [41]. Hence, IPMCs is an actuator as well as being a sensors.

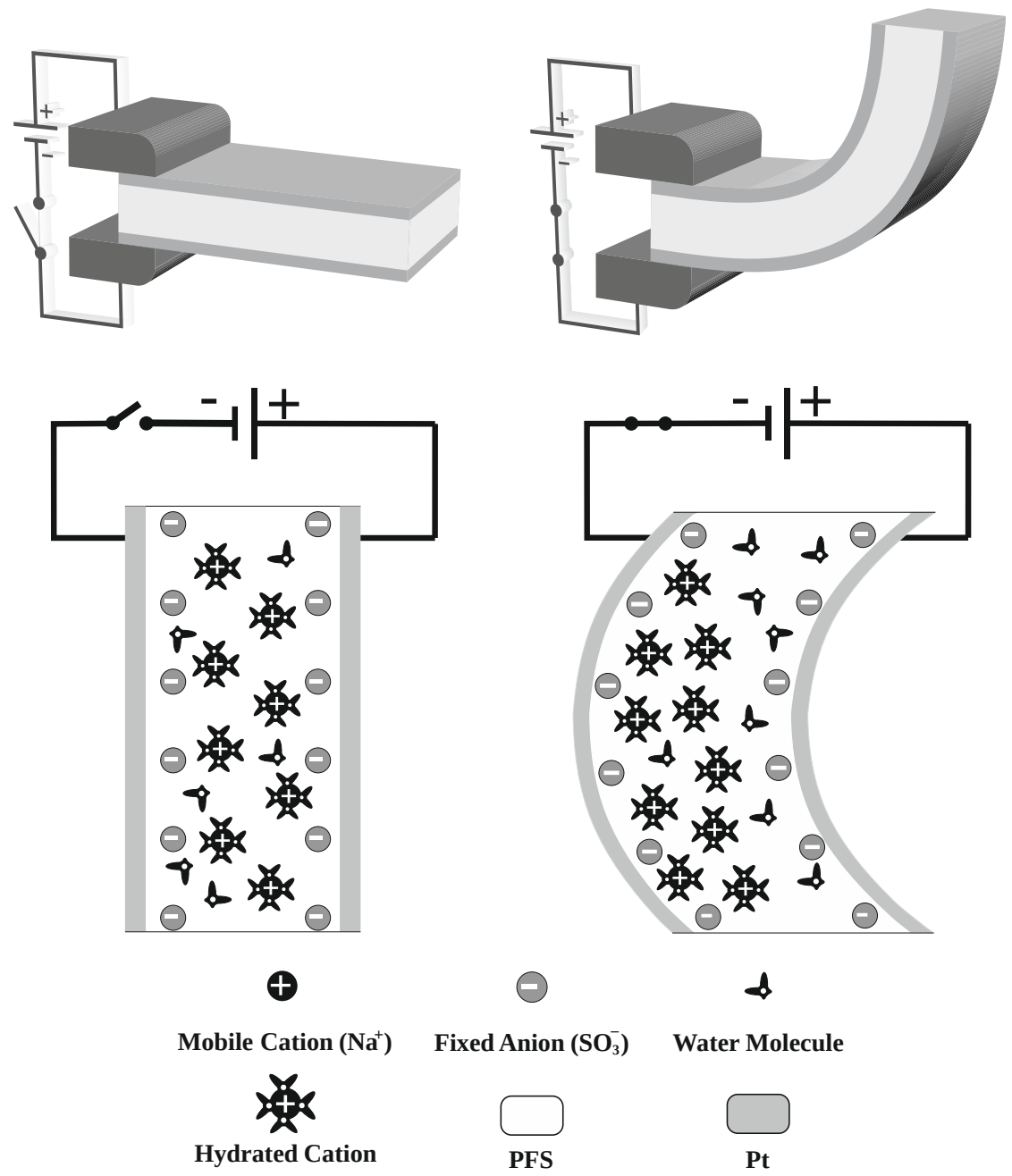

**Fig. 2. IPMC actuator's working principle: (Left) before applying voltage; (Right) after applying voltage[41].**

## A. *IPMC Fabrication*

There are several fabrication procedures for IPMC devices that here we want to talk about the most conventional procedure for a Nafion based and Pt electrodes IPMC. At first, the top and bottom of the membrane of the devices should be roughened by microblasting techniques or simply using small size and soft sandpaper in order to increase the surface roughness and consequently increasing the surface area. After that, to clean the reminding particles of the roughening process, the membrane should be cleaned ultrasonically, treated with boiled HCl (2N) and deionized water for around 30 min (**Fig.3(a)**). Then the membrane should be soaked for about 10 hours in the tetraammineplatinum chloride hydrate salt in order to exchange protons for Pt ions (**Fig.3(b)**). The initial step in electroless plating of IPMCs is the surface electrode coating process [37]. In the initial compositing process, the ions of platinum saturate the polymer membrane to improve the adhesion between the electrodes and the electrolyte membrane (**Fig.3(c)**). There is another point in the surface electrode coating process, as it increases the electrical conductivity of the Pt electrodes and protects them from oxidation (**Fig.3(d)**). Finally, the ion exchange polymer is immersed in a salt solution to allow platinum-containing cations to diffuse via the ion exchange process (**Fig.3(e)**) [37, 59].

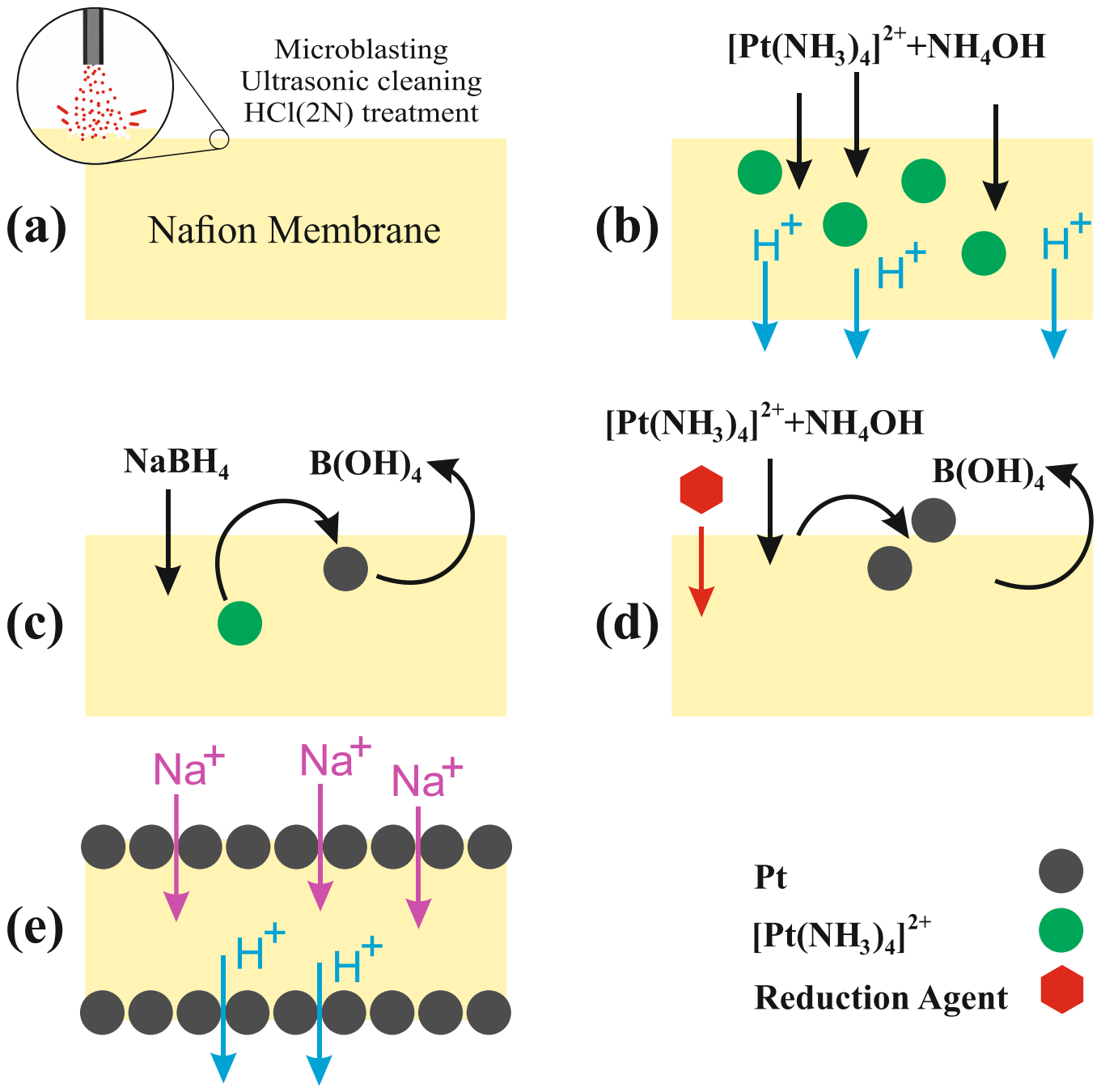

**Fig. 3. The conventional fabrication process of IPMC actuators: (a) pre-treatment, surface roughening, and cleaning, (b) ion-absorption, (c) primary electrode plating using reduction, (d) secondary plating (reduction again) and (e) ion-exchange.**

## B. *IPMC Analytical Models*

IPMCs' model can be categorized into two main groups. The first group is system identification based models that model the dynamic behavior of IPMCs using experimental datasets only by finding a mathematical relationship between input (usually applied voltage to IPMC) and output (usually tip displacement of IPMC)[41, 60-64]. The second group is related to analytical models that we study more about them here. Analytically there is a verity of the models to describe the dynamic behavior of IPMC sensors and actuators. For example, some researchers use distributed Resistor-Capacitor (RC) equivalent circuits [65-71], and some others use Partial Differential Equations (PDEs), Finite element methods, COMSOL modeling, etc. [72-75]. But some models are based on the physical and multi-physical ion, and water transport approaches which are categorized into three categories as; thermodynamics of irreversible process (TIP) models, frictional models (FR), and Nernst-Planck PDE based (NP) models [76]. Between three main multi-physical approaches NP models are the most straightforward approach to interpret ion transport [76] and they are very compatible with the physics of operation of ionic content in the IPMCs. As the first time of using NP PDE for IPMCs, it was presented for charge distribution of it by Nemat-Nasser [77]. The main idea behind the NP model is finding a solution for Nernst-Planck PDE in order to find tip displacement of a IPMC device. The required Nernst-Planck PDE for IPMC boundary problem is as follows:

$$\mathbf{J} = -d\left(\nabla C^{+} + \frac{C^{+}F}{RT}\nabla\phi + \frac{C^{+}\Delta V}{RT}\nabla p\right) + C^{+}\mathbf{v} \qquad (1)$$

Where some electro-physio-chemical phenomena like diffusion, ion migration, and convection are involved in this PDE. In Eq.1 $\mathbf{J}$ is the spatial vector of the cations flux, $C^+$ is the concentration of cations, $T$ is the absolute temperature, $d$ is the ionic diffusivity, $p$ is the fluid pressure, $R$ is the gas constant, $\mathbf{v}$ is the free solvent velocity field, and $\Delta V$ is the volumetric change of the membrane [78]. In order to find the output of this model (usually tip displacement of IPMC), most of the researchers try to find the electric charge density ( $\rho$ ), and then mechanical stress ( $\sigma$ )by involving some physical assumption and solving some PDEs. In order to find mechanical stress, most of the researchers [65, 78-84] use the Nemat-Naser coupling relation as follows [77]:

$$\sigma = \alpha \rho \quad (2)$$

where $\alpha$ is the constant coupling coefficient. In the next steps, by using the definition of bending moment and applying linear Euler-Bernoulli beam theory and also some mathematical efforts, we can calculate the tip displacement of IPMCs. For example, by a similar approach, Chen and Tan [78] have presented the following relation for tip displacement of the IPMC ( $\delta_{Tip}(s)$ ) in small deformation situation.

$$\delta_{Tip}(s) = -\frac{1}{2}\frac{\alpha_0 W}{YI}\frac{K\kappa_e\left(\gamma(s)-\tanh\left(\gamma(s)\right)\right)}{\left(\gamma(s)s+K\tanh\left(\gamma(s)\right)\right)}\times \\ \times\frac{V(s)L^2-4\int_0^L\int_0^z\int_0^{z'}\left(r_1'/Wi_s(\tau,s)\right)d\tau dz'dz}{1+r_2'\theta(s)/W} \quad (3)$$

The details of this relation and the definition of unknown functions and parameters have been described in [78]. The proposed governing NP PDE can cover the thermal effect, but most of the multiphysical models ignore it in their PDE. Some papers also have been claimed that the behavior of IPMC actuators in microfluidic devices are sensitive to the thermal effects and it is essential to understand the thermal properties of the device [85]. To cover this sensitivity they have proposed an NP-based multiphysics modeling of an IPMC for microfluidic control device [85].

IPMC's promising potentials and its capability to be miniaturized have made it to be a good candidate for MEMS [86] devices like some of the active microfluidic devices. In the rest of this section, we want to have a complete review in the application of IPMC actuators in microfluidic devices and their related components. Most of the microfluidic-based applications of IPMCs are related to micropumps, but we can find some other applications of IPMCs for making microvalves and micromixers. Hence three subsections have been defined here for reviewing the application of IPMC in micropumps, microvalves, and micromixers for microfluidics devices.

*C. IPMC-based microfluidic micropumps*

***IPMC-based Micropump 1***

The first idea in using of IPMC actuator as an active element of micropumps goes back to the work of Guo et al. in 1996 [87], and they republished the improved version of their work in 2003 [88]. Their proposed micropump has two one-way valves and a circular diaphragm that they made from IPMC (**Fig.4**). Like the general form of the pumps, this micropump also has a chamber and a diaphragm on the diaphragm which is a circular IPMC as mentioned before. When we apply an electric voltage to this IPMC, it bents into the anode side. Then the volume of the pump chamber will be increased, and the result is inflowing of liquid from the inlet to the chamber. Now if we change the polarity of the applied voltage or decrease its amplitude, the volume of the pump chamber will be decreased, and the result would be inflowing of liquid from the chamber to the outlet. Hence to have a continuous pumping, a sine voltage drives the IPMC diaphragm. As mentioned earlier, this micropump also has two one-way valves the structure of which is shown in **Fig.5**. These valves have a taper shape and have a through hole leading from the tip to the outside. By application of a voltage to the IPMC actuator, it will be bent into the flow direction when driven by using the same sine voltage that was applied to the diaphragm. Hence the fluid can flow into the valve and then flows out to the outlet. As depicted in **Fig.5-b,** using this valve prevents back streaming of the fluid to the chamber from the outlet because the IPMC actuator closes through the hole.

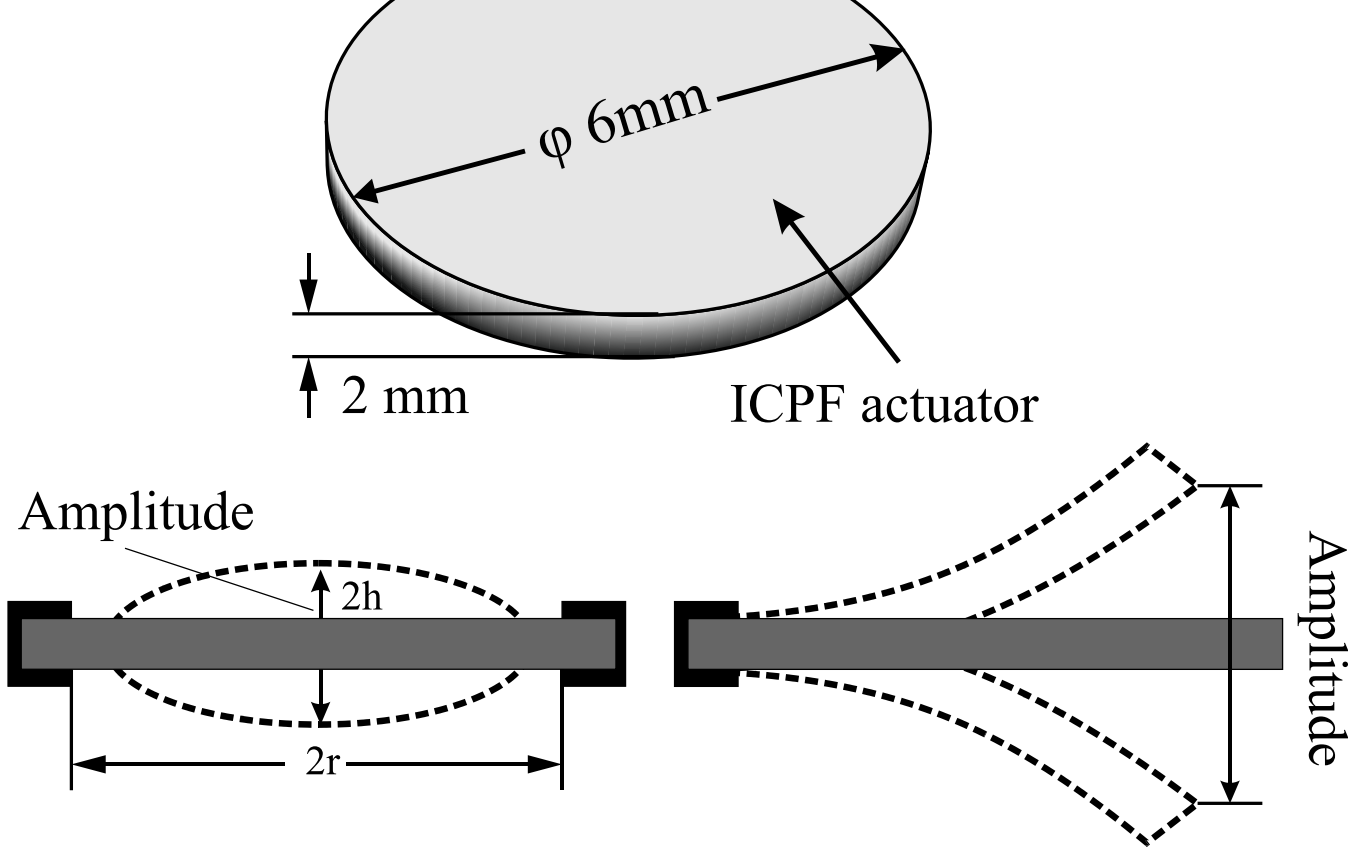

**Fig. 4. The required IPMCs (ICPF here) in the proposed micropump of [87, 88]. (Up) for Pump chamber and (Down) for one-way valve.**

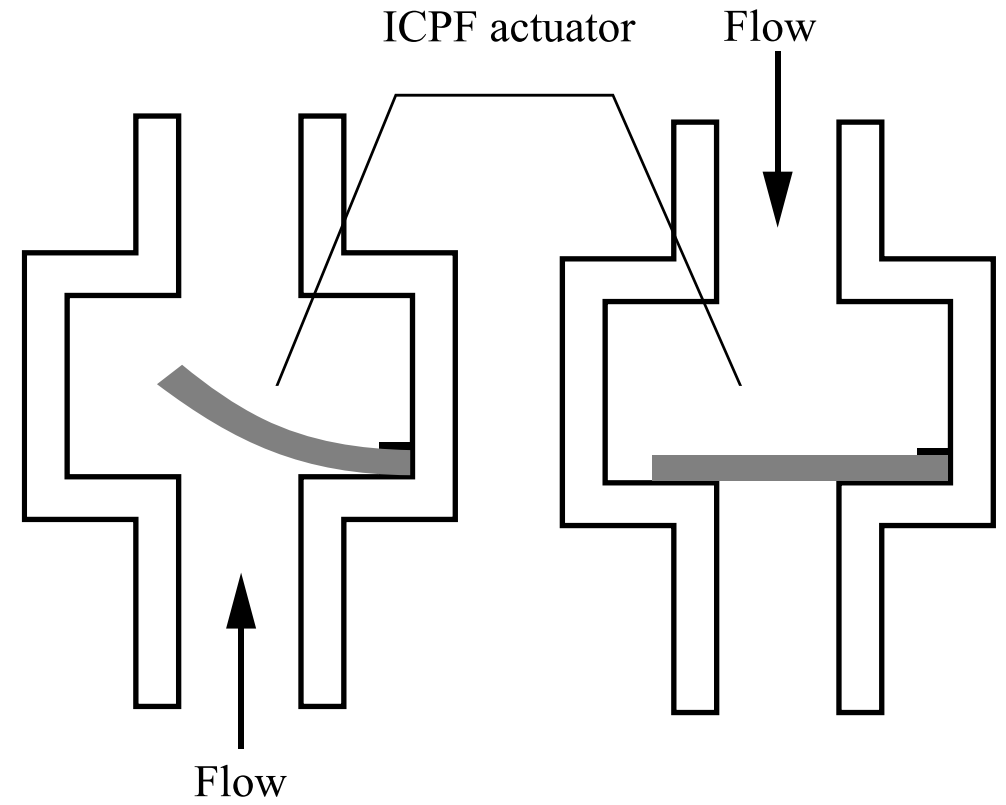

**Fig. 5. The required IPMCs (ICPF here) in the proposed valves of [87, 88]. (Left) Feed forward and (Right) back forward.**

The proposed micropump has been made in the form of a cylinder with the size of 12 mm in diameter and 20 mm in length using stainless materials. Two circular Nafion based IPMC actuators were used as the diaphragm pump, and under experimental tests, this IPMC-based micropump can pump the fluids in the range of 3.5-37.8 μl/min in response to a 1.5 Volt sine wave with the frequencies in the range of 0.1-15 Hz.

### *IPMC-based Micropump 2*

Another type of IPMC based micropump was presented in 2004 by Pak et al. [89]. This micropump is a device combined of PDMS (Poly dimethylsiloxane) microchannel and IPMC actuator. In the PDMS microchannel, there is a nozzle/diffuser structure which makes the fluids flow from the inlet to the outlet of the device (**Fig.6**). When IPMC is deformed in response to the input applied voltage, it can play the role like a diaphragm to pump fluid. As it is shown in **Fig7-b**, this device can work properly, and its flow rate is 9.97 μl/min in response to an 8 $V_{P-P}$ and 0.5 Hz input voltage. The used IPMC has been fabricated by commercial Nafion 112 as membrane and polypyrrole conductive polymer as electrodes where in order to have the best performance, it is needed to have 17-19 μm thick electrode[89].

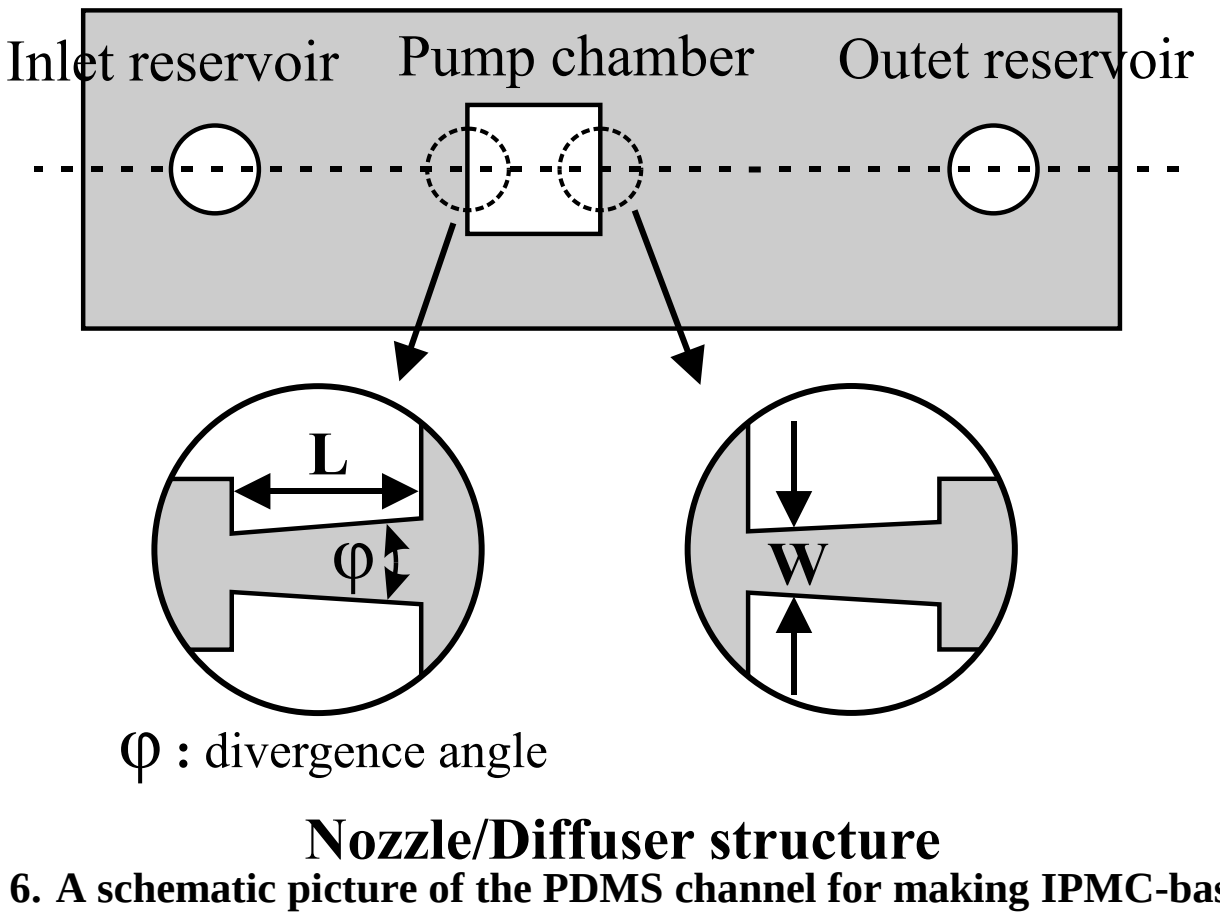

**Fig. 6. A schematic picture of the PDMS channel for making IPMC-based micropump [89].**

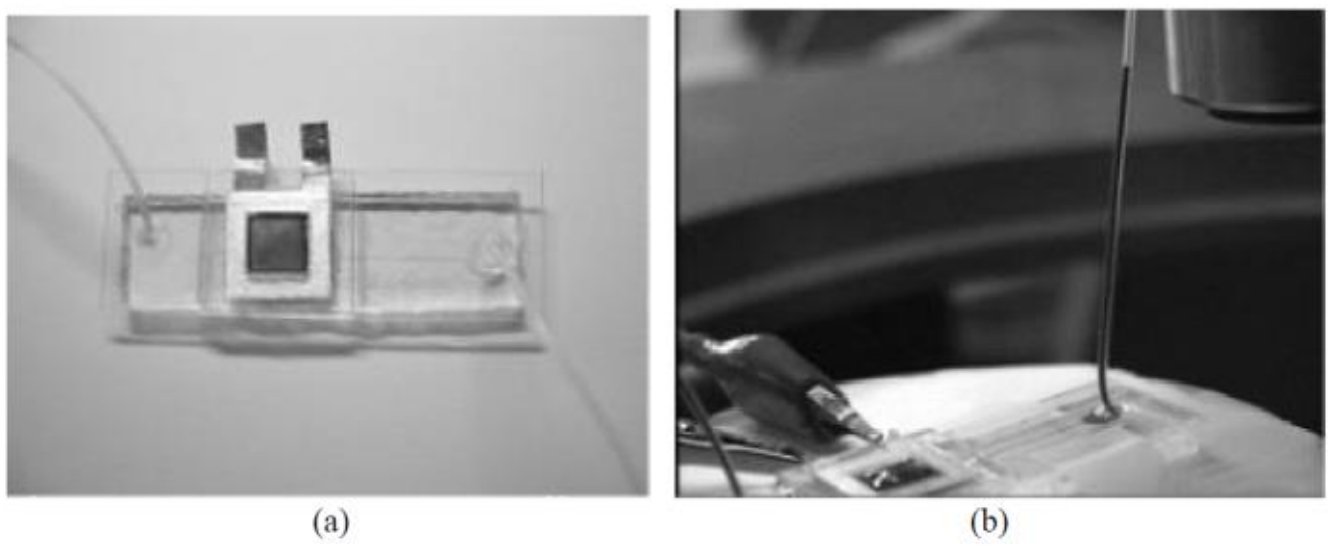

**Fig. 7. Fabricated IPMC-based micropump,(a) Before test. (b)Under testing[89].**

### *IPMC-based Micropump 3*

In continuing the Pak et al. idea [89], in another paper Lee and Kim [90] tried to model the behavior of the IPMC disk (**Fig.8**) as a micropump diaphragm. They believed that due the easy and low-cost manufacturing process, IPMC is a good choice for fabrication of micropump in competition with other technologies. A finite element analysis (FEA) was used to find the shape of IPMC diaphragm and its electrodes and also the estimation of its stroke volumes ($\Delta V$). Based on the optimum designed nozzle/diffuser structure (**Fig.9**) and the estimated stroke volume of the IPMC diaphragm, the flow rate of the IPMC-based micropump was estimated at a low Reynolds number (about 50) and it means that this IPMC-based micropump is an appropriate choice for microfluidic devices because in microfluidic devices we face with low Reynolds number fluids [90].

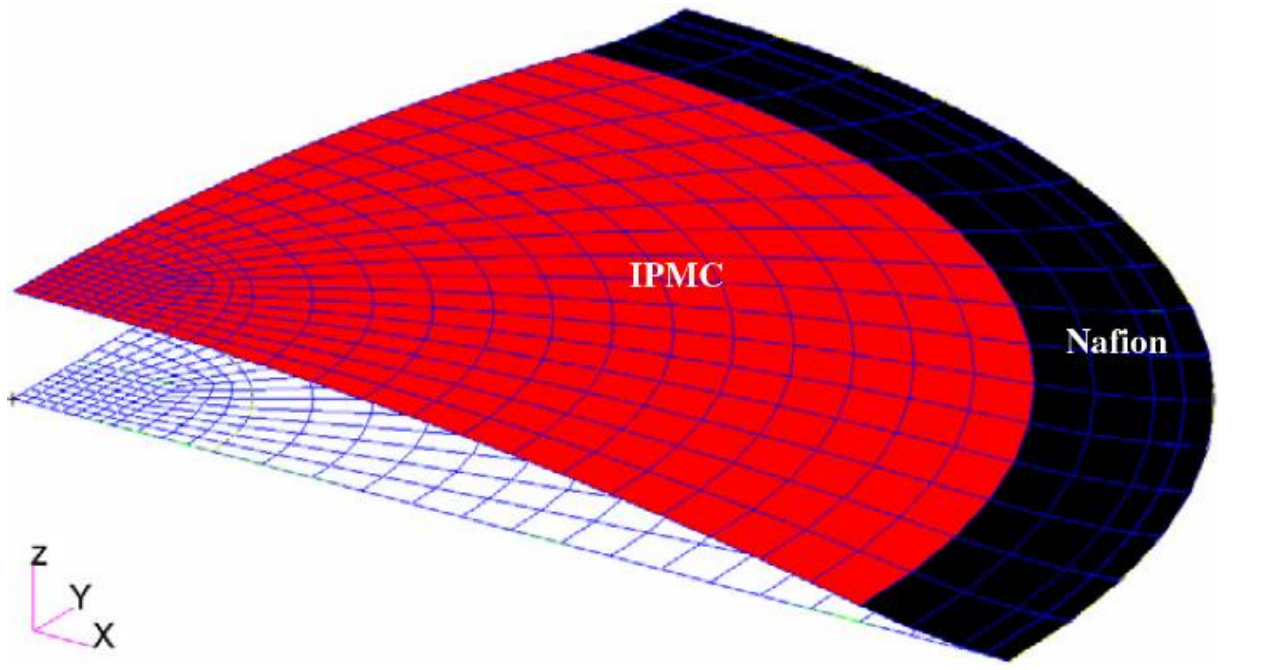

**Fig. 8. Bended form of IPMC diaphragm (Radius of the electrode (red circle) = 8.5 mm)[90].**

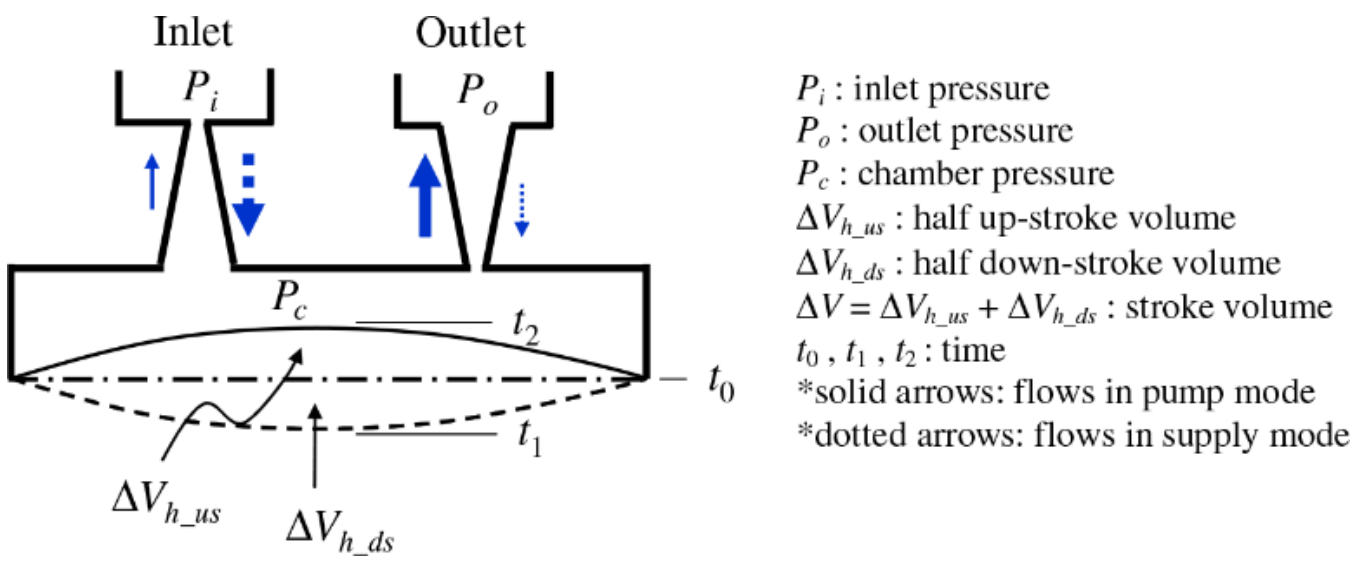

**Fig. 9. A schematic diagram of the IPMC-based micropump with a nozzle/diffuser structure [90].**

The assumed IPMC for this model had a Nafion membrane with *Pt* electrodes and based on the calculated results for the IPMC diaphragm, the optimum center displacement (the best result to have maximum pumping rate) was 0.966 mm, in which the radius of the electrode was 8.5 mm. Based on the numerical result of this study, in response to a 2 V and 0.1 Hz input voltage, the flow rate of this micropump is 8.2 μl/sec (492 μl/min).

### *IPMC-based Micropump 4*

A flap valve IPMC-based micropump is presented in [91, 92] (**Fig.10**). In this device, a multilayered IPMC has been fabricated with a consecutive recasting method, which consists of a Nafion/modified silica layer as a membrane sandwiched between two thin layers containing Nafion, layered silicate, and conductive material particles (Ag nanopowder) as electrodes. The developed modified multilayered IPMC has better blocking force without making any constraint and problem for its bending displacement. Same as previous works the device was used as the diaphragm of the micropump. The main material using in the flap valve was PDMS (due to its flexibility and solving the sealing problem), but the whole of micropump is a composition of several PMMA (Poly(methyl methacrylate)) layers that is equipped with IPMC diaphragm and PDMS flap valve (**Fig.11**). In this work, it is concluded that the designed flexible support is a useful structure for the IPMC diaphragm, particularly from the aspect of the IPMC displacement versus the blocked pressure diagram and the pumping performance. This device also can work properly and better than the previous ones with a higher flow rate. Its flow rate is 760 μl/min in response to a 3 V and 3 Hz input voltage. Even though the proposed IPMC-based micropump has some advantages such as low required input voltage, no leakage problem, and simple fabrication process, but the durability of the IPMC is still an important challenge that it should be solved if we want to use it in practical in microfluidic micropump applications.

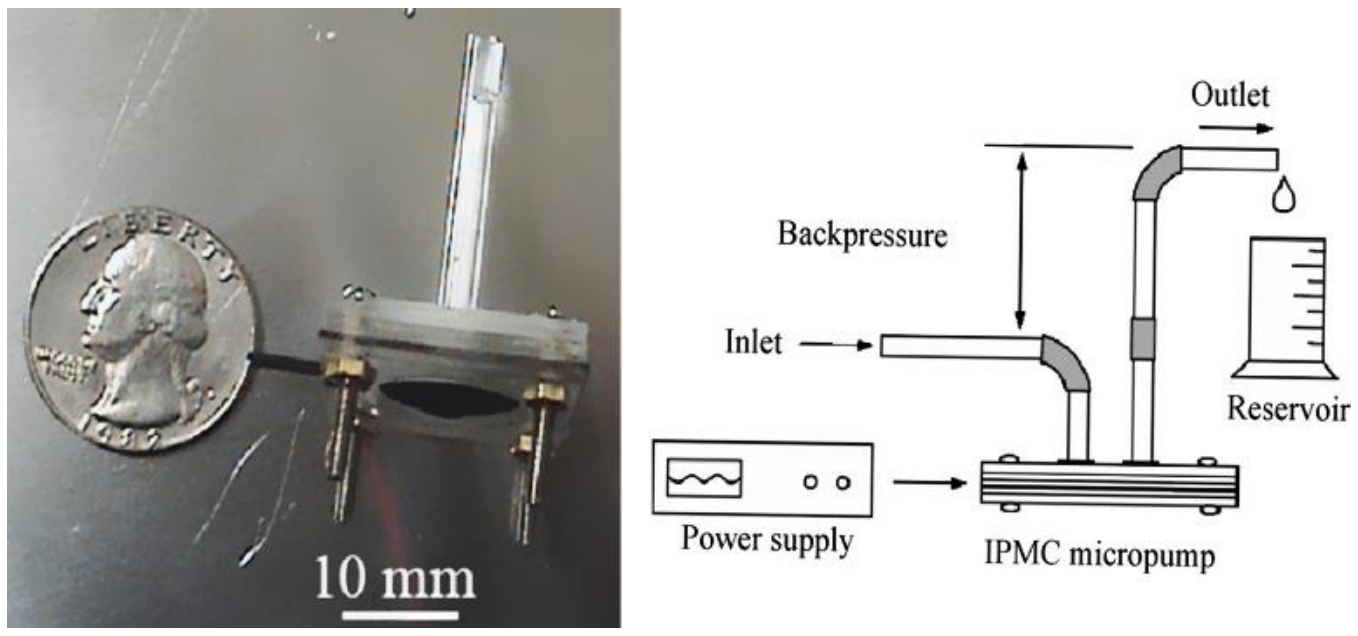

**Fig. 10. Left: Photograph of the fabricated IPMC-based micropump. Right: Experimental setup for testing the flow rate of the micropump[91, 92].**

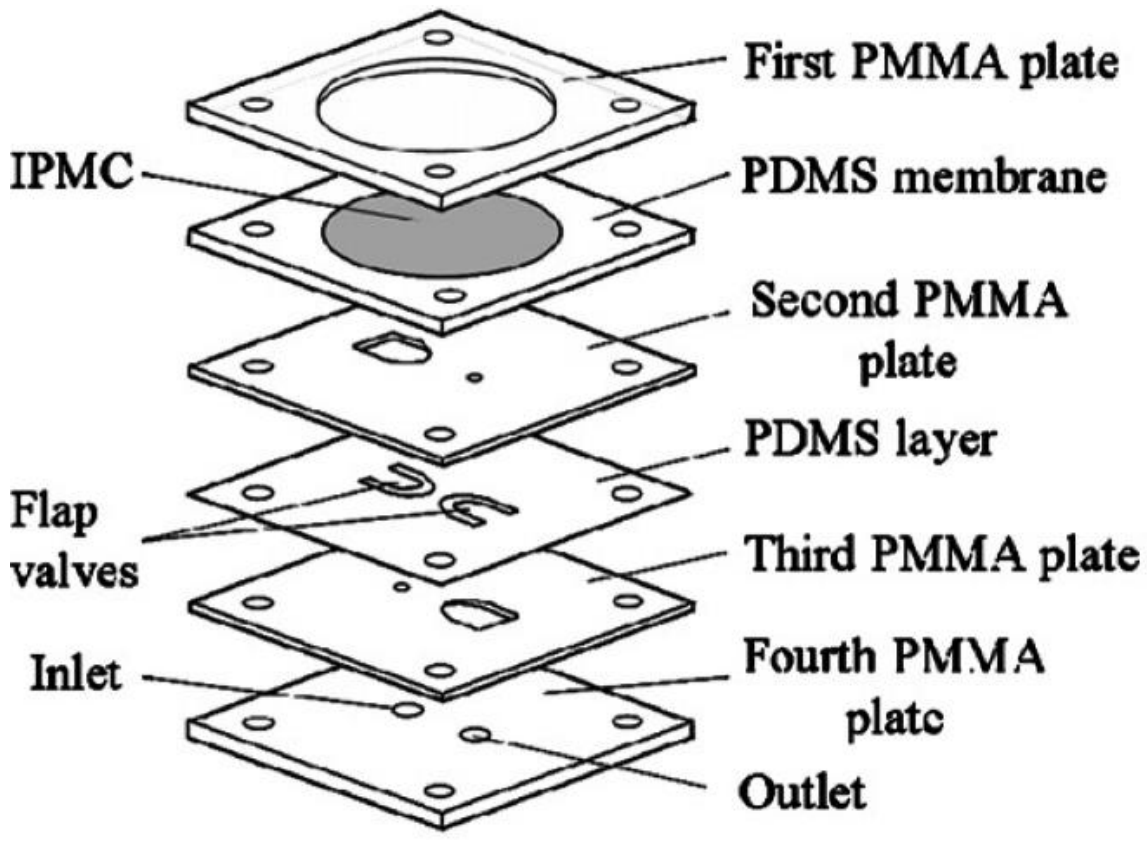

**Fig. 11. The exploded view of the IPMC-based micropump structure[91, 92].**

### *IPMC-based Micropump 5*

In 2010 Santos et al. [93] also presented a new kind of IPMC-based micropump. The main structure of this device also was similar to the previous ones, and an IPMC diaphragm has been used to make pumping pressure. In this work, two different IPMC diaphragms have been fabricated and experimentally tested. These two types of diaphragms are similar to each other, and only the way of applying voltage is different on them. As it is shown in **Fig.12**, device no.2 has four additional branches in comparison with the device no.1 where we can apply a voltage to the IPMC by two thin copper rings attached to these branches (**Fig13.b**) while in no.1 two copper wires are soldered to the center of the Pt electrodes in both sides (**Fig13.a**). The IPMC diaphragm of device no. 2 is more practical and durable and hence the micropump has been fabricated by this diaphragm. The proposed micropump is a four layers nozzle/diffuser pump as depicted in **Fig.14** with two Acrylic layers on top and down, with a layer of IPMC diaphragm covered by a layer of Teflon. This pump was fabricated (**Fig.15**) and the stroke volume of both IPMC diaphragms was tested with different values of supplied electric current. The results showed an asymmetric behavior of upward and downward displacements for both diaphragms. Stroke volumes up to 80 μl were measured and in response to a 0.1 Hz input voltage (amplitude was not reported), the flow rate of the fabricated micropump with IPMC diaphragm of device no. 2 is reported to be about 8.02 μl/s (481.2 μl/min).

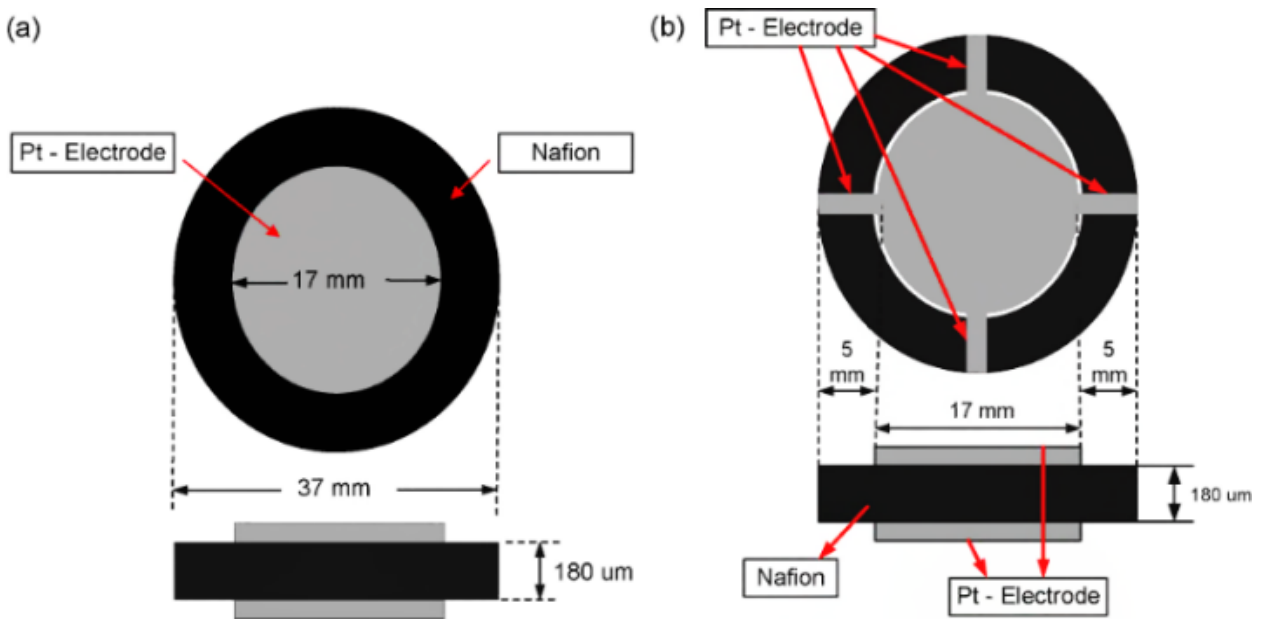

**Fig. 12. Structure of the IPMC-based diaphragms. (a) IPMC diaphragm no. 1. (b) IPMC diaphragm no. 2 [93].**

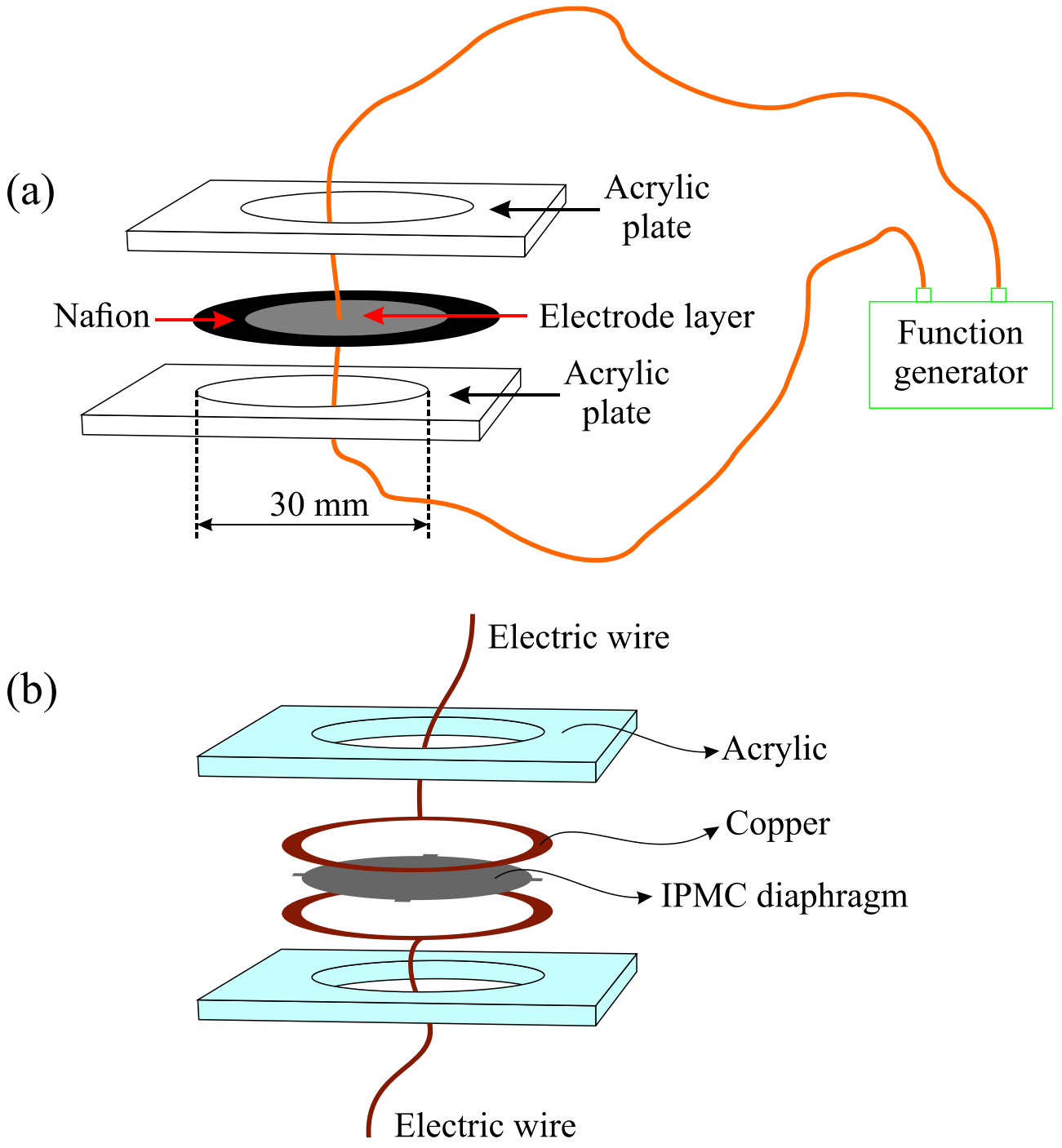

**Fig. 13. The method of applying the voltage to IPMC and the supporting structure of IPMC-based diaphragm. (a) IPMC diaphragm no. 1. (b) IPMC diaphragm no. 2. Redraw from [93].**

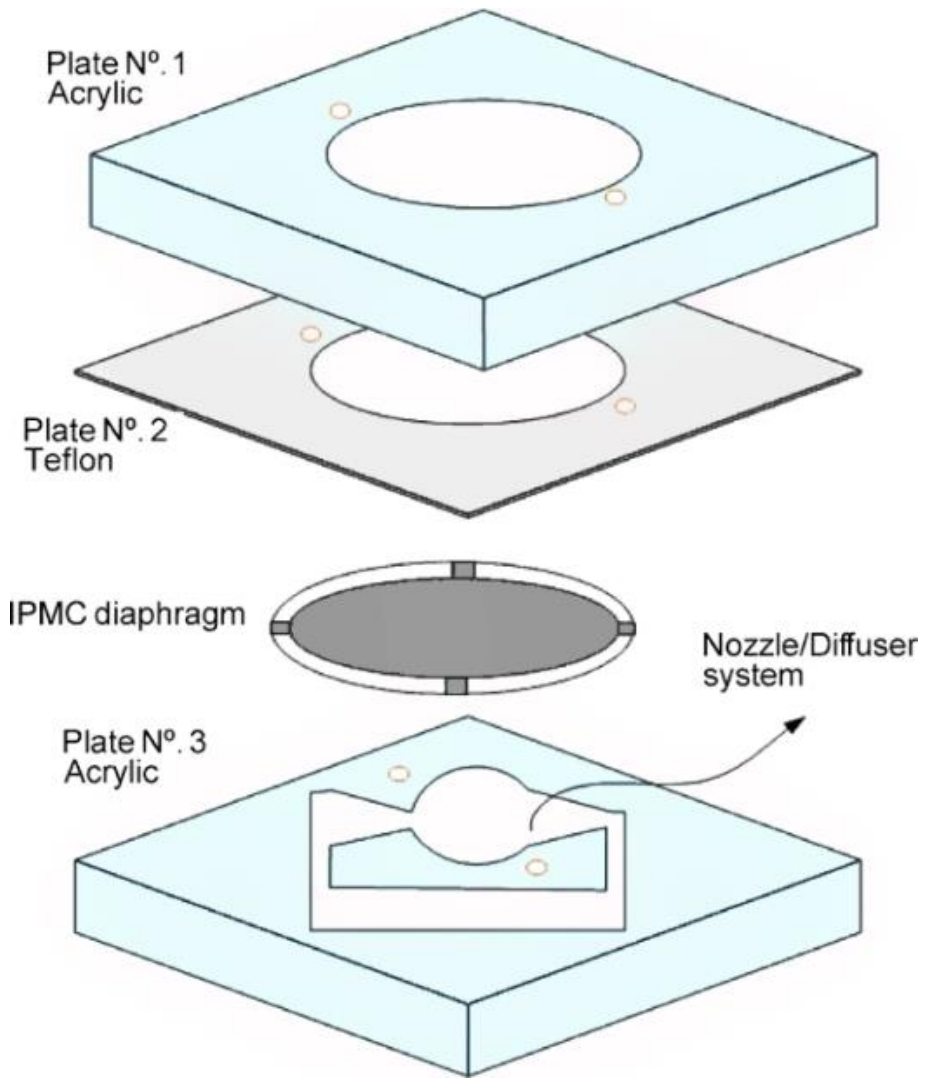

**Fig. 14. The layers of the micropump device in [93].**

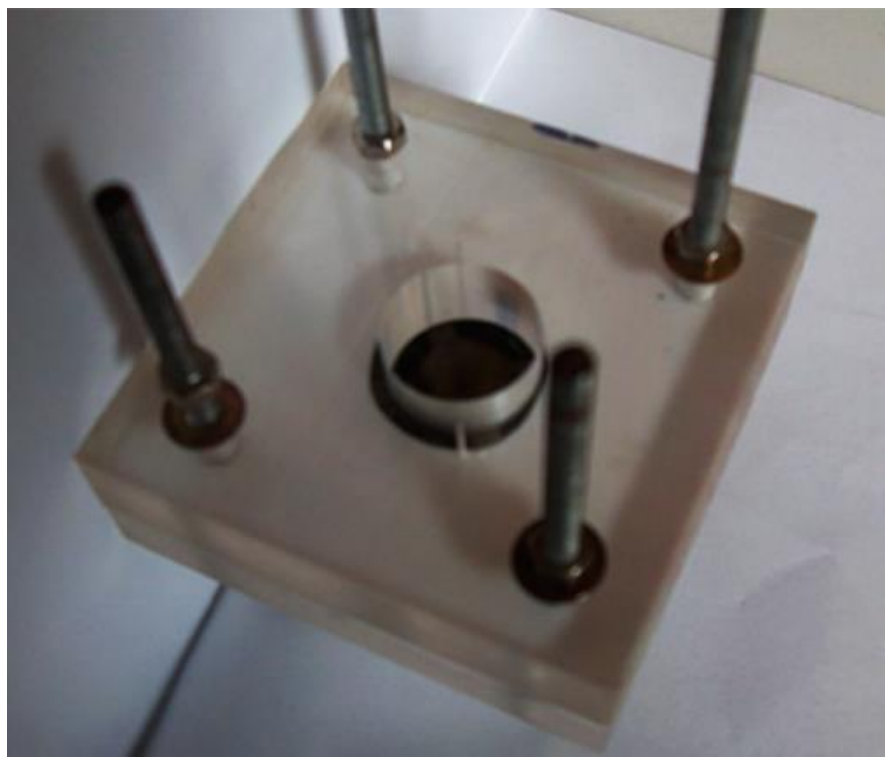

**Fig. 15. Fabricated micropump in [93].**

### *IPMC-based Micropump 6*

Fourteen years after his work as the first IPMC-based micropump [87], Guo presented a new kind of IPMC-based micropump with several improvements [94]. He found, or better to say he adopted an idea from Fang and Tan work [95], that use of a single disk-shaped piece of IPMC (like all previously presented micropumps) is not an optimum choice as a diaphragm. This kind of IPMC disks are clamped at all edges, which makes an obstacle for IPMC's deformation, i.e., it decreases the IPMC bending, and consequently, the pumping ability of IPMC-based micropump will be decreased significantly. In order to solve this problem and increasing the deformation and pumping ability of the micropump, he and Wei [94] presented a new form of IPMC diaphragm using four pieces petal-shaped IPMC actuator instead of a disk-shaped one. This petal-shaped structure has been depicted in **Fig.16** where four quarter-disk-shaped IPMC actuators are attached on a thin PDMS diaphragm. In this structure, the IPMCs are only clamped on one edge, and they are more free to bend and making an up-down movement in the PDMS diaphragm so that the pumping operation occurs.

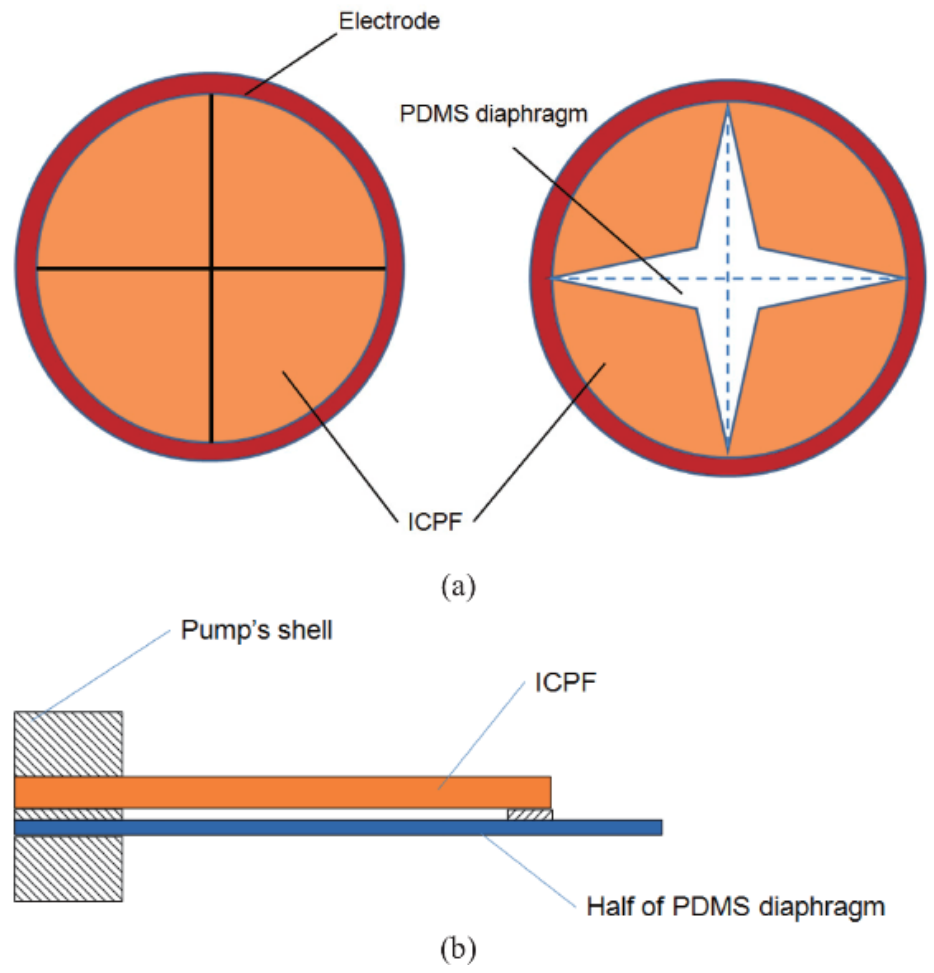

**Fig. 16. (a) Schematic of a petal-shaped IPMC actuator (Here ICPF) in top view. Left is before applying voltage and right is after applying voltage. (b) The cross-sectional schematics of the bonding of PDMS diaphragm and IPMC actuator.**

The 3D schematic of this micropump is shown in **Fig.17**; this micropump consists of a petal-shaped IPMC actuator in a sealed actuating chamber, and two check valves with a PDMS diaphragm attached to the IPMC actuator. Under applying voltage, the IPMC and then PDMS diaphragm will be moved up and down. Thus this movement makes the pumping pressure and the fluid will be pumped from the inlet to the outlet. The functionality of the microvalves of this micropump is keeping the directional flow during the pumping process. The microvalves are fabricated using PDMS diaphragms. The working principle and duties of these valves are as follows [94]:

1. When a pressure is applied from the bottom, the outlet diaphragm will be lifted to allow the fluid flow out through the outlet.
2. When a pressure is applied from the top, the inlet diaphragm will be pushed down and allows the fluid flow into the pump through the inlet.

The proposed micropump has been made in the form of a cube with the size of 30mm×30mm and 27mm (height). A petal-shaped IPMC actuator is used as the active element of the pump with a PDMS diaphragm and under experimental tests, this IPMC-based micropump was able to pump the fluids by 202 μl/min flow rate in response to a 5-volt 2-Hz applied voltage. In comparison to their previous works [87, 88], they could improve the flow rate of the micropump more than five times, and their new pump might be a better candidate for some of the microfluidic application like drug delivery.

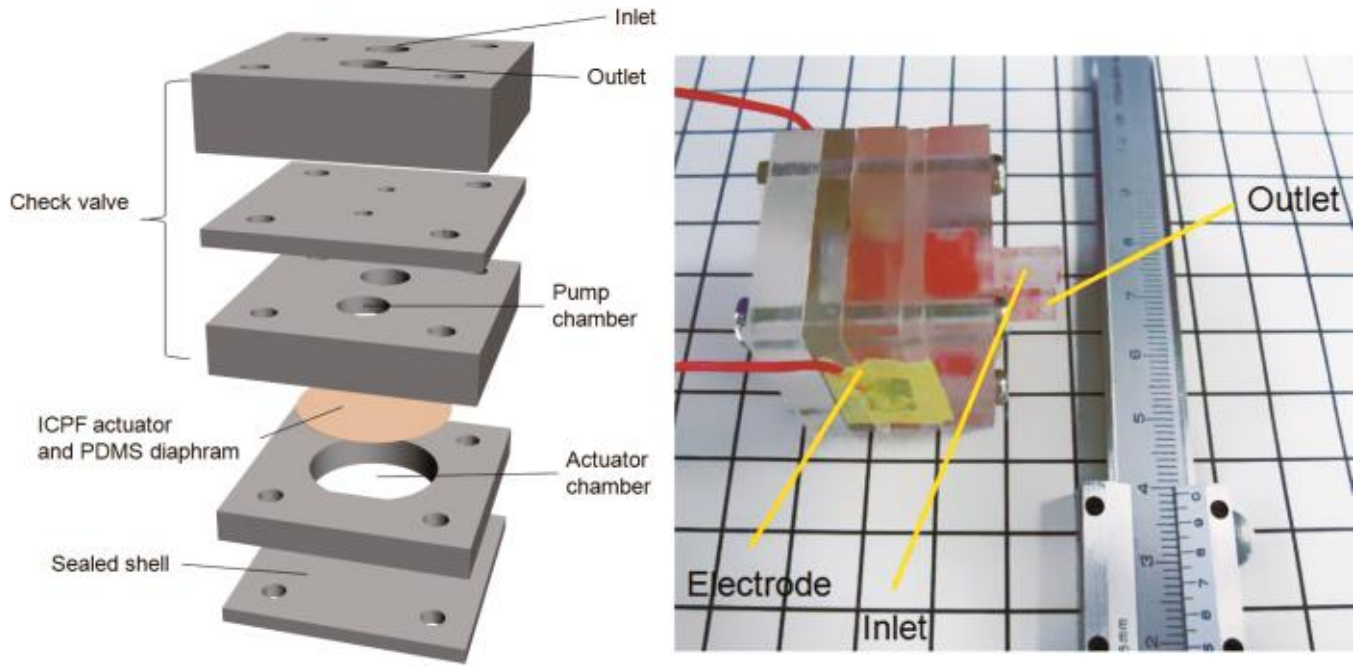

**Fig. 17. The presented micropump in[94], (left) exploded 3D schematic, (right) integrated version.**

### *IPMC-based Micropump 7*

The main idea of this type of micropumps is not novel, and Fang and Tan [95] and Wei and Guo[94] previously proposed it. But it has an additional optimization stage in comparison to previous works that we discuss it here. In order to increase the displacement of the IPMC actuator and the pumping capability of the micropump, Nam and Ahn [96] also have proposed an idea like Fang and Tan [95] and Wei and Guo[94]. They have presented a petal-shaped IPMC diaphragm using four pieces of IPMC actuators instead of a single piece. This structure for a disk shape has been depicted in **Fig.18** where four quarter-disk-shaped (diced disk) IPMC actuators are attached by two thin elastic films (Polytetrafluoroethylene(PTFE)) in both sides. In this structure, the IPMCs are only clamped on one edge, and they are free to bend and making an up-down movement in the elastic layers. This structure can avoid the wasting of energy due to stretching of the membrane, and by an ANSYS analysis, it has been proven that the diced disk is better than a disk as a pump diaphragm. This structure causes a lower strain ratio, and it is more durable with larger displacement, higher volumetric change, and hence higher pumping ability. Moreover, it was also shown that a diced square-shaped diaphragm like that in **Fig.19** could work better than diced disk diaphragm, meaning that at the lowest strain ratio it can generate the highest displacement and volumetric change.

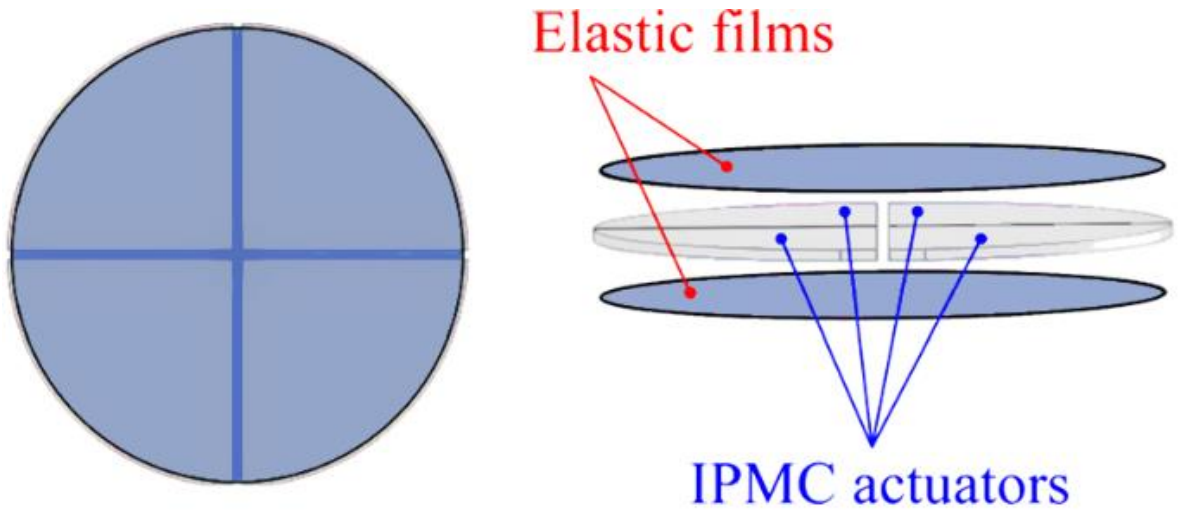

**Fig. 18. Disk-shaped IPMC-based diaphragm. Left: Top View. Right: exploded view[96].**

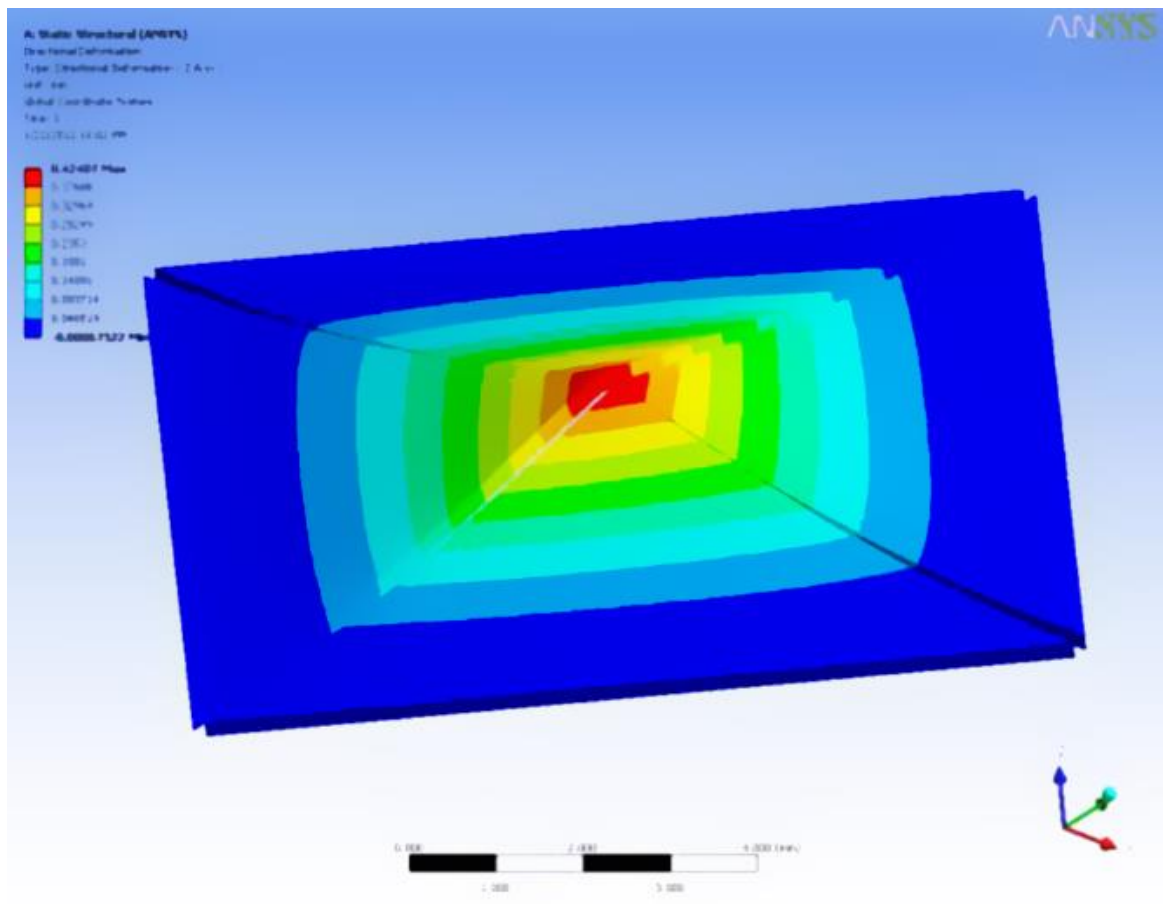

**Fig. 19. Diced square -shaped diaphragm[96].**

In this work, using the proposed IPMC-based square-shaped diaphragm, a micropump has been created. Structure of this micropump is shown in **Fig.20** where it has two inlets and two outlets, and by bending of the IPMC-based diaphragm the volumetric size of the chamber will be changed, and so the pumping is resulted. The photograph of the fabricated version of this micropump is shown in **Fig.21,** and it has been tested for the verity of the input voltages (1-5 V) and frequencies (0.05-9 Hz). As another IPMC-based application, this micropump also works better in low frequencies (Here less than 3 Hz), and it has enough potential to be a micropump for microfluidic devices.

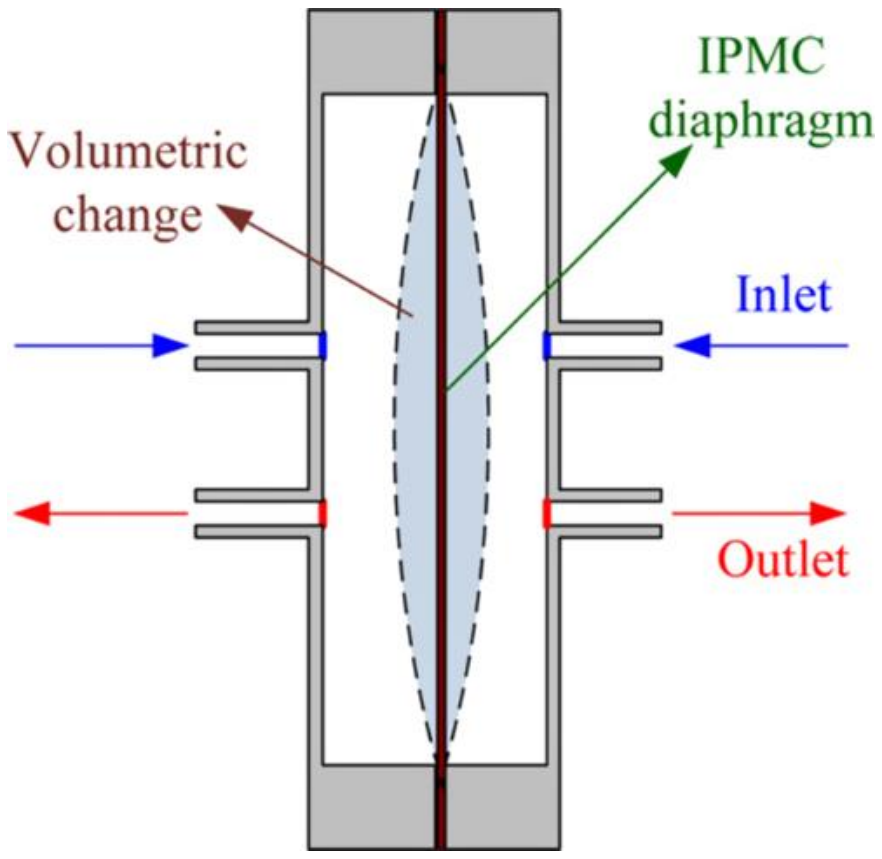

**Fig. 20. Schematic diagram of the micropump presented in [96].**

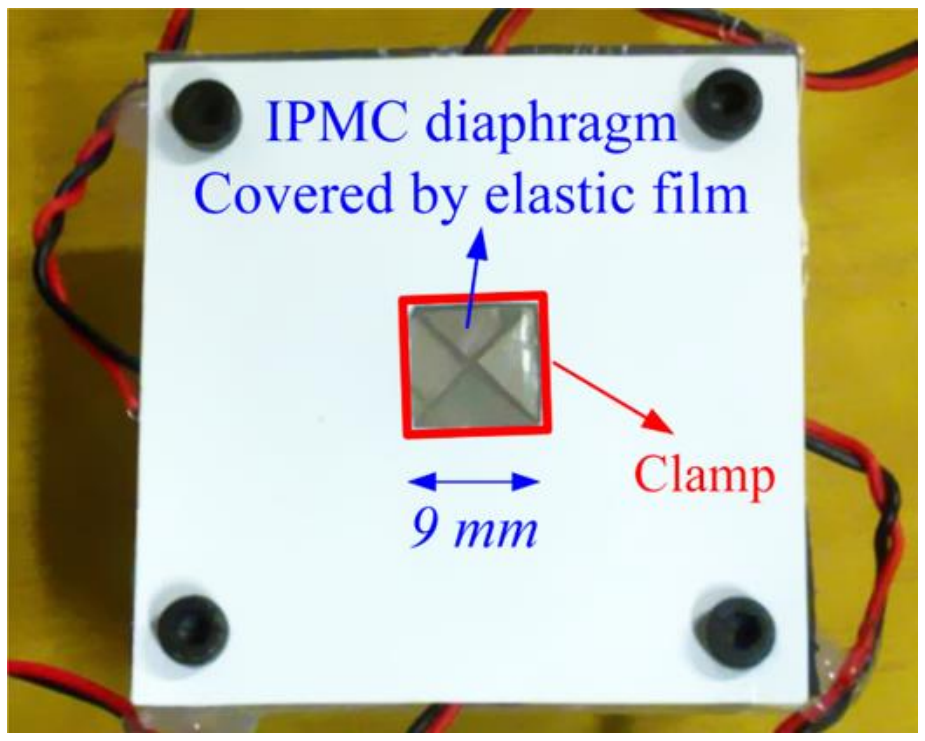

**Fig. 21. Fabricated diced square-shaped diaphragm[96].**

### *IPMC-based Micropump 8*

In all of the previous IPMC-based micropumps, a disk-shaped or petal-shaped IPMC was used as diaphragm or active element of the micropump, but McDaid et al. [97] presented a new version of IPMC-based micropump using the IPMC in its standard cantilever form. The authors believe the major issue in integrating IPMC in microfluidic devices is its control in order to have a reliable and durable actuation over a long period and many cycles. Talking about the control part of the device in [97] is not of consideration in this study, and we just discuss the type of the microfluidics micropump that they have presented. The structure of the proposed micropump is shown in **Fig.22**. As shown in the figure, there is no diaphragm, and the pumping actuator is an IPMC in its cantilever beam form. The device has been fabricated by four layers of Perspex (Poly(methyl methacrylate)) for making the inlet channel, the outlet channel, the pump chamber where the IPMC actuator is embedded in it, and the bottom layer as the tank that allows the IPMC remains hydrated throughout all experiments and hence it can work durable over a long time. A thin layer of latex also has been attached to the bottom of the pump chamber, so when the IPMC is deforming, the volume of the pump chamber also will be changed. Hence, a pressure is produced, and pumping of the fluid is obtained. In order to have a continuous pumping actuation, a sinusoid IPMC displacement is needed. By an open loop test, the optimum frequency to have the highest pumping rate was obtained around 0.1 Hz where this result is in line with [93]. Of course the high pumping rate is not the desire of all applications. For example, in some drug delivery microfluidic chips, the lower pumping rates are needed. Hence the proposed controller can work by lower frequencies too, for example, 0.05 Hz.

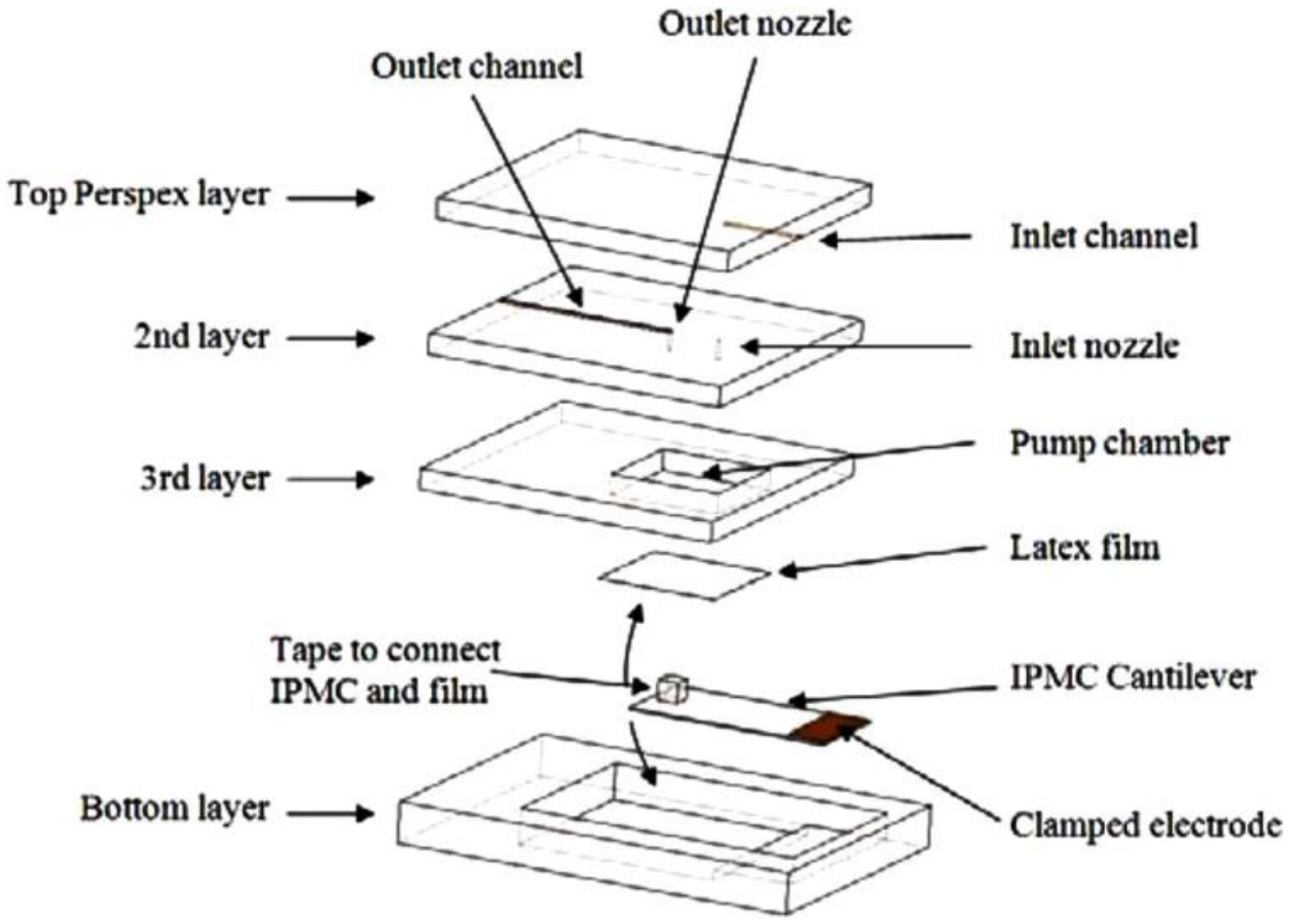

**Fig. 22. Layers and components of designed micropump in [97].**

### *IPMC-based Micropump 9*

In another paper, shoji has presented an idea like the previous discussed ones [98]. He has fabricated a disk-shaped IPMC diaphragm again and used it in a micropump. There is not any novel point in the diaphragm that it has presented in this paper but the proposed structure of the micropump is different, and due to its valve-based performance it is more reliable. However, this micropump is more complicated to fabricate. As shown in **Fig.23** this micropimp was designed by thirteen layers of Acrylic in order to make a 3D flow path inside the body of the micropump. The disk-shaped IPMC diaphragm has been embedded at the bottom of the micropump, and as depicted in **Fig.24,** it is clamped by two stainless steel flat gaskets (Like the method presented by Santos et al. [93]). If we apply a square waveform with amplitudes of +2.0 V and −2.0 V to it, the micropump can work by the flow rate of ~5 µl/s (~300 µl/min) with no water leakage.

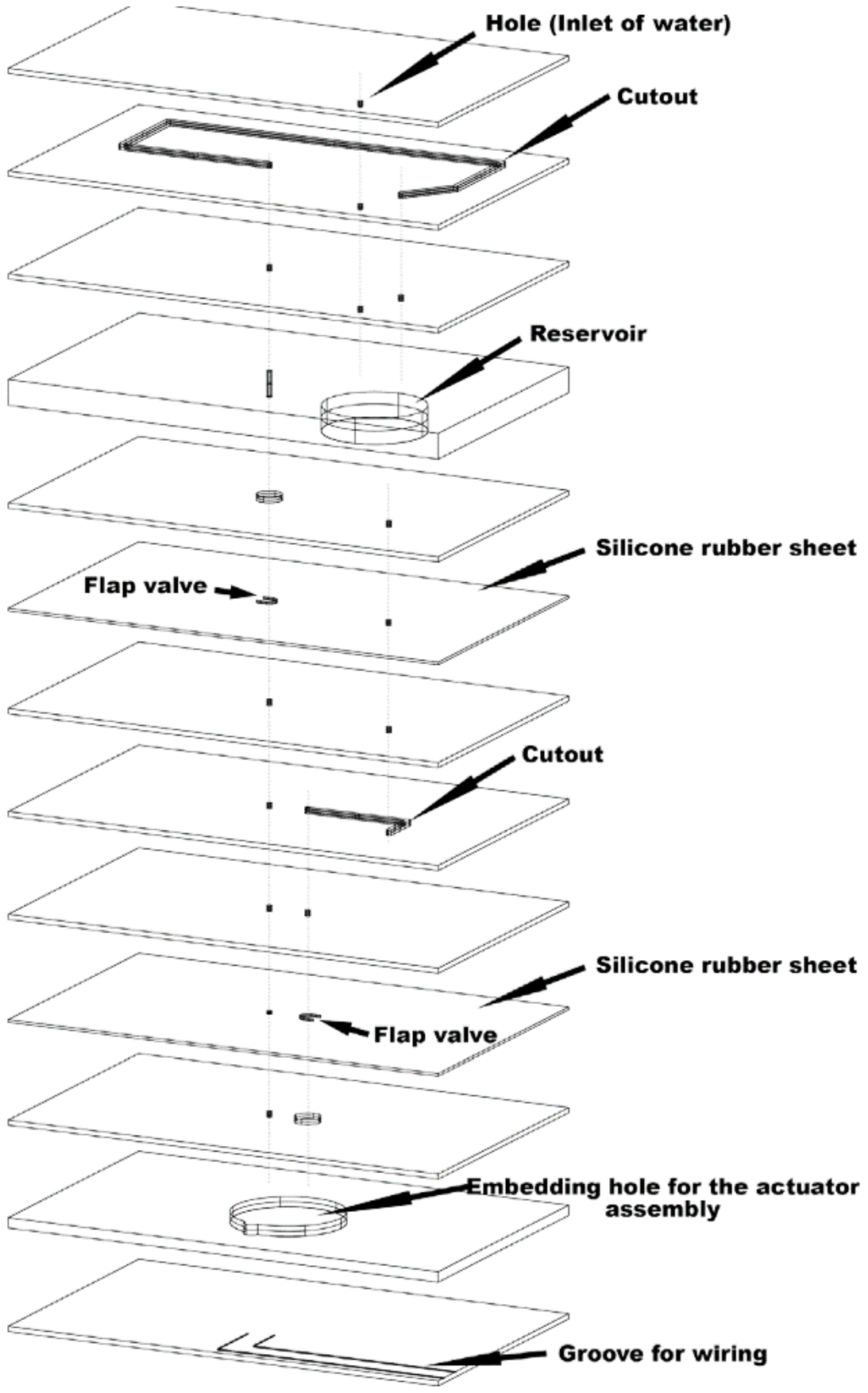

**Fig. 23. Thirteen layers of Acrylic to make a three-dimensional flow path for micropump. The IPMC diaphragm is embedded in this structure[98].**

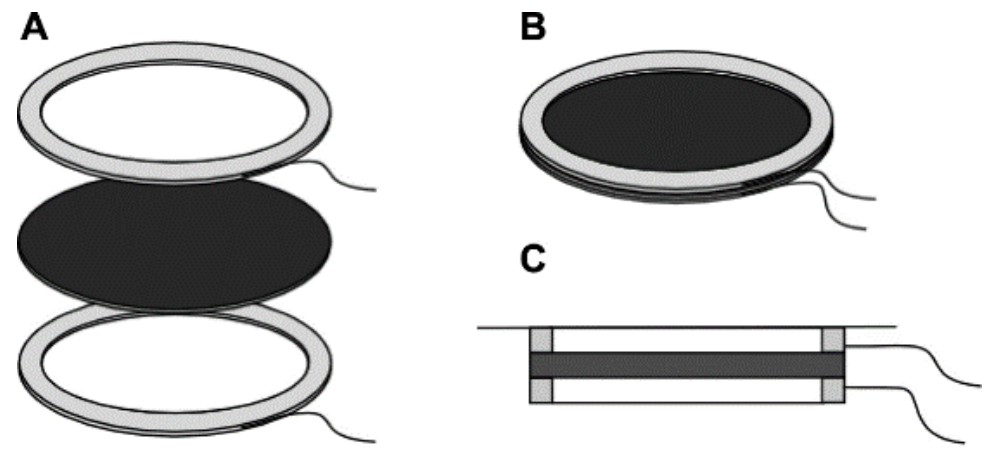

**Fig. 24. Sandwiched structure of the IPMC diaphragm. The black disk is IPMC diaphragm where it has been clamped between two stainless steel flat gaskets (gray gaskets ). (A) exploded view, (B) compact view, (C) side view[98].**

### *IPMC-based Micropump 10*

In order to design an efficient micropump for drug delivery usage, Wang et al. [99] have presented a different type of IPMC-based micropump. This pump is the first experimentally investigated double chamber valve less IPMC-based pump that doesn't have any conventional IPMC diaphragm unlike to the most of the previous samples. It has a gold coated cantilever beam formed IPMC as an actuator, and by bending this actuator into the chambers, it can make the required pumping pressure. The expanded view of this pump is depicted in **Fig.25-a**. The pump's diaphragm has been made from PDMS and the IPMC actuator (**Fig.25-c**) makes the required deflection of this diaphragm. The pump has two inductive coil sensors that have been fabricated by two layers of the printed circuit board or the same PCB (Green). There are also two chambers that are made using the fused deposition method (FDM) by 3D printer (Yellow). To apply the electric voltage to the IPMC by the standard connectors, two electrodes with soldering ability are needed that can be made simply by PCB (Red). The diffuser/nozzle elements (Green) are very sensitive, and it is necessary to make them precisely. Hence they have been printed separately with stereolithography (SLA) method that its resolution is 50 micron which is good enough for diffuser/nozzle elements. The dimension of the diffuser/nozzle also is critical to have optimum pumping ability, and to find this dimension the authors calculated it in their previous simulation works [100, 101]. Based on their simulation they have found the dimension of the diffuser/nozzle of this micropump as written in the **Fig.25-b**.

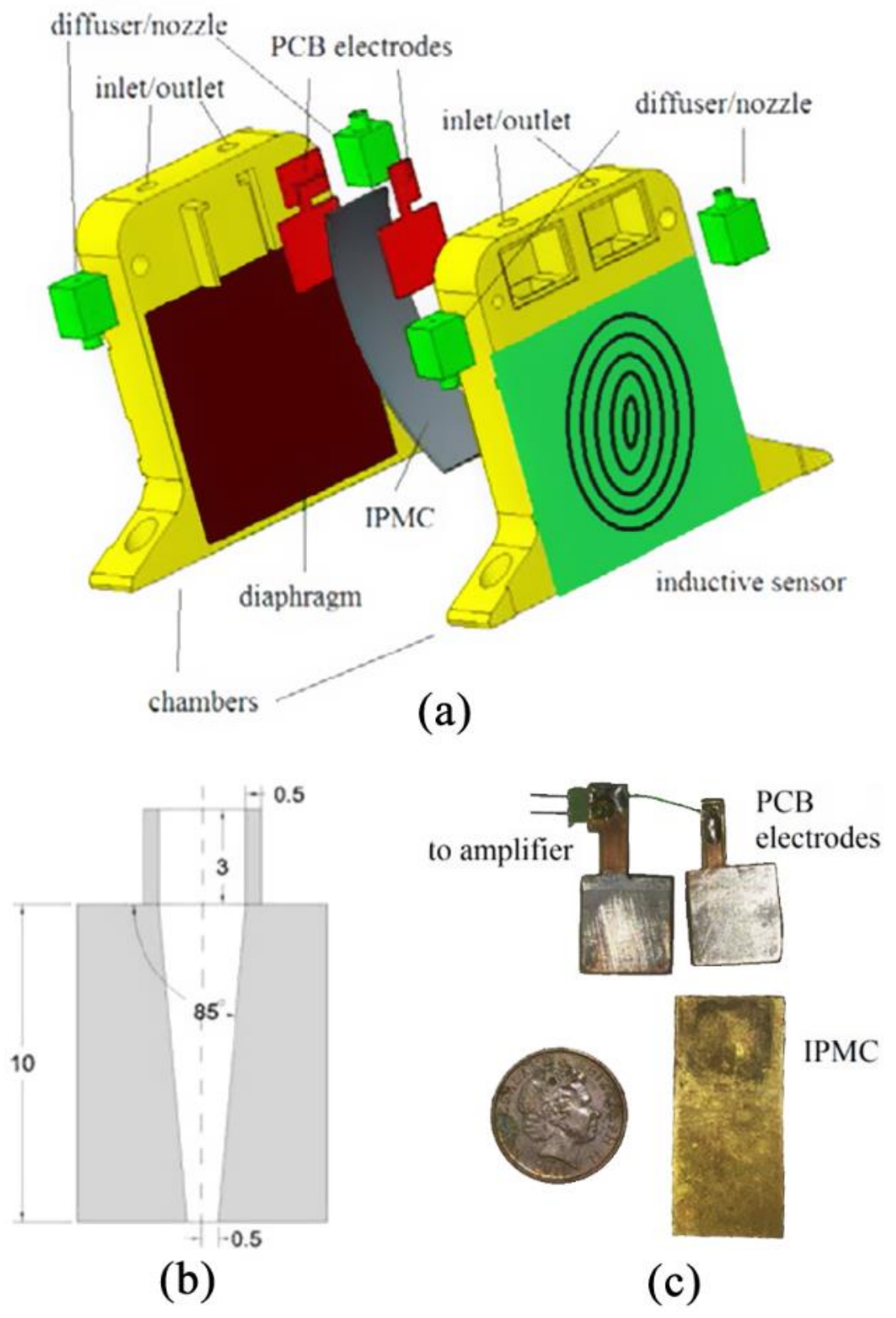

**Fig. 25. (a) 3D schematic of the micropump structure, (b) diffuser dimension in mm and (c) IPMC and its PCB electrodes[99].**

Although IPMCs are promising actuators, they also have some constraints like dehydration, hysteresis, and back-relaxation [56, 57] effects that without solving them we cannot use IPMCs in most practical applications for a long period. To overcome these drawbacks, two adaptive controllers, a PID controller with iterative feedback tuning (IFT), and repetitive controller (RC) has been designed for this micropump. The experimental results of this work show that the micropump can work continuously (for ~2 hours) with the maximum flow rate of 780µl/min (In response to 2V and 0.1 to 0.5 Hz) which is suitable for long time drug delivery usage.

***IPMC-based Micropump 11***

The last and newest IPMC-based micropump is made by Wang and Sugino [102]. As is shown in **Fig.26,** their idea has a new inner petal-shaped IPMC as the actuator of the pump's diaphragm. This actuator can make the required deformation of the diaphragm to provide the appropriate flow rate and the consistent back pressure for self-contained microfluidic chips. This work is a new idea that there is not yet a detailed article for it and we can only find the main idea and some general data about it in a book chapter [102]. The IPMC actuator of this micropump has been fabricated by Nafion 117 membrane and Pt electrodes, and to apply the electric voltage to it a pair of copper electrodes were clamped to both sides of the inner petal-shaped IPMC. As shown in **Fig.26-a,** the structure of the micropump has four layers of a transparent polymeric material (the type of material has not been reported but based on the photographs of the fabricated version it seems to be made of PMMA or PDMS). In the two bottom layers, the channels, the pump chamber, and the flap valve were engraved, and the two top layers are the sandwiched supporter for the inner petal-shaped IPMC actuator. It should be noted that the inner petal-shaped IPMC is a hollow actuator and inevitably it has been attached to a thin and soft layer of a material (maybe PDMS or latex) as the diaphragm. The experimental results of this work show that this micropump can pump the fluids by the flow rate of 162 to 1611 µl/min in response to 0.5 to 3 V at 1 Hz. The back pressure on the micropump also was measured as high as 71 mm-$H_2O$ in response to 3 V at 1 Hz.

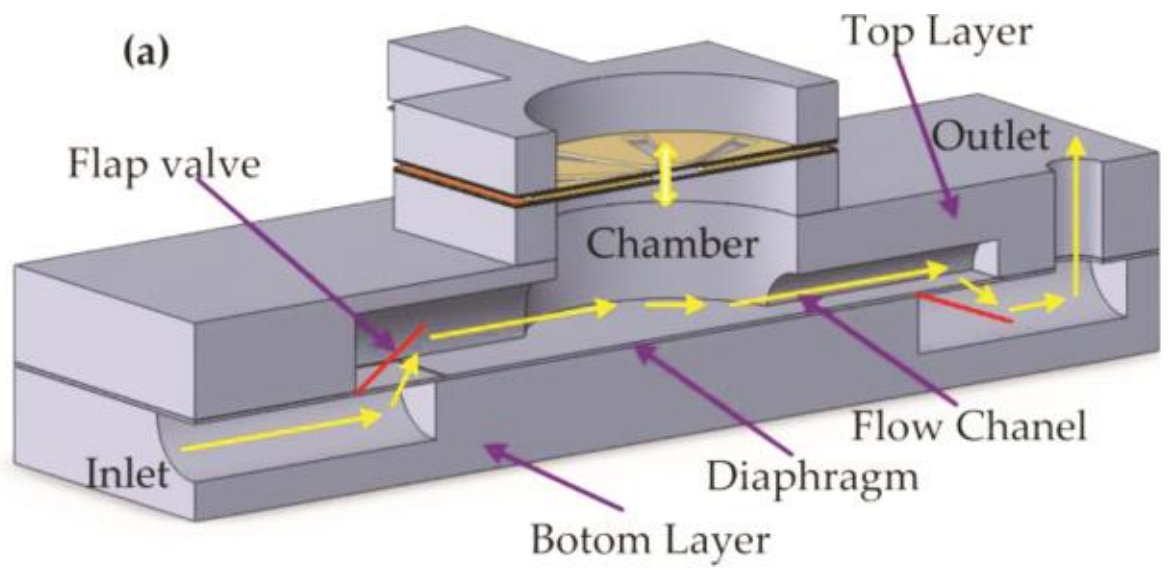

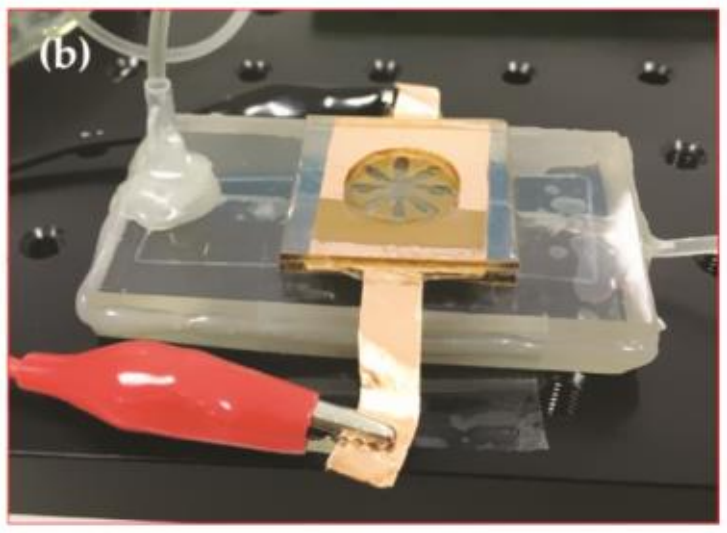

**Fig. 26. (a) 3D schematic of proposed micropump in [102] and (b) its fabricated version.**

*D. IPMC-based microfluidic microvalves*

As discussed before, most of the microfluidic-based applications of IPMC are related to micropumps, but it also has promising potential to play the role of a microvalve for microfluidic devices. So far, just four papers have been published about IPMC-based microvalves. In this work we review all the presented research on this topic as follows.

***IPMC-based Microvalve 1***

Previously in the section of *"IPMC-based Micropump 1"* we talked about two papers on IPMC-based micropumps [87, 88] that they were related to IPMC-based microvalves too. Hence we can say that the first IPMC-based microvalve was presented in that part and we don't describe it again here. In continue, we discuss about two other works that are entirely related to IPMC-based microvalves.

***IPMC-based Microvalve 2***

The first IPMC-based microvalve was presented by Yun et al. [31]. The idea behind this microvalve is simple, and as shown in **Fig.27,** an IPMC actuator has been embedded in a microfluidic chip, which can move a piece of PDMS in the microchannel. IPMC actuator is attached to this PDMS part, and in the normal state, when the input voltage of the IPMC is off, this part is into the microchannel and it means that the channel is closed. But if we apply an input voltage by a proper polarity, the IPMC will be bent up, and then the valve piece will be pulled up meaning that the channel is open.

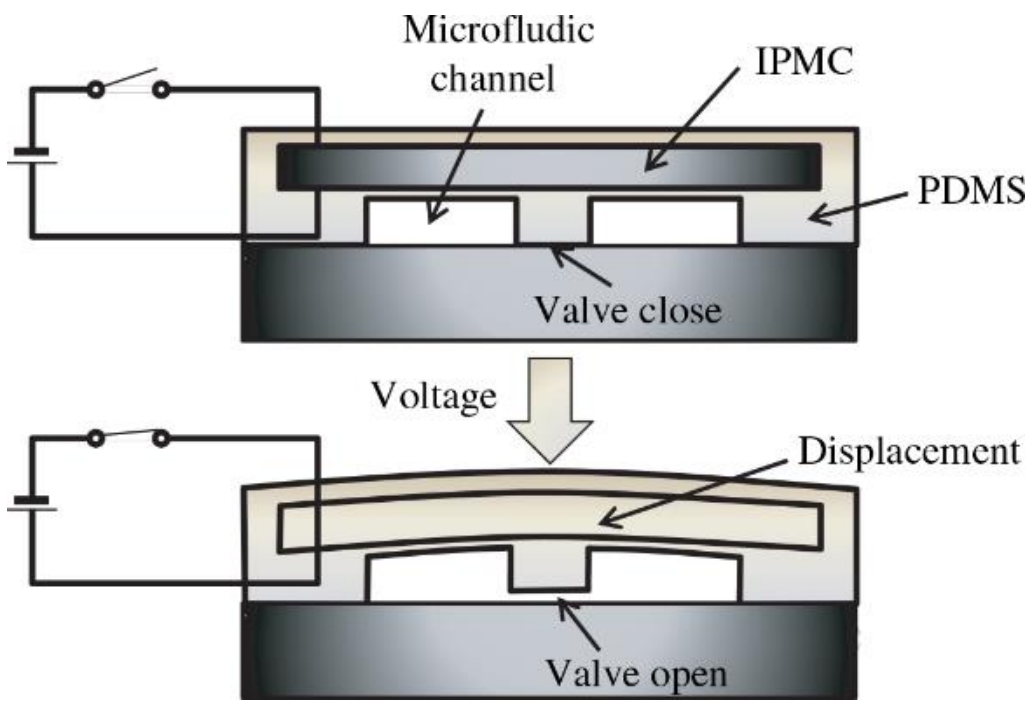

**Fig. 27. The working principle of IPMC-based microvalve systems in[31].**

The used IPMC in this work is a Nafion based cantilever beam (Nafion 1110) formed IPMC that has Pt electrodes, and during the routing fabrication procedure of microfluidic chip it will be embedded into the chip. The fabrication procedure of this microvalve has been depicted in **Fig.28**. As it is seen here, the four first stages are exactly the same routine fabrication process of the microfluidic chips (SU-8 based photolithography, baking, PDMS curing, plasma treatment, etc.). After making the required microfluidic chip, the IPMC actuator should be fixed in its proper area, and then an additional PDMS mixture should be poured onto the IPMC and keeping it in the oven for 6 hours at 65$^{o}$C to have a compact and seamless device.

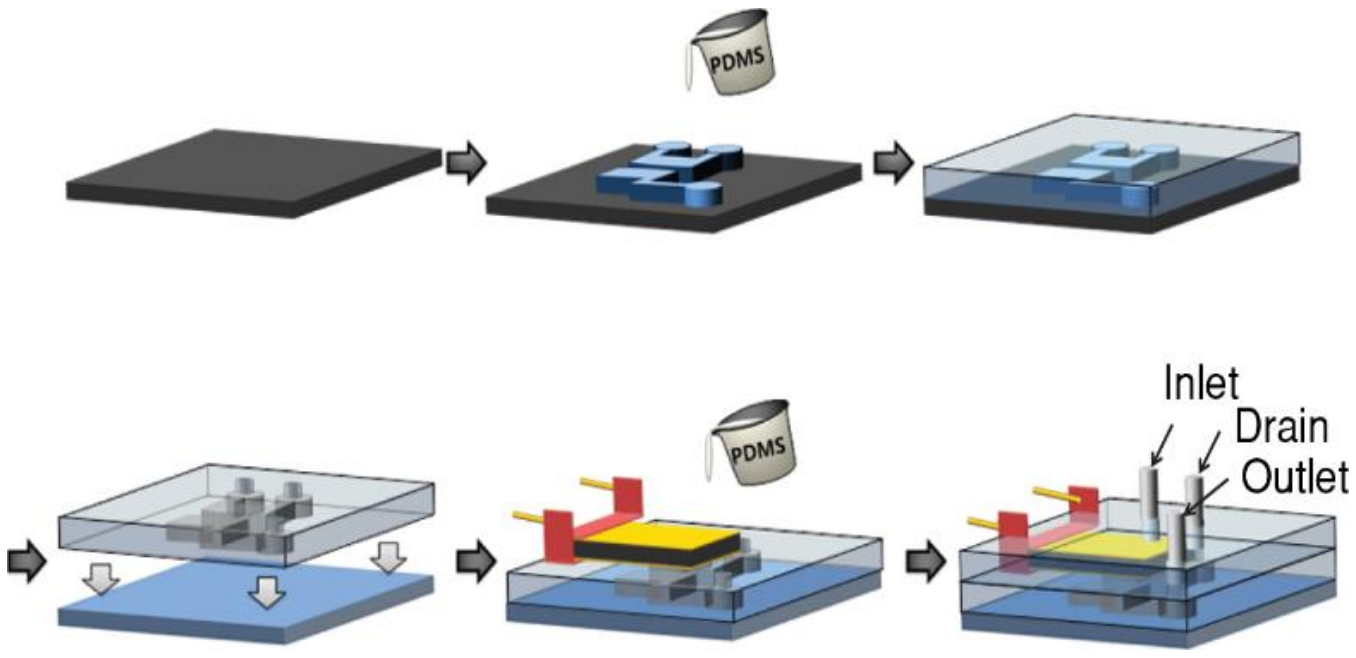

**Fig. 28. Fabrication procedure of proposed IPMC-based microvalve in[31].**

As shown in **Fig.29**, this microvalve should pass the fluid from the Inlet to the Outlet when the valve is open and should send the fluid to the drain when the valve is closed. This operation has been experimentally proven, and the microvalve can work properly. In the fluorescence images of polystyrene bead flow (0.01 ml/min) taken from the microvalve operation (**Fig.30**), we can see the results are matched with the assumption, and the microvalve is working properly. The proper input voltage amplitude for this microvalve is 5 V, and in response to this voltage the IPMC displacement and consequently the valve displacement is in the range of 80-150 μm. This is while the depth of the microchannel is about 10 μm and it means that the IPMC displacement is absolutely enough for this aim and the valve's operation is reliable for the microfluidic applications. Also, its open/close switching is good enough, but to have better and more durable microvalve; it is necessary to improve the current version of the IPMC, especially looking for the improvements in its back relaxation, it's on/off responding speed, and its capability to work for a long time.

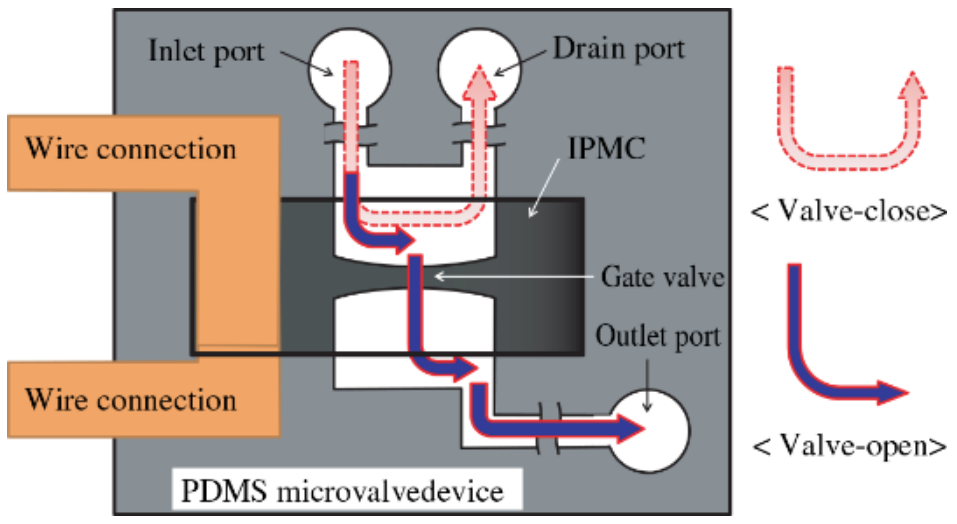

**Fig. 29. The schematic working principle of the microvalve in the microfluidic system[31].**

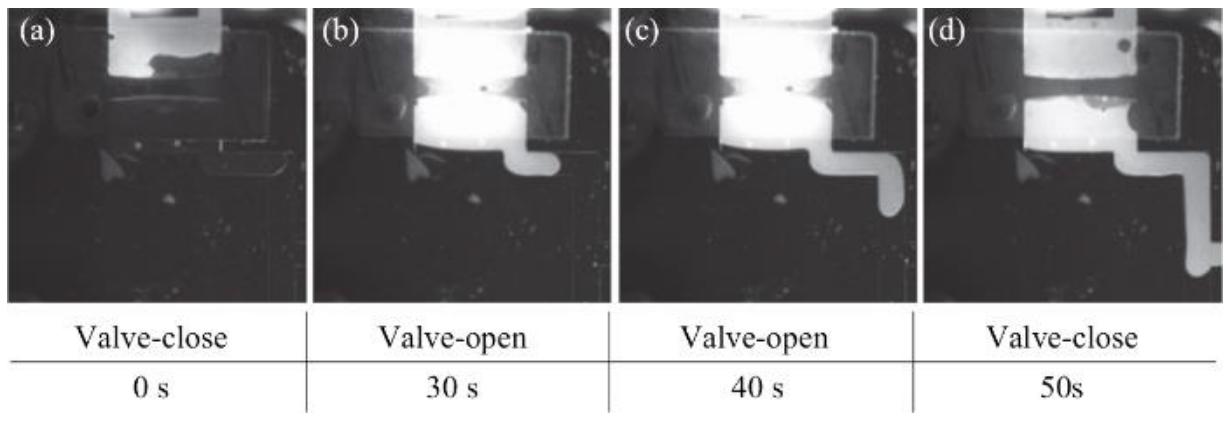

**Fig. 30. Fluorescence image of polystyrene bead flow (0.01 ml/min) in the microfluidic chip during microvalve operation[31].**

### *IPMC-based Microvalve 3*

Another IPMC-based valve [103] is the more complete version of the previous one that by the same team was published in 2013 as a US patent. As you can see in **Fig.31,** the working principle is exactly the same as previous one that was presented in [31].

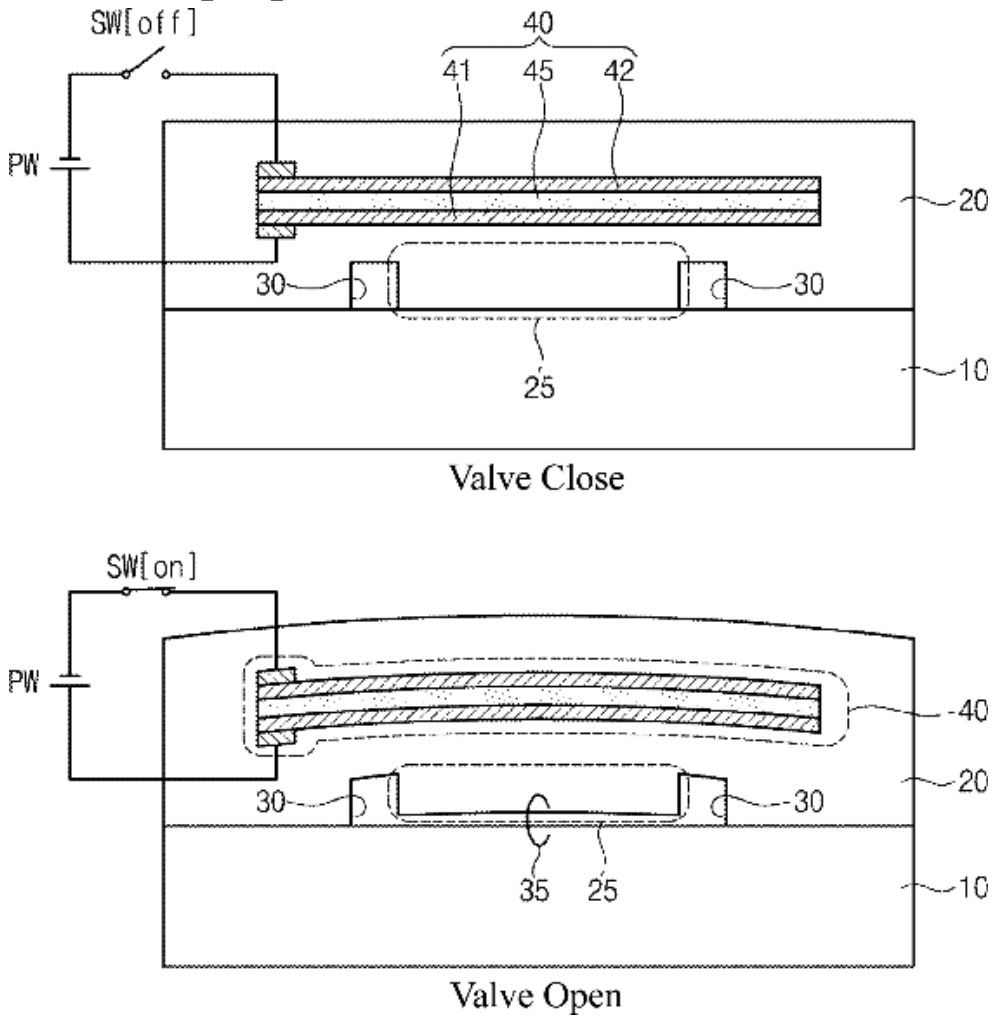

**Fig. 31. The working principle of IPMC-based microvalve systems in [103]. (10 and 20 ) PDMS layers, (25) PDMS ball valve,(40) IPMC actuator, (41,42) Pt electrodes, (45) Nafion membrane, (30) Microchannel[103].**

The main difference of this work with [31] is that the authors have developed the previously presented microvalve for multivalves microfluidic and Lab-on-a-chip (LOC) devices. For example, they have presented a two valves lab-on-a-chip device as shown in **Fig.32**. Referring to this device, the top PDMS layer (20), which makes two channels (301 and 302) spaced apart from each other, can be formed on a substrate (10). Besides, this PDMS layer (20) can be formed as the third channel (303) between channels 301 and 302 where the path of this channel (303) is controlled by two IPMC-based microvalves embedded in the device. They also believe this IPMC-based microvalve can work in the very complex microfluidic and Lab-on-a-chip devices. For example as depicted in **Fig.33,** they have developed a 16 valves LOC device where 16 IPMC-based microvalves (42) can control the flow of fluids between 16 microchannels as well as the inlet and the outlet. To control these microvalves a control unit (90) has been attached to the LOC device where it sends its control signals to the microvalves through a conductive transmission line (71) [103].

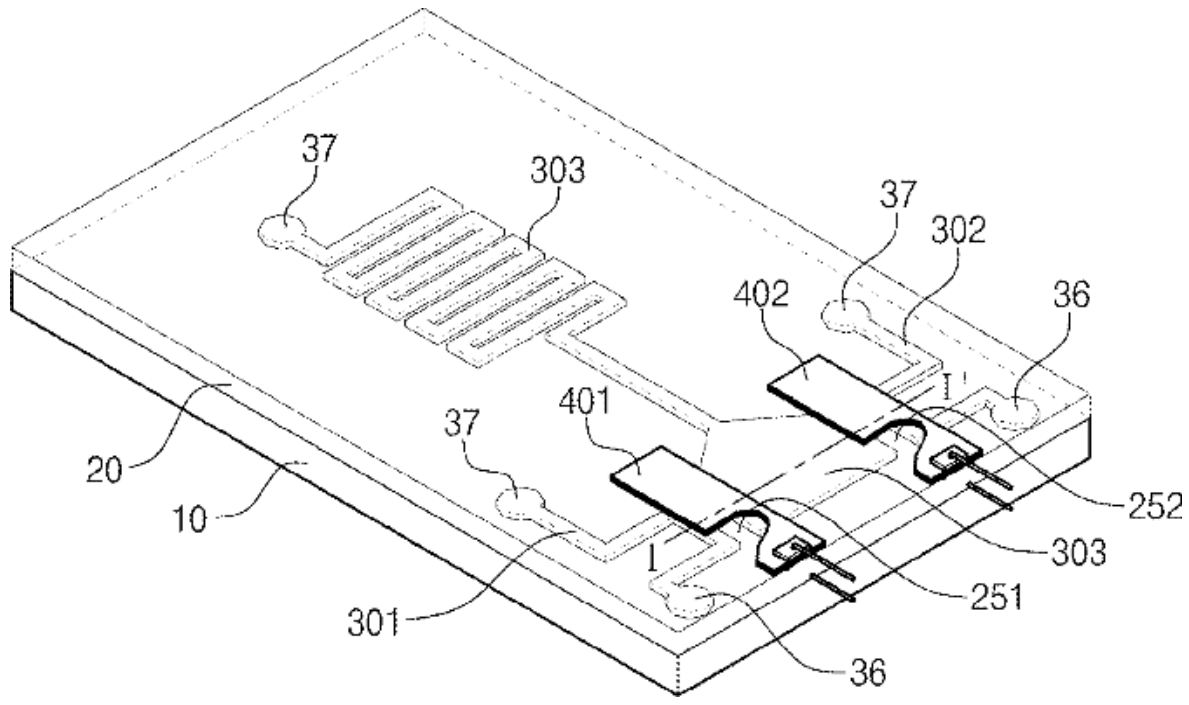

**Fig. 32. Two valves LOC device using two IPMC-based microvalves[103].**

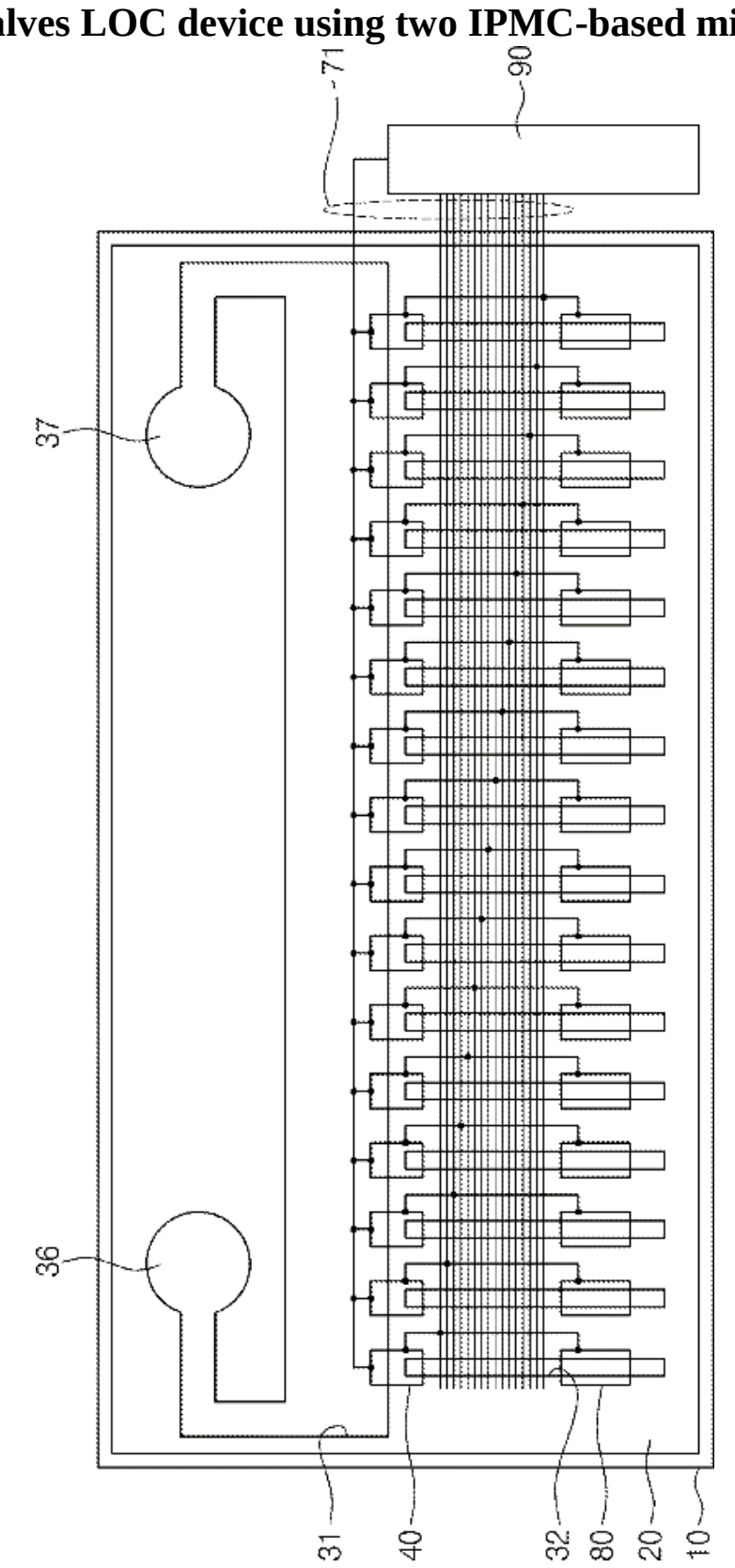

**Fig. 33. Sixteen valves LOC device using 16 IPMC-based microvalves (40) [103].**

## *E. IPMC-based microfluidic micromixers*

For the IPMC-based micromixers, the same as for microvalves, and we cannot find more than two works [104, 105]. But it also has great potential to be an appropriate choice for active microfluidic micromixer, especially if we can miniaturize the IPMC to a thinner and smaller actuator. In this part, we describe the ideas behind the two works that they have ever been presented.

### *IPMC-based Micromixer 1*

The work in [104] presents a novel active laminar mixing method for improving mixing procedure in the fluids that have low Reynolds numbers. The main target of this paper is using the proposed active mixer for microfluidic devices, but the size that they have validated in their work is bigger than the microfluidic device. But the idea is feasible and maybe it can be developable to make practical microfluidic device. This active mixer is the first mixer that has used Ionic Polymer Transducer (IPT) or the same IPMC actuator as a mixing element. As shown in **Fig.34,** the working principle of this active mixer is that a cantilevered form of an IPMC actuator is embedded in the channel. Then by applying some periodic or varying voltage like pules train, sinusoidal wave, and etc., we can make a bending swing in the IPMC and it can mix the laminar fluids [104].

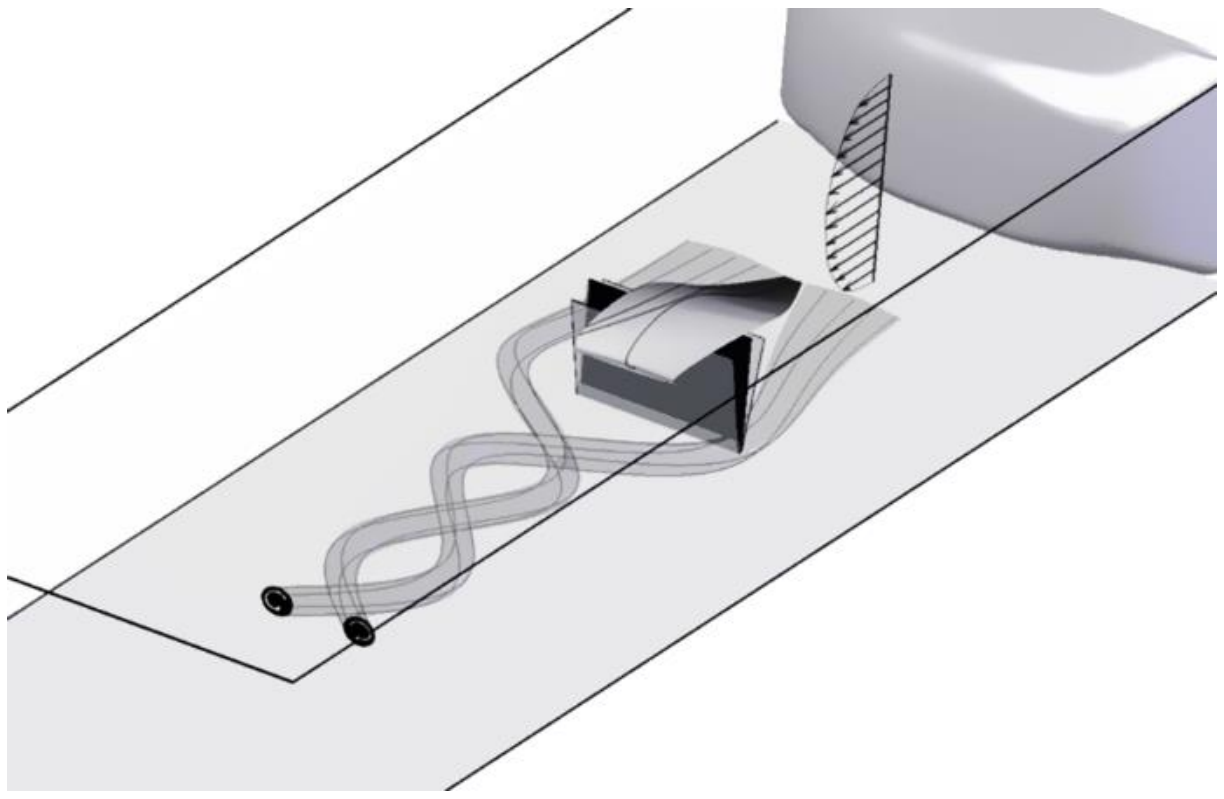

**Fig. 34. Conceptual schematic of stream tubes flowing past an orthogonally oriented IPMC[104].**

As depicted in **Fig.35**, IPMC has been embedded into the channel in two orientations, parallel and perpendicular to the incident flow, and both states were tested in a straight rectangular channel at a Reynolds number 10. The used IPMC in this study was an improved Nafion based cantilevered beam of 6.5mm height and 8.5mm width, that it has been embedded in a bigger channel. It is obvious that this size is much bigger than a normal microfluidic chip. This work is presenting an idea for millifluidic systems. By the way, in response to a 1.5 V amplitude at 1 Hz frequency square wave, the mixing rate and flow characteristics of this system were quantified using planar Digital Particle Image Velocimetry (DPIV). The results were measured at several locations downstream of the IPMC. Mixing potential was measured in the range of 150% to over 375% when compared to a baseline case. This shows that the proposed IPMC mixer has enough potential to be a mixer for laminar fluids. The current version of this mixer is not usable for the microfluidic systems, but if we miniaturize the IPMC to a thinner and smaller actuator, it may be a feasible idea for microfluidic systems [104].

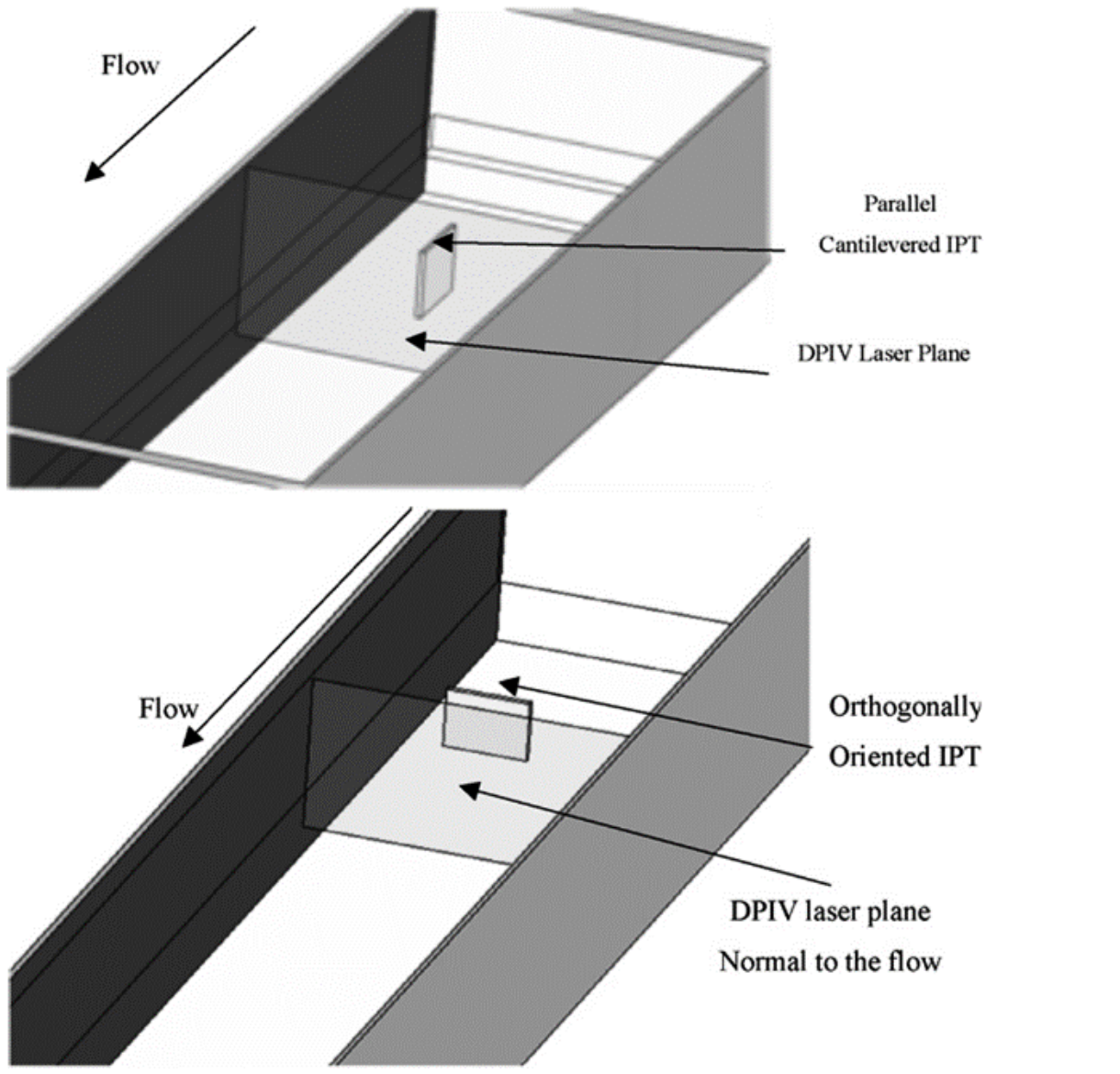

**Fig. 35. The schematic view of the relative orientation of the IPMC concerning the laser plane for, (Up) the parallel IPMC and (Down) the orthogonal IPMC cases [104].**

### *IPMC-based Micromixer 2*

In the work of C. Meis et al. [105], an active microfluidic mixer prototype is proposed using Nafion based IPMC as artificial cilia. The working principle of this micromixer is also almost like the previous one and to mix the low Reynolds number fluids, the IPMC makes a perturbation on them, and this perturbation creates a localized flow pattern disruptions in their laminar flow regime. The 3D schematic of the device is shown in **Fig.36** that is fabricated by PDMS (Sylgard 184 – Dow Corning) using easy and fast "ProtoFlo" fabrication method [105]. In this device, a straight microchannel with a T-junction was designed (5000 μm width × 400 μm height). The utilized IPMC for this work is different, and it is very thin, and some improvements have been applied to it. For example instead of widely used Nafion 117 (~200 μm thick), the Nafion 211 (~25 μm thick) were used as ionomer (membrane). Of course, two layers of polycations and polyanions, poly-(allylamine hydrochloride) (PAH) and gold nanoparticles, were also deposited respectively to both sides of Nafion 211. Another improvement was about the enrichment of the ionomer with 1-ethyl-3-methylimidazolium trifluoromethane sulfonate (EMI-Tf) ionic liquid. Finally, instead of Pt layers, a 50 nm thick highly conductive gold leaf was hot-pressed on either side for electrodes. In order to integrate the IPMC actuator into the microchannel, it is necessary to cover the IPMC against sustaining damage due to the fluids, which also protects from the water absorption and consequently avoiding expansion of the Nafion membrane since the Nafion expansion causes the detaching of the gold leaf electrodes. The protective cover was developed using a thin HDPE/LDPE blend sheet (Product #618574, Berry Plastics Corporation, Evansville IN). In order to assemble the micromixer, the encapsulated IPMC was sandwiched between two activated (using oxygen plasma treatment) PDMS layers. The IPMC actuator was inserted to the channel approximately 3/4 of its width [105].

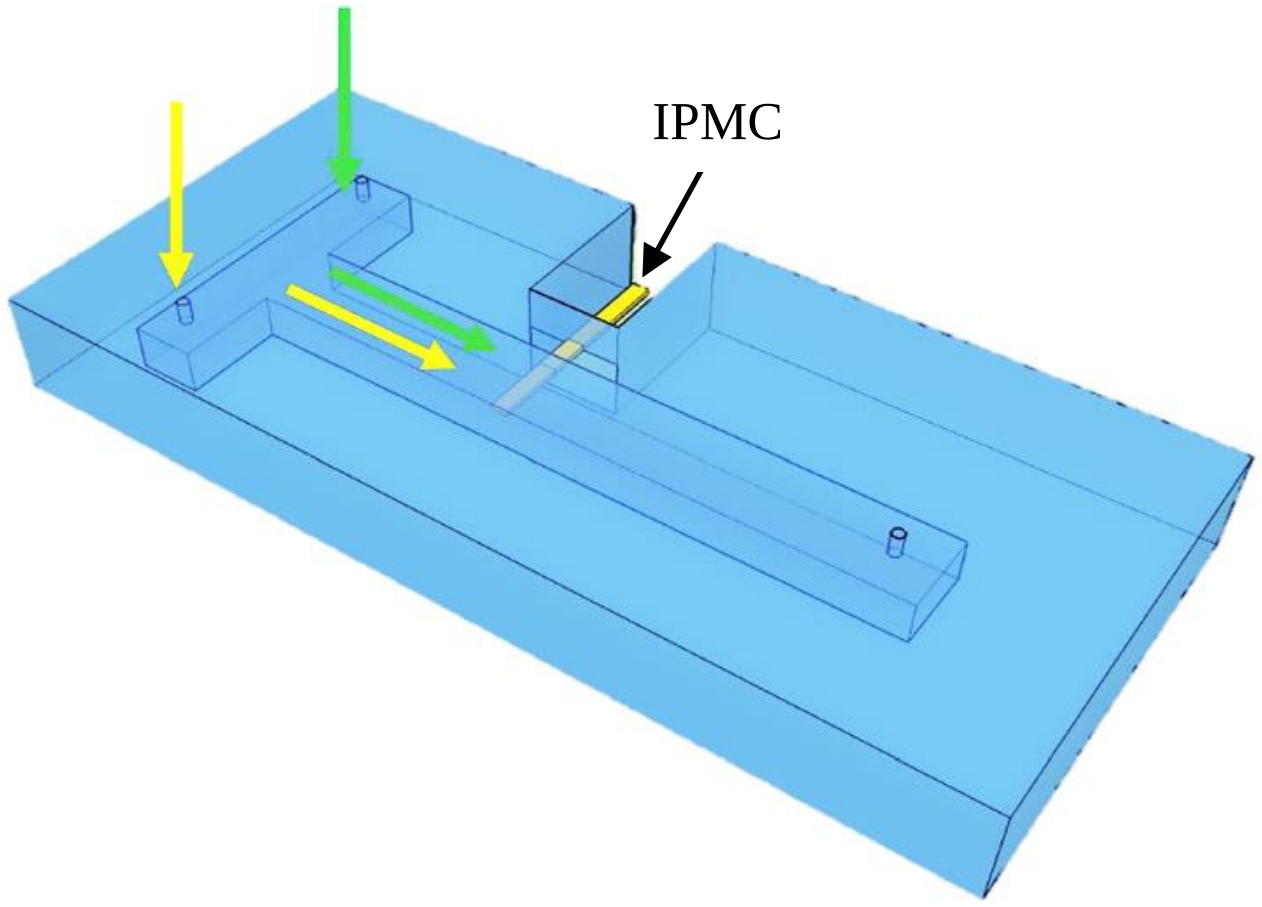

**Fig. 36. 3D Schematic of the proposed microfluidic micromixer presented in[105] .**

For testing the performance of this micromixer, the laminar flow in two different colors was injected into each side of the T-junction, and a square wave function of 4.5 V (9 Vpp) amplitude at 1 Hz frequency was applied to the IPMC. Simultaneously with the laminar flow into the microchannel, the vibration of the IPMC actuator made the fluids mix together where the mixing rate of them was recorded with a charge-coupled device (CCD) camera at a rate of 30 frame per second (fps). Using Adobe Photoshop CC, the recorded video was split to its frames, and by ImageJ (Image Processing and Analysis in Java) free software selected frames were analyzed, and the mixing rate was extracted. The selected frames from t=0s until t=10s (denoted as T0 to T10) and using ImageJ the surface plots of pixel intensity and mean histogram values were produced. The mean histogram values for T0, T5 and T10 have been shown in **Fig.37**. The current IPMC-based micromixer needs to be optimized in order to solve some problems on it especially for reducing the mixing time. But this results showed that IPMC actuators and other similar *i*-EAPs are the viable alternative active microfluidic mixing methods, and they can contribute to the progression of the microfluidic and lab-on-a-chip devices [105].

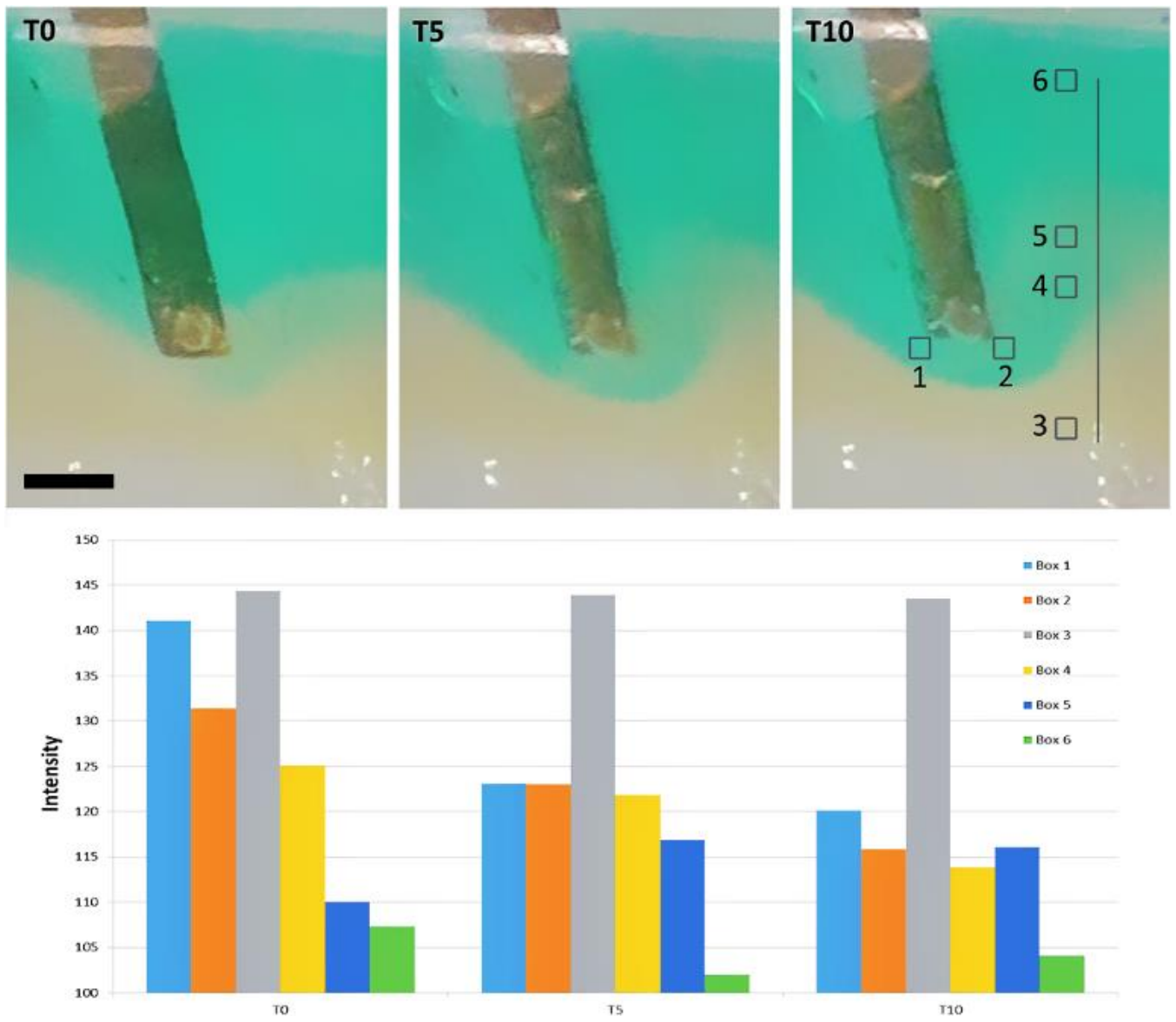

**Fig. 37. Top right: Reference image at T10 of selected pixel regions used for analysis. The plot of mean intensity values generated from histograms of the labeled boxes 1–6 (10 × 10 pixel areas) shown in T10. Lower intensity values correspond to darker colors (green) in the images, and higher intensity values correspond to brighter colors (yellow). Reference images for T0 and T5 are also shown; identical box scheme was used for analysis. Scale bar is 1 mm[105] .**

## III. Conducting / conjugated polymer Actuators

Conducting or sometimes called conjugated polymers (CPs) are groups of semiconducting polymers that will be conductive when they are doped with donor or acceptor ions [33, 106]. As shown in **Fig.38**, polypyrrole (PPy) and polyaniline (PANI) are two main conducting polymers. Conducting polymers have a variety of applications, for example, energy storage devices, LEDs, and electrochromic windows are some of these applications. [33, 106]. Besides the mentioned application for CPs, they also have great potential to be use in *i*-EAP actuators or artificial muscles. In 1996 Baughman et al. [39] presented the CP artificial muscles or the same CP actuators (CPAs) and now they are among mature and practical EAPs, and we can find a variety of novel ideas for CPAs.

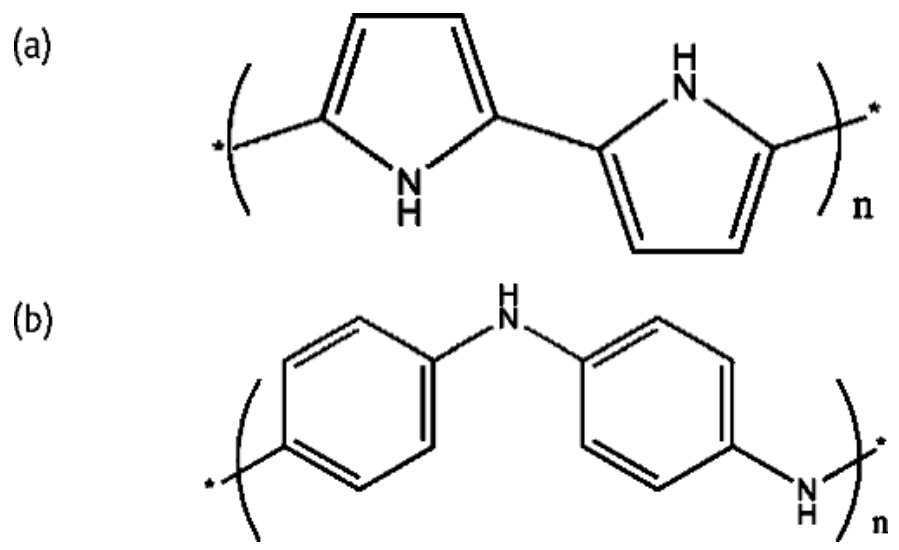

**Fig. 38. The chemical structure of (a) polypyrrole, and (b) polyaniline conducting polymers [33].**

In a general definition, CPA is the composite of several conducting polymers (ionically and electrically conducting) and some ionically (not electrically) conductive layers like PVDF that are sandwiched together. One of the most common CPAs is trilayer CPA. As depicted in **Fig.39,** a normal trilayer

CPA has an ionically conductive membrane (electrically insulator separator) that has been sandwiched between two electrically and ionically conducting polymer electrodes [107]. The mechanism of trilayer CPAs is based on the motion of the ions between the electrodes through the membrane. When we apply an electric voltage to CP electrodes, due to the reduction-oxidation (redox) process and because of the ionic conductivity of the membrane (separator) ions move from one electrode to another one through the membrane. This ion distribution causes a contraction in the ion-donor CP electrode, and an expansion in the ion-accepter CP electrode, and consequently this contraction-expansions make a mechanical bending in the CPA [107].

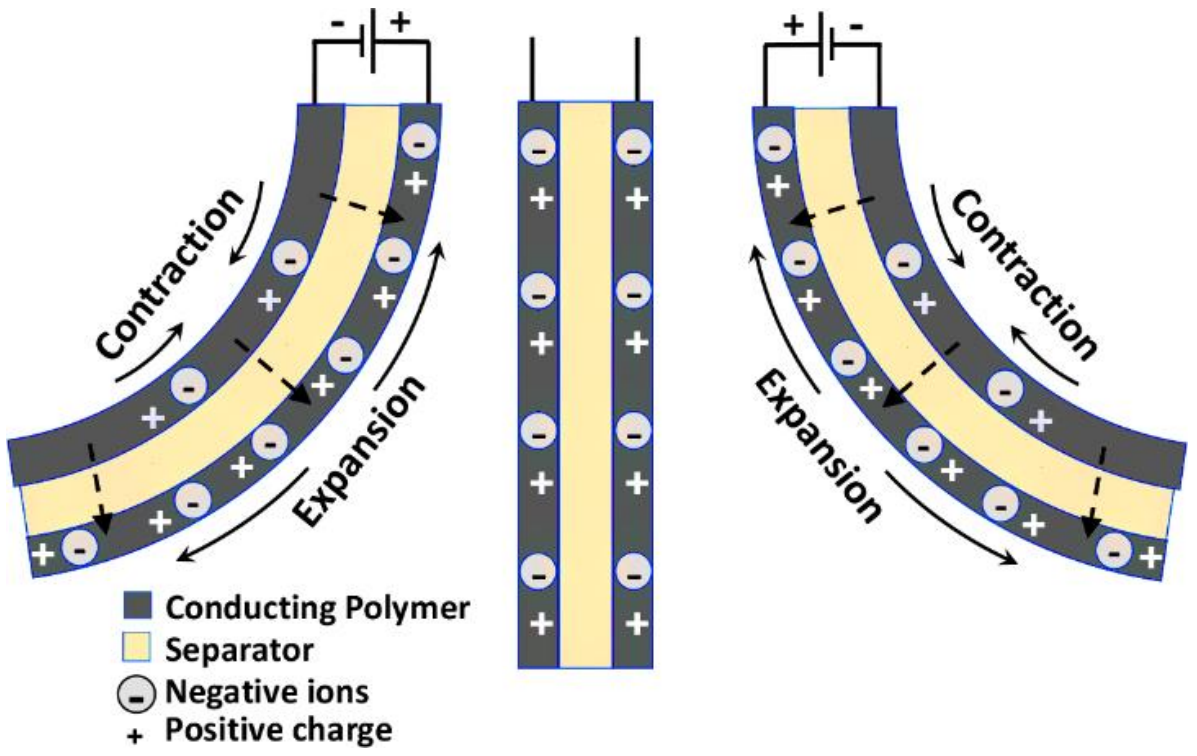

**Fig. 39. Working principles of a conventional trilayer CPA.**

The most common conductive polymer, separator, anion, and electrolyte are PPy, PVDF, $TFSI^-$ (bis(trifluoromethanesulfonyl)imide), and lithium TFSI ($Li^+TFSI^-$) respectively. For these materials the working principle of a trilayer CPA is as follows:

This trilayer CPA has a PVDF membrane (separator) with two electrochemically deposited layers of PPy on both sides that have been doped with $TFSI^-$ anions. PVDF is an amorphous and porous polymer which for trilayer CPA is a substrate as well as a storage tank for the electrolyte. After applying the voltage to the PPy electrodes, the anode side is oxidized (big anions are attracted to the oxidized PPy layer and make it expand) while the cathode one is reduced (anions leave the reduced PPy layer and it makes a contraction) [108]. The redox process for the mentioned material is as follows [95]:

**Oxidation :** $PPy + TFSI^- \rightarrow PPy{+}TFSI^- + e^-$
**Reduction :** $PPy{+}TFSI^- + e^- \rightarrow PPy + TFSI^-$

Whereas shown in **Fig.39,** the different volume changes in the two PPy electrodes (due to anions displacement) cause the bending of the trilayer CPA. This redox process and then ions movement between electrodes involves some electrical properties like diffusive impedance, double layer capacitance, and charge transfer resistance [109]. Hence to interpret these properties and also to find the relationship between the input electrical energy and the output mechanical deformation of the CPAs, some models have been developed. Like the IPMC's models, some of the CPA's models are based on multiphysics approaches [110], and some of them are system identification based [111], while a large group of them model the CPAs based on RC distributed lines [107, 109, 112]. For example, Takalloo et al. [107] by using an RC distributed network have proposed the following relation for the curvature of a trilayer CPA ($k_{(f)}$) in response to input voltage ($V_{(f)}$).

$$k_{(f)} = \frac{12(m+1)\varepsilon_{i(f)}}{h_p\left(nm^3+6m^2+12m+8\right)V_{(f)}} \qquad (4)$$

Where the details of this relation and the definition of unknown functions and parameters have been described in [107].

## A. *CPA Fabrication*

In order to fabricate a trilayer CPA by the mentioned materials, two layers of PPy should be electrochemically deposited on both sides of the separator (PVDF here). Electrochemical deposition or electrodeposition, also known as electroplating, is a method for coating of a material onto a conductive surface from an electrolyte solution. As it is shown in **Fig.40**, there is a setup consisting of a potentiostat and three electrodes, working electrode (WE), reference electrode (RE), and counter electrode (CE). The three electrodes are deeped in a container containing electrolyte solution and are connected to the potentiostat from their top parts. The potentiostat can control the deposition procedure. The WE is the target that should be coated, the CE is for closing the electric circuit, and the RE makes a fix reference point for the potentiostat. When a constant electric potential is applied between the WE and the RE, the electrolyte substances are prefered to be deposited on the surface of the WE rather than remaining in the electrolyte, and this is the total working principle of the electrochemical deposition.

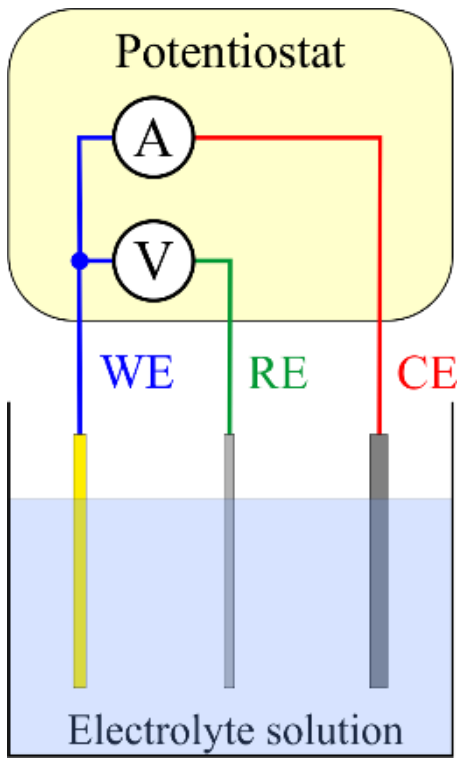

**Fig. 40. Schematic of a hardware setup for electrochemical deposition.**

For trilayer CPA with the mentioned materials, the electrolyte solution is usually a composition of pyrrole and $Li^+TFSI^-$ salt which are mixed with propylene carbonate (PC) as the solvent. The WE is a porous PVDF separator that in order to have an electrically conductive surface, should be sputtered by a thin layer of gold. This gold layer makes a highly conductive surface that it ensures a high-quality electrochemical growth of PPy. As referenced in some papers [95, 113] and shown in **Fig.41**, two stainless steel meshes are used as the CE where the WE is kept between them in order to ensure the symmetric coating of PPy on both sides of the PVDF.

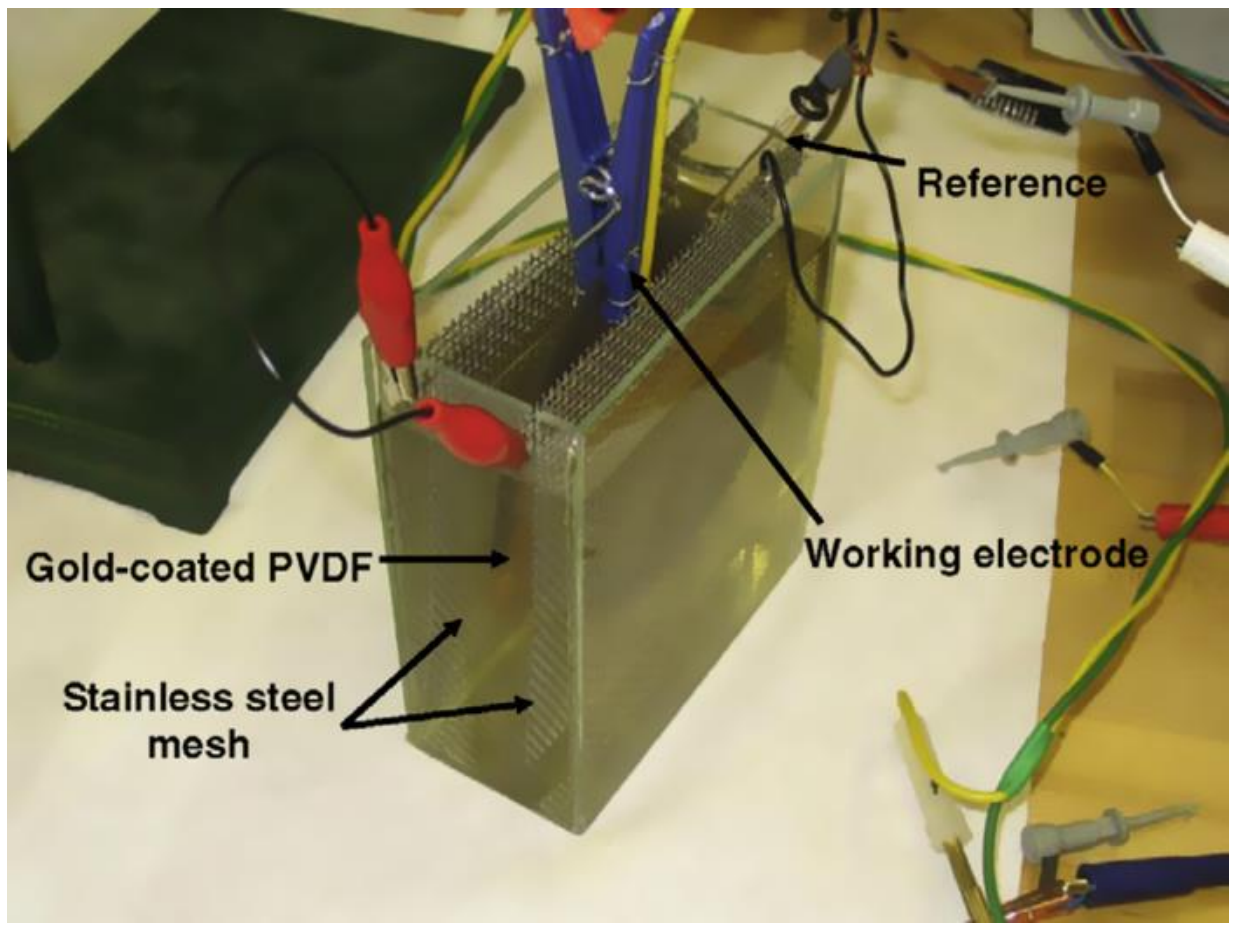

**Fig. 41. Photograph of a hardware setup for electrochemical deposition procedure for fabrication a trilayer CPA [95].**

Like IPMCs, CPAs also have a variety of promising features such as; working in the air, lightness, large bendability, ease of fabrication, low required input voltage, and especially their ability to be a MEMS element [114]. Hence these features make them another great actuators candidate for some of the active microfluidic devices. In this part, we review the applications of the CPAs in microfluidic devices and their related components. The trend of the applications of the CPAs in microfluidic devices is similar to the IPMC part, and the most of them are related to micropumps. We can also find some other applications of the CPAs in making microvalves, but there are not any CPA-based micromixers. Hence two subsections have been defined here for reviewing the application of CPAs in micropumps and microvalves in microfluidic devices.

*B. CPA-based microfluidic micropumps*

***CPA-based Micropump 1***

The first CPA-based micropump was proposed in 2004 by Wu et al. [115]. Their proposed micropump is a kind of peristaltic pump that utilizes a hollow cylindrical polypyrrole based CPA that is a good candidate for microfluidic applications. The layers of this actuator that they have named TITAN (The Tube In Tube Actuator Node), is shown in **Fig.42**. As is shown in the figure, TITAN is a multilayer CPA with the following layers:

**A:** The porous PVDF fiber that maintains the micropump cylindrical shape.

**B:** Metalized polyurethane tube that has the role of the working electrode for TITAN which has been wrapped with platinum wire and coated with PPy.

**C:** Inert PVDF membrane which is used as an inert electrochemical separator to hold the supporting electrolyte.

**D:** Metalized PVDF membrane with a deposited layer of PPy connected to a stainless steel mesh which is used as the TITAN counter electrode.

**E:** A Plastic tube for covering the electrode assembly.

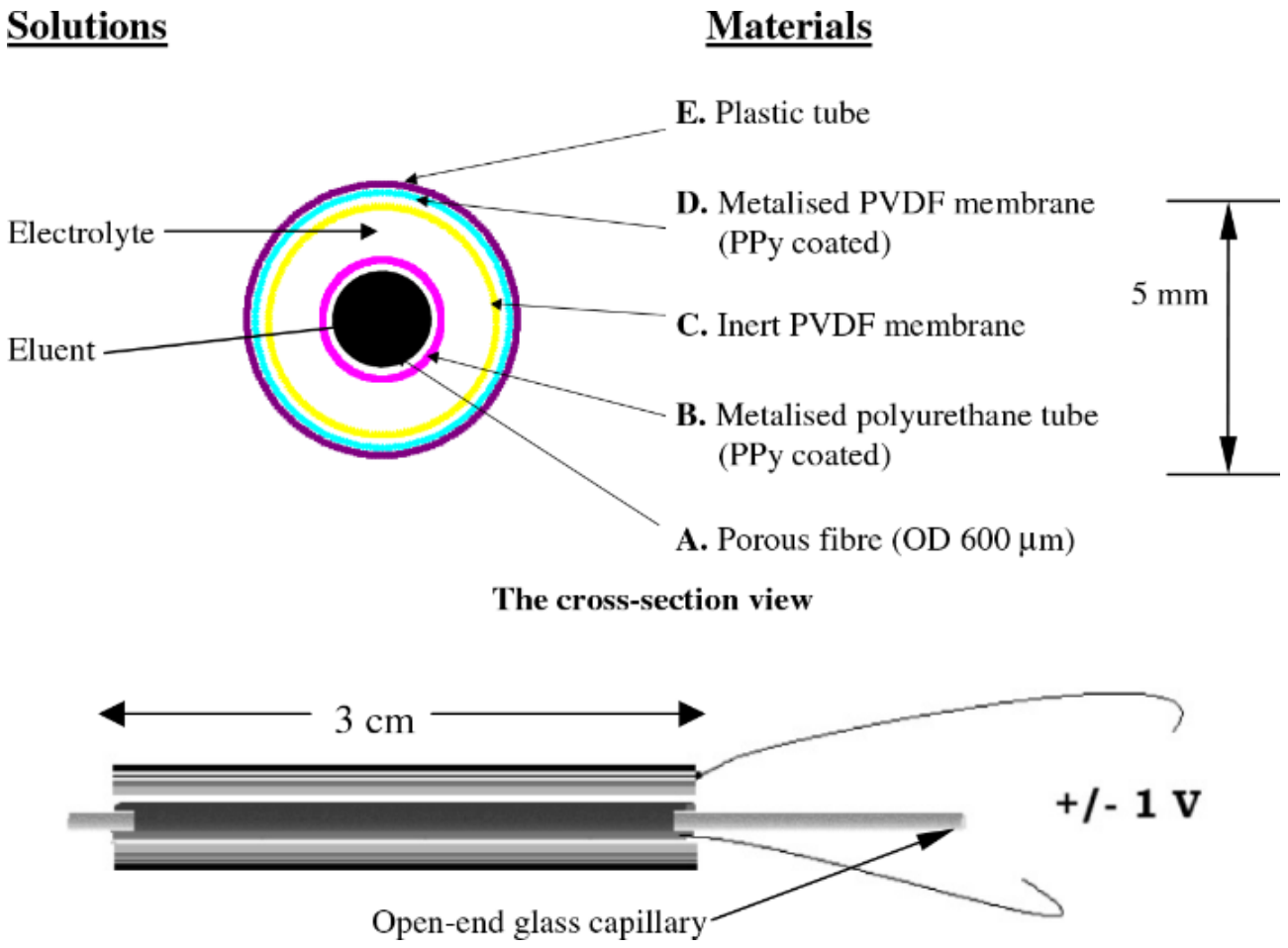

**Fig. 42. Schematic diagrams of the TITAN layers[115].**

As seen in **Fig.43**, to use this micropump, an open-end glass capillary must be connected to TITAN (placed into the TITAN), and a stainless steel wire should be placed into it in order to provide the means of closing off the tube when the pump was fully contracted. The working principle of this micropump is based on the contraction and expansion of TITAN where this working principle has been described in **Fig.44**. TITAN was tested in the real situation, and in response to only 1 V ( 8.7 mW), it can make the flow rates of up to 2.5 µl/min by back pressure of 50 mBar, which is enough to pump fluids in a glass capillary channel [115].

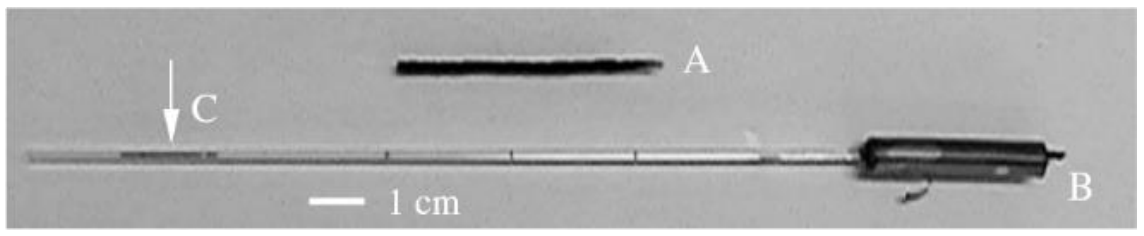

**Fig. 43. The TITAN micropump assembled on a glass capillary. (A) TITAN working electrode (B) TITAN CPA, (C) the glass capillary filled with organic PC solvent; an aqueous green food dye plug was placed inside as a marker[115].**

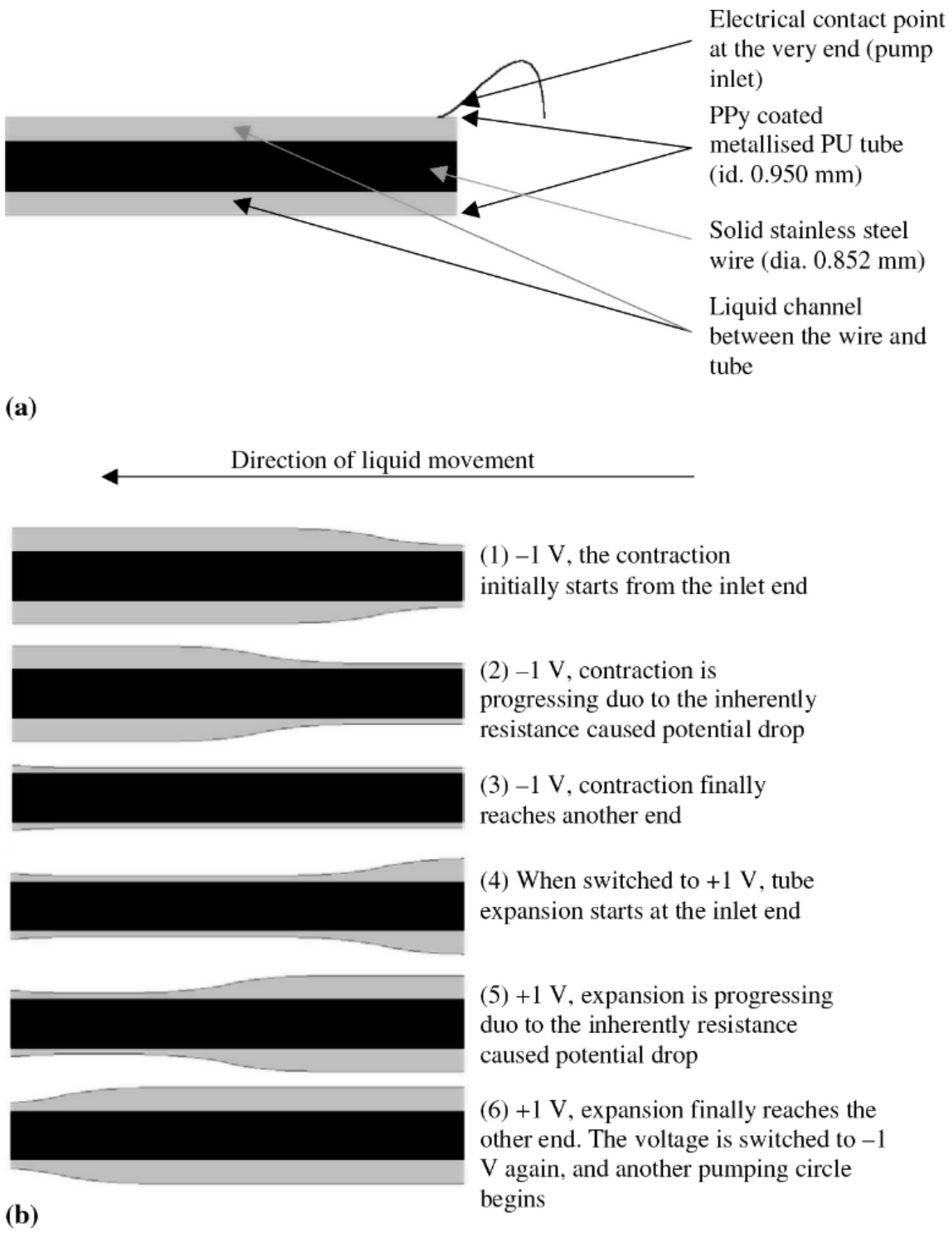

**Fig. 44. (a) The main layers of assembled TITAN and (b)its working principle at oxidized state (expansion) to show pumping sequences[115].**

### *CPA-based Micropump 2*

In another work, Kim et al. [116, 117] presented a low-power and low-cost PPy-CPA microfluidic pump that has two one-way push button check valves. As illustrated in **Fig.45** and **Fig.46**, this micropump has been fabricated by a PPy-based CPA and a PDMS diaphragm that have been packaged into a PMMA body. The used CPA of this work is a bilayer actuator with PPy electrode and silicon PDMS membrane. This CPA is in the form of a rectangular that has been attached to PDMS diaphragm using silicon glue at five anchoring points, four at the corners and one at the center (**Fig.46**).

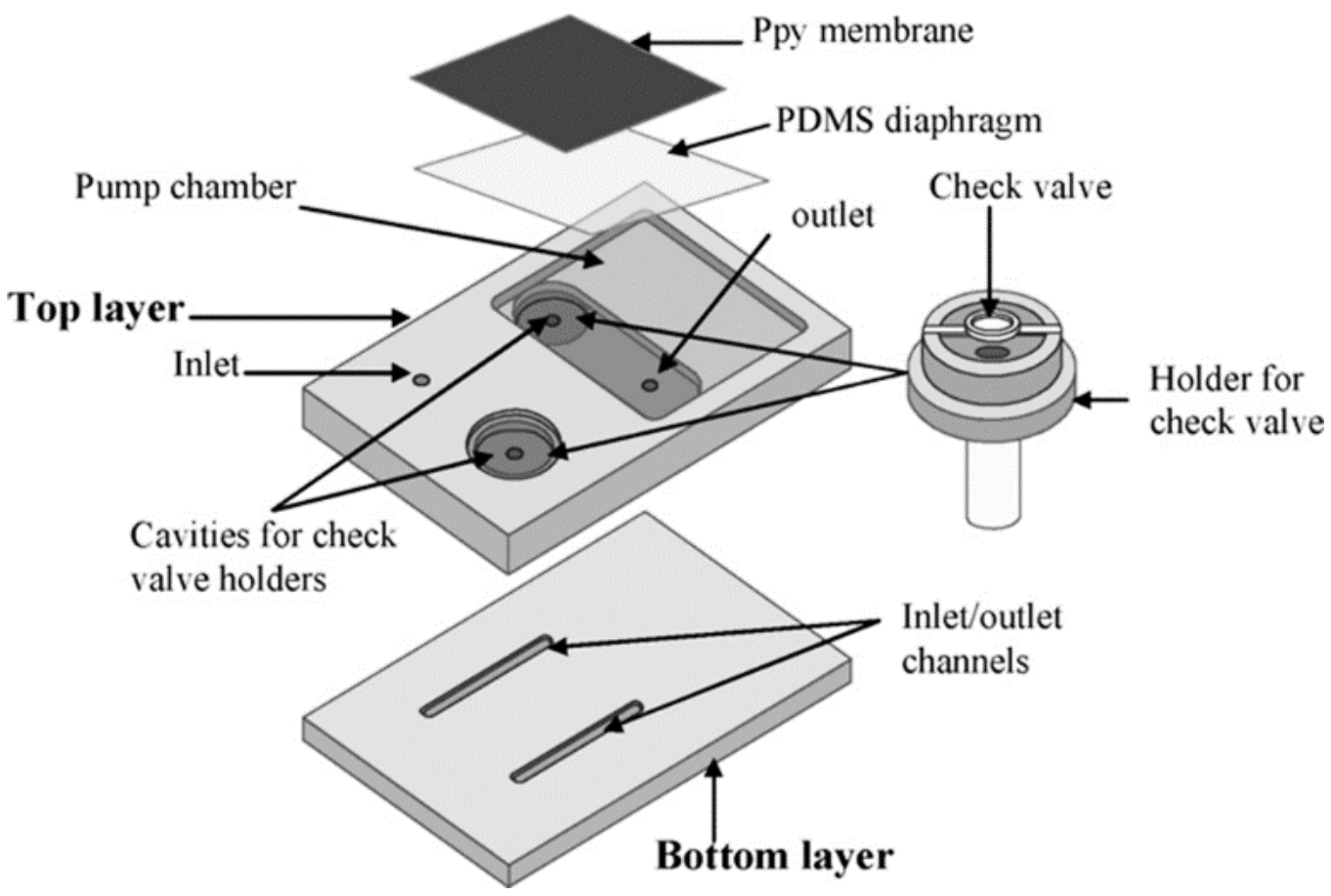

**Fig. 45. Exploded view of the proposed micropump in [117].**

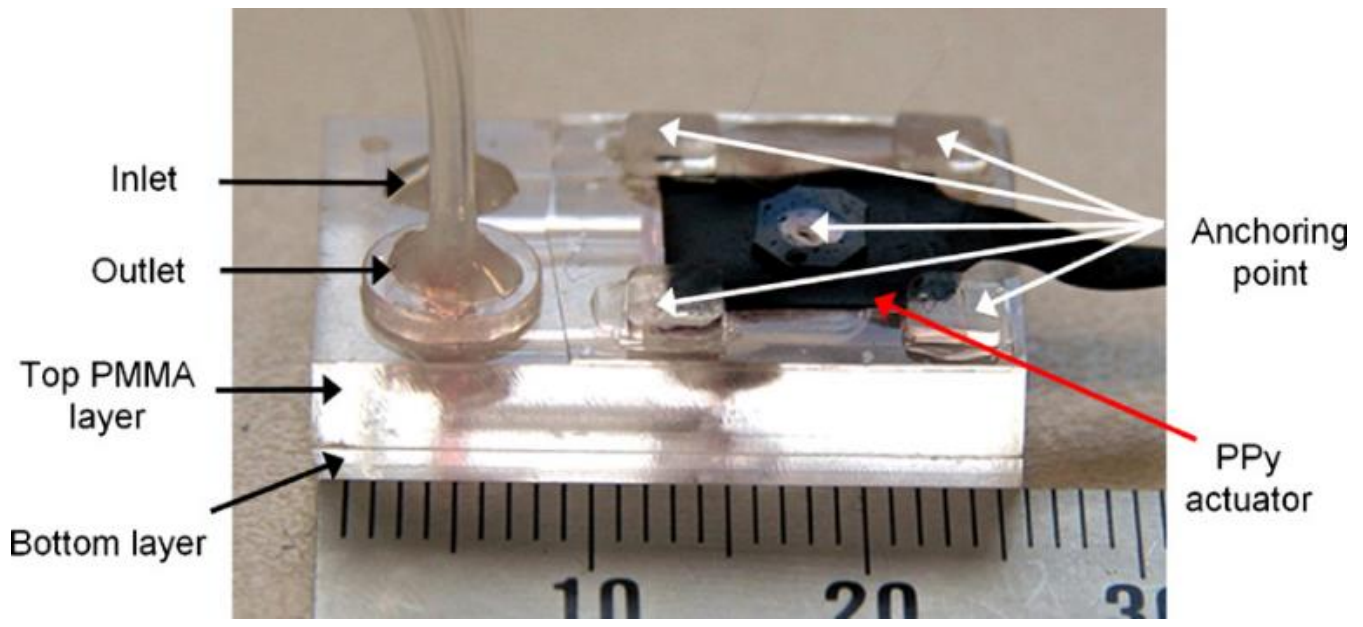

**Fig. 46. The photograph of the fabricated version of the proposed micropump in [117].**

Under applying the voltage, the CPA will be deformed, and it can make the same deformation as in the PDMS diaphragm. As a result, when a ±1.5V square-wave is applied to the CPA, it and so the diaphragm are bent in the form of convex and concave so that this bending causes the required push or pull actions of the pump for pumping rate of 18-52µl/min (11mBar back pressure) with a lifetime of greater than 20,000 cycles. This kind of CPAs can produce large deformation, but their produced pressure are not as large as the pressure that conventional solenoid pumps produce. To solve this low-pressure problem, two check valves were fabricated as are shown in **Fig.47**. During the prime-mode of the pump, the upward bending of the diaphragm generates a negative pressure to close the outlet check valve and lift the inlet check valve to suck the fluid into the pump chamber. But during the second-mode of the pump, a positive pressure is generated by downward bending of the diaphragm that forces the inlet check valve to close and outlet check valve to open. Hence the fluid will be pumped to the outlet from the chamber.

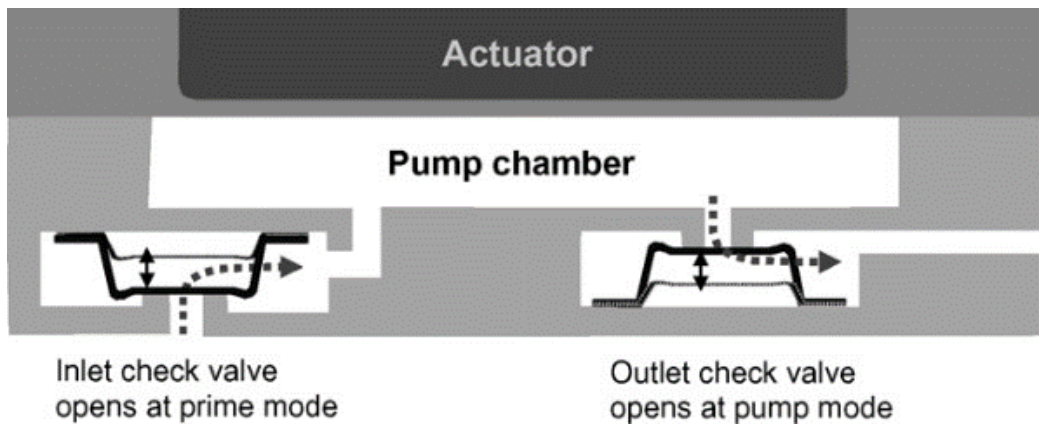

**Fig. 47. The working principle of check valves in [117].**

The same group in another paper [118] used their proposed pump in a practical microfluidic chip for chemical analysis and detection of Fe ions. As shown in **Fig.48,** the setup for detection of Fe ions have a microfluidic chip, proposed PPy-based CPA micropump, two Petri dishes for stock and wasted solutions, as well as the optical detector module which also has an optical cuvette with LED and a photo-diode optical sensing module.

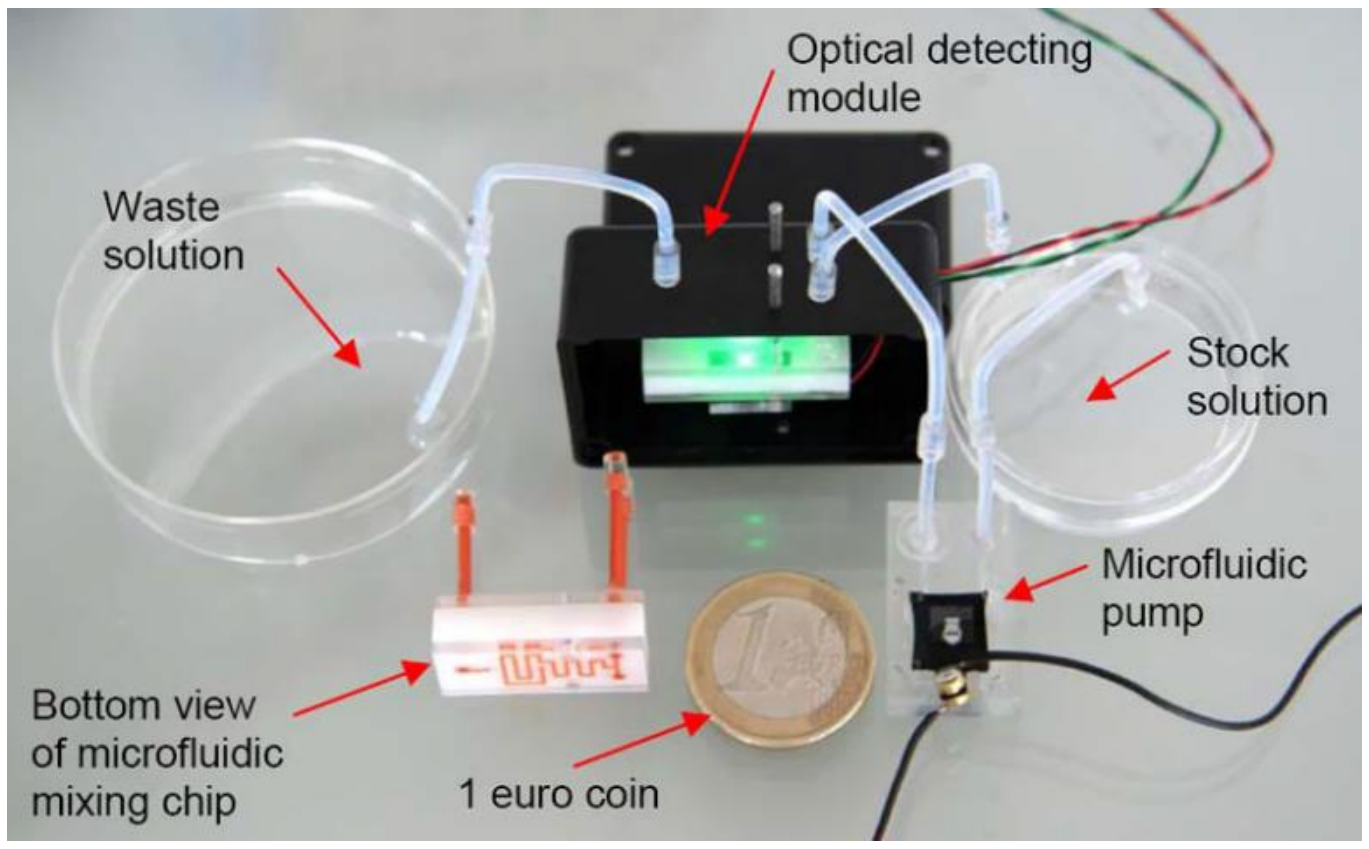

**Fig. 48. Experimental setup for detecting the concentration of Fe ions using a microfluidic chip, PPy-based micropump, and optical sensing module [118].**

### *CPA-based Micropump 3*

In [119] the development and assessment of a biomimetic pump based on soft CPAs have been reported. The main idea or hypothesis of this work is the fabrication of a tube-shaped CPA that by its contraction/expansion, it can provide the driving force for liquid movement. The authors have also explained this micropump to be integratable in the microfluidic channels and that it can have functionality like the blood vessels. There is some experiment in this paper toward the proposed idea, but the presented micropump is not integrated into a microfluidic channel. They have fabricated the micropump independently and used it for pumping two solutions in a microfluidic micromixer chip. Talking about the mentioned microfluidic chip is not our purpose, and we discuss about the proposed CPA-based micropump. Also as shown in **Fig.49**, the microfluidic chip is a kind of mixer that has been fabricated by PDMS where the Griess-Ilosvay reagent is pumped through its upper channel and the nitrites solution is pumped through its lower channel.

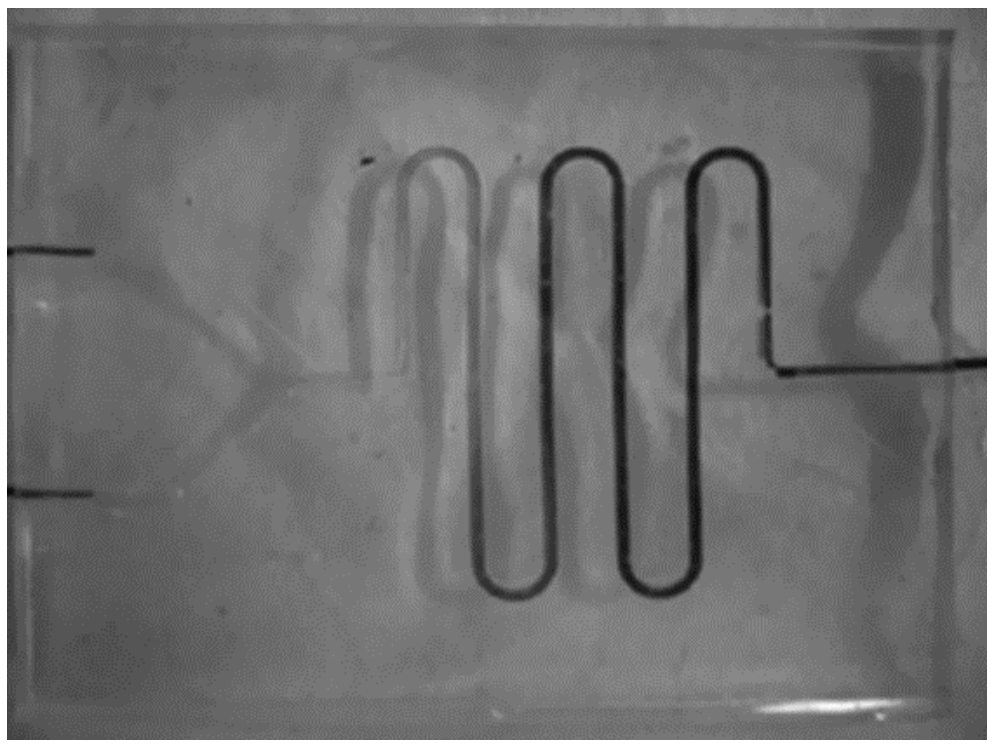

**Fig. 49. Utilized microfluidic chip in [119].**

The constructed micropump was presented in three types; Nafion-PPy CPAs, and two other methods. The first method is using two Nafion-PPy CPAs in the form of a tweezer for deforming a PDMS chamber (**Fig.50**). This configuration can make sufficient deformation in order to produce the required pressure in pumping actions. The Nafion-PPy CPA is a kind of trilayer CPA that has a Nafion membrane instead of a PVDF where here a Nafion 117 film was used. This method for making the CPA is very similar to the fabrication of IPMC that were talked about previously in the IPMC part, and the result was cantilevered formed CPA. But in the second approach, the PPy electrodes were deposited onto polyurethane tubes using the same conventional approach for fabrication of CPA using electrochemical deposition that was described in part "*CPA fabrication*." Finally in the third method, in order to have a PPy tube, the polypyrrole was electro synthesized onto a platinum wire where a thiner platinum wire was wrapped around it to facilitate electrical connection (**Fig.51**). The first and the second approaches have been tested here. In the first approach in the response to 3V, the micropump could pump the fluid with the flow rate of 0.4 μl/s (24 μl/min). The second method has been tested under 1 V applied voltage, but no numerical index for its flow rate is reported.

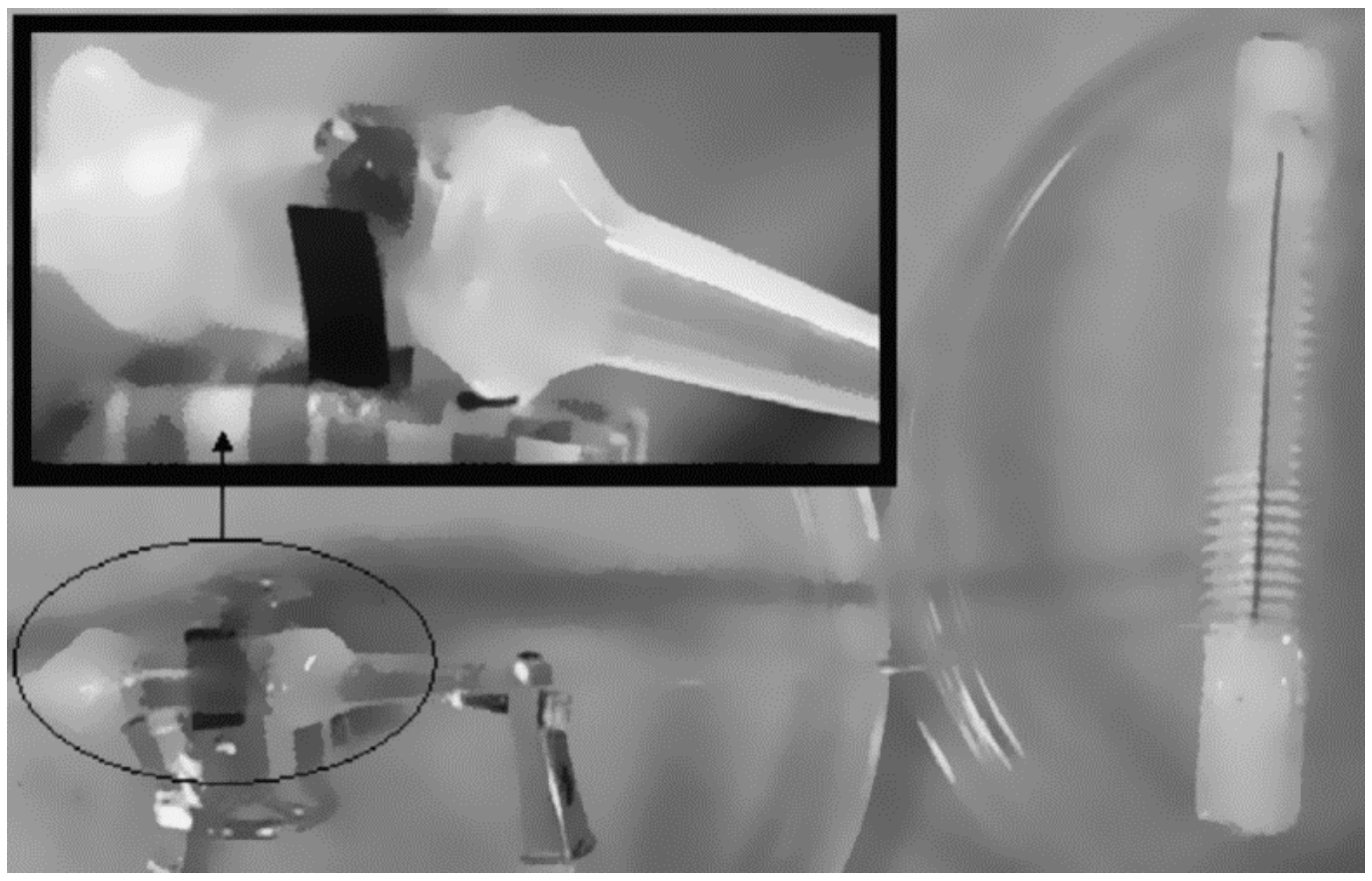

**Fig. 50. A photograph of the first micropump developed in [119].**

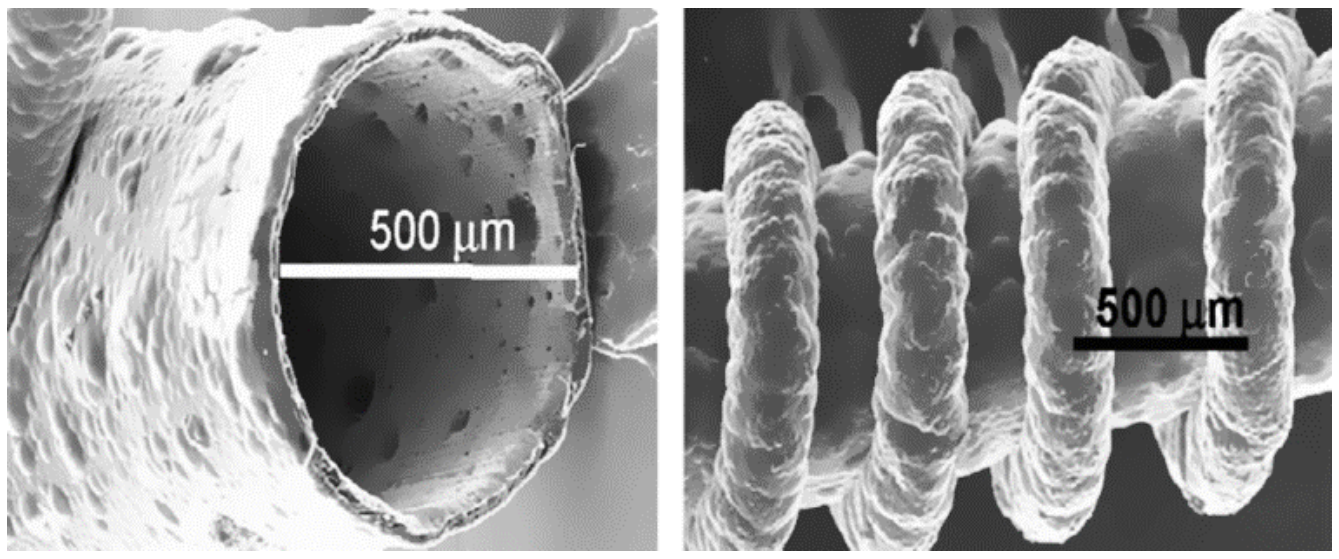

**Fig. 51. Scanning electron microscopy (SEM) images of a PPy tube obtained by electrodeposition of PPy onto a platinum wire [119].**

### *CPA-based Micropump 4*

In another work [120], a planate bimorph CPA was developed based on PPy and two types of acids, such as p-phenol sulfonic acid (PPS) and dodecylbenzene sulfonic acid (DBS). The method of PPy deposition was the same electrodeposition explained before, and for working, counter, and reference electrodes, a titanium (Ti) plate, a platinum (Pt) plate, and a silver wire were used, respectively. Also the electrolytes were the same PPS and DBS. The presented planate CPA could only deform its central part locally, and this ability makes it a useful actuator for playing the role of a diaphragm in a micropump. As depicted in **Fig.52,** to create micropump, the presented CPA was integrated with a tank and a flow channel.

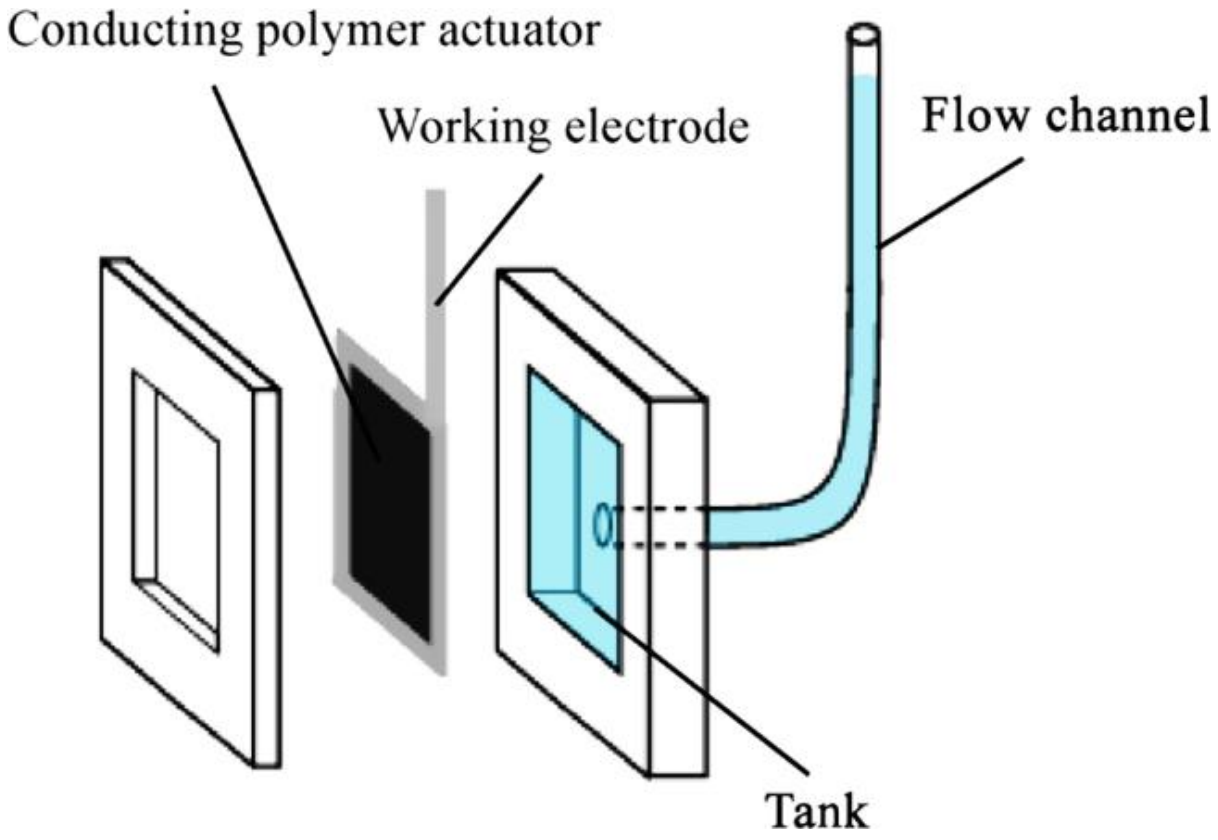

**Fig. 52. The exploded view of the micropump presented in [120].**

The proposed micropump was tested (**Fig.53**) and the water level in the micropump flow channel showed that the reciprocating motion of the bimorph CPA (±2mm) can pump the fluid by the flow rate approximately equal to 28 µl/min. The most of microfluidic chips and insulin pumps need the flow rate in the range of 1 to 50 µl/min and so the proposed micropump can work as a microfluidic pump. The changing of the pump volume and its actuation can be controlled by application of electrochemical potential into an electrolyte tank that is a constraint for this micropump because we cannot integrate it into an individual chip. Hence for its usage, we need to have an electrolyte tank and counter (CE) and reference (RE) electrodes (**Fig.54**).

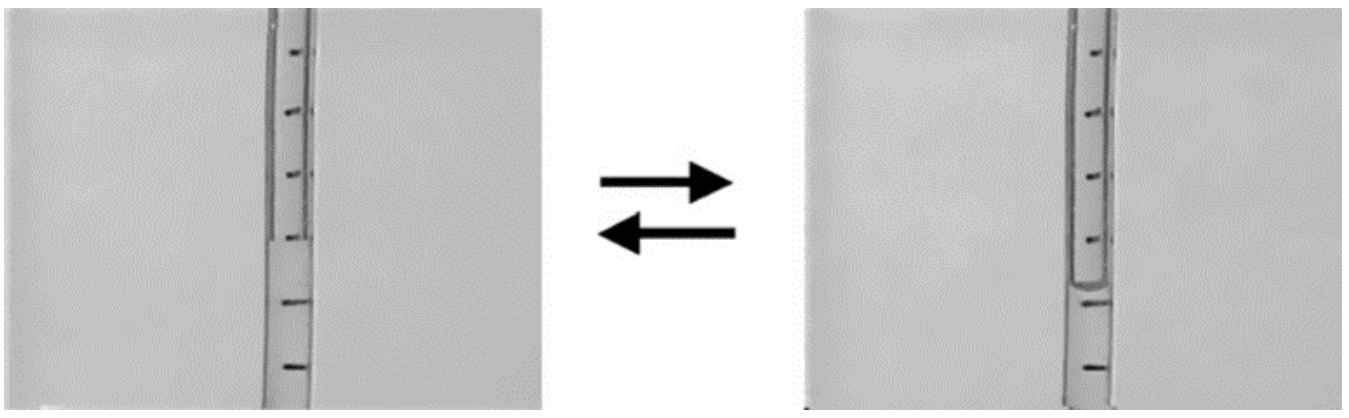

**Fig. 53. Flow channel of micropump presented in [120] and the change of water level in it.**

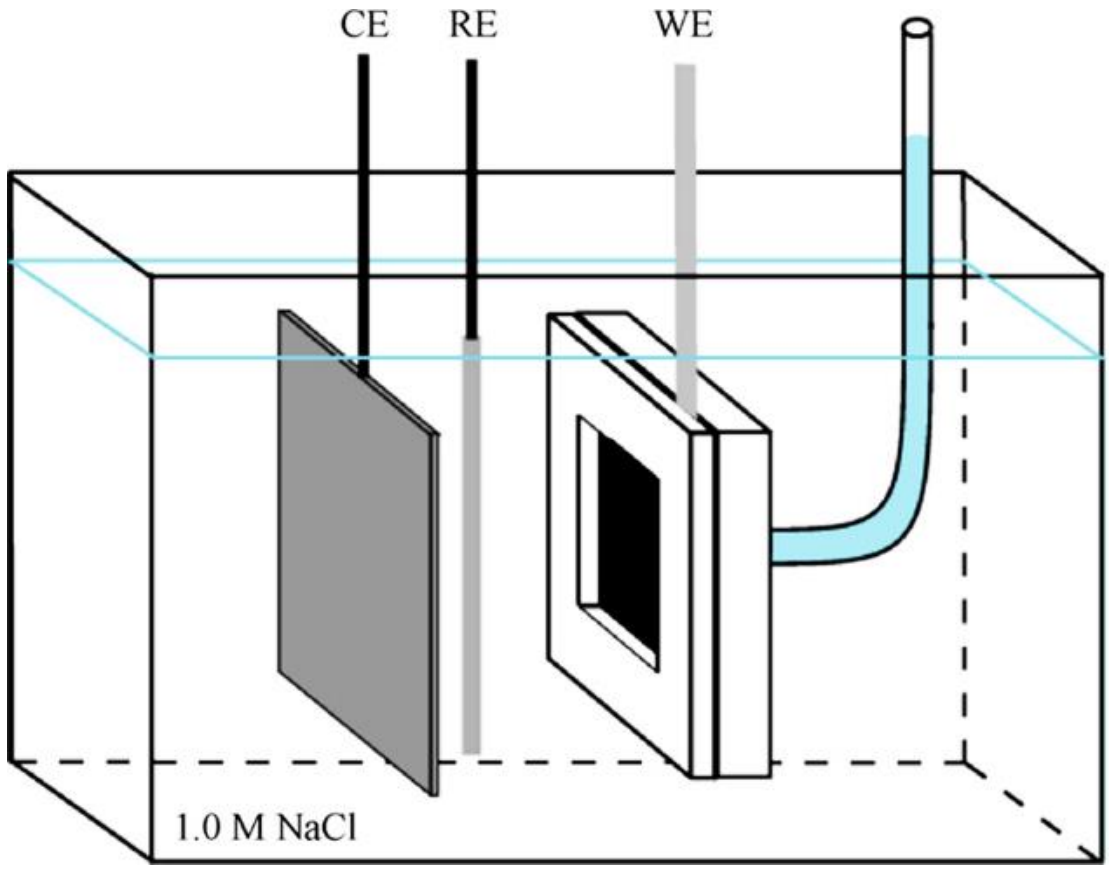

**Fig. 54. Using presented planate CPA micropump in[120].**

***CPA-based Micropump 5***

The first idea for using the petal-shaped *i*-EAP actuators in micropumps was published by Fang and Tan in 2010 [95]. As explained previously, regarding the IPMC part, use of a single disk-shaped piece of IPMC is not a proper choice as a diaphragm for micropumps, and this choice is not appropriate for CPA-based actuators either [95]. Like IPMCs, this kind of CPA disks are clamped at all edges, and this form of clamping makes important constraints for CPA's bending, i.e. it restricts the CPA deformation and so the flow rate of the CPA-based micropump will be decreased remarkably. To find a solution for this problem and to improve the bending and flow rate of the micropump, Fang and Tan [95] for the first time presented a new form of CPA diaphragm using petal-shaped CPAs instead of a single piece of CPA as explained in part "*IPMC-based Micropump 6"* earlier. This petal-shaped structure has been depicted in **Fig.55** where four quarter-disk-shaped CPAs are attached on a thin elastic diaphragm. In this structure, the CPAs are only clamped on one edge, and they are free to bend and making a reciprocating movement in the elastic diaphragm facilitating the pumping operation to be performed.

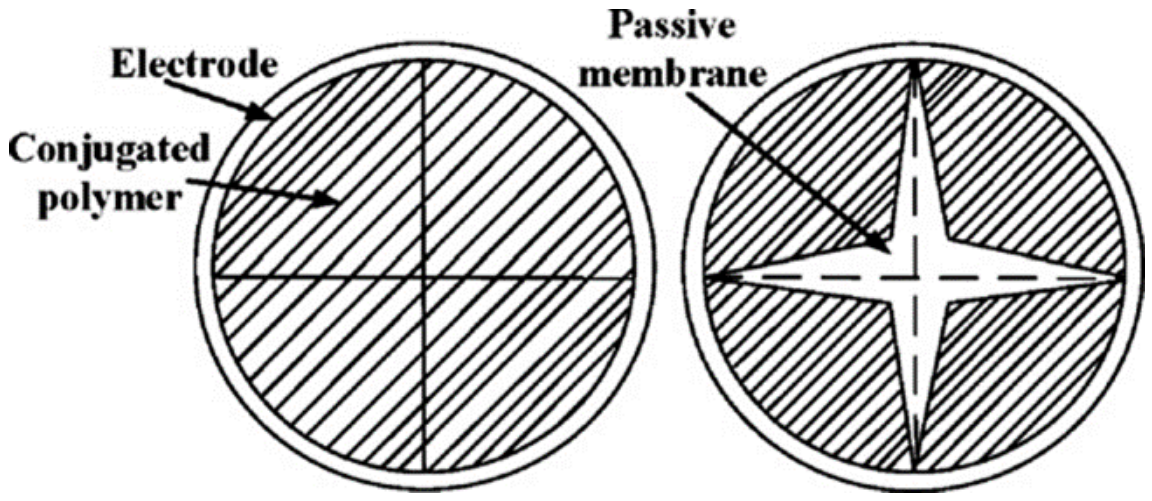

**Fig. 55. Schematic of a petal-shaped CPA in top view. Left is before applying voltage and right is after applying voltage[95].**

The exploded 3D schematic of this micropump is illustrated in **Fig.56**; as it is shown here, this micropump is consisted of a petal-shaped CPA, an elastic diaphragm attached to the CPA, a layer for the chamber, four layers for check valves, and two layers as bodies all made by PDMS. Under applying the voltage, the CPA and so elastic diaphragm would move up and down. Thus this reciprocating displacement makes the pumping pressure, and fluid will flow from the inlet to the outlet.

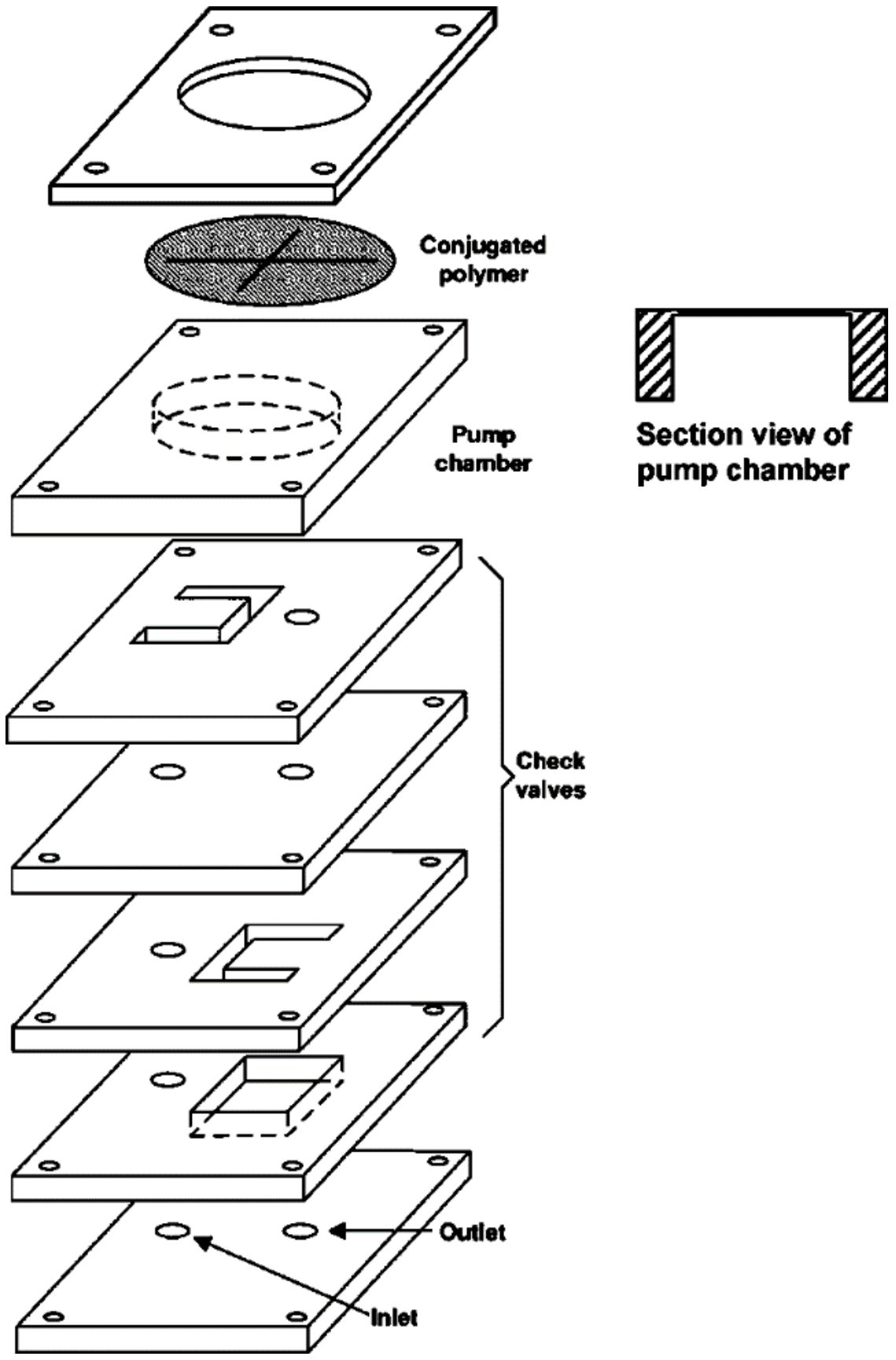

Fig. 56. The exploded 3D schematic of presented micropump in [95].

The functionality of the microvalves of this device is keeping the directional flow during the pumping process. The microvalves are fabricated using four layers of PDMS, and as depicted in **Fig.57,** their working principles and duties of these valves are as follows:

1. When a pressure is applied from the bottom, the outlet diaphragm will be lifted to allow the fluid flow out through the outlet.
2. When a pressure is applied from the top, the inlet diaphragm will be pushed down and allows the fluid flow into the pump through the inlet.

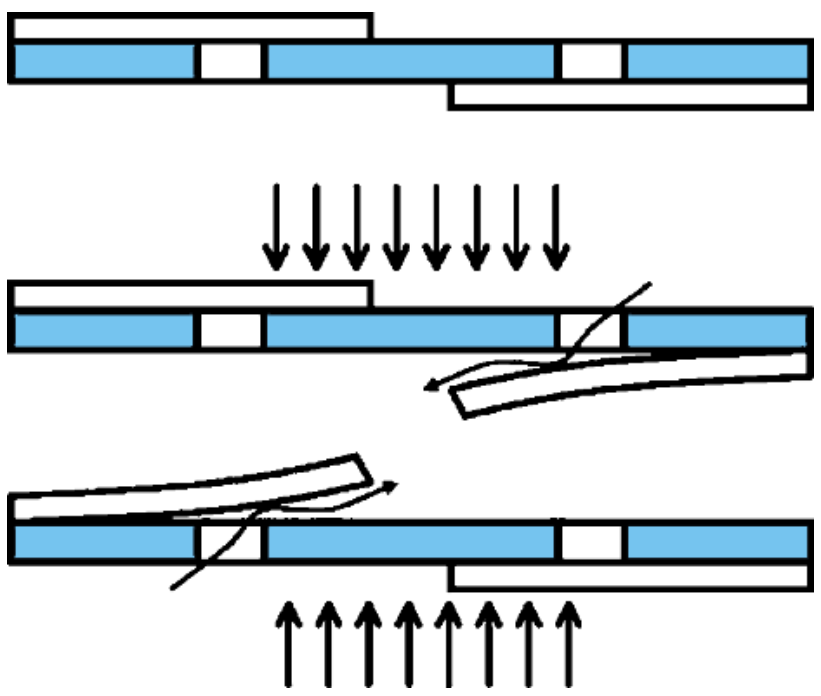

Fig. 57. The mechanism of flap check valves presented in [95].

As the conclusion for this section, we can say that the proposed micropump in [95] has been fabricated by several layers of PDMS in the form of a cube as shown in **Fig.58**. A petal-shaped Ppy-based trilayer CPA with a PVDF membrane was used as the active element of the micropump with an elastic diaphragm, and under experimental tests, this CPA-based micropump was able to pump the fluids by a rate of 1260 µl/min in response to a 4 Volts and 0.5 Hz applied signal. This results show that in comparison to all *i*-EAP based micropumps presented in this review paper, this type has the maximum flow rate and maybe it is a better candidate for some of the microfluidic application.

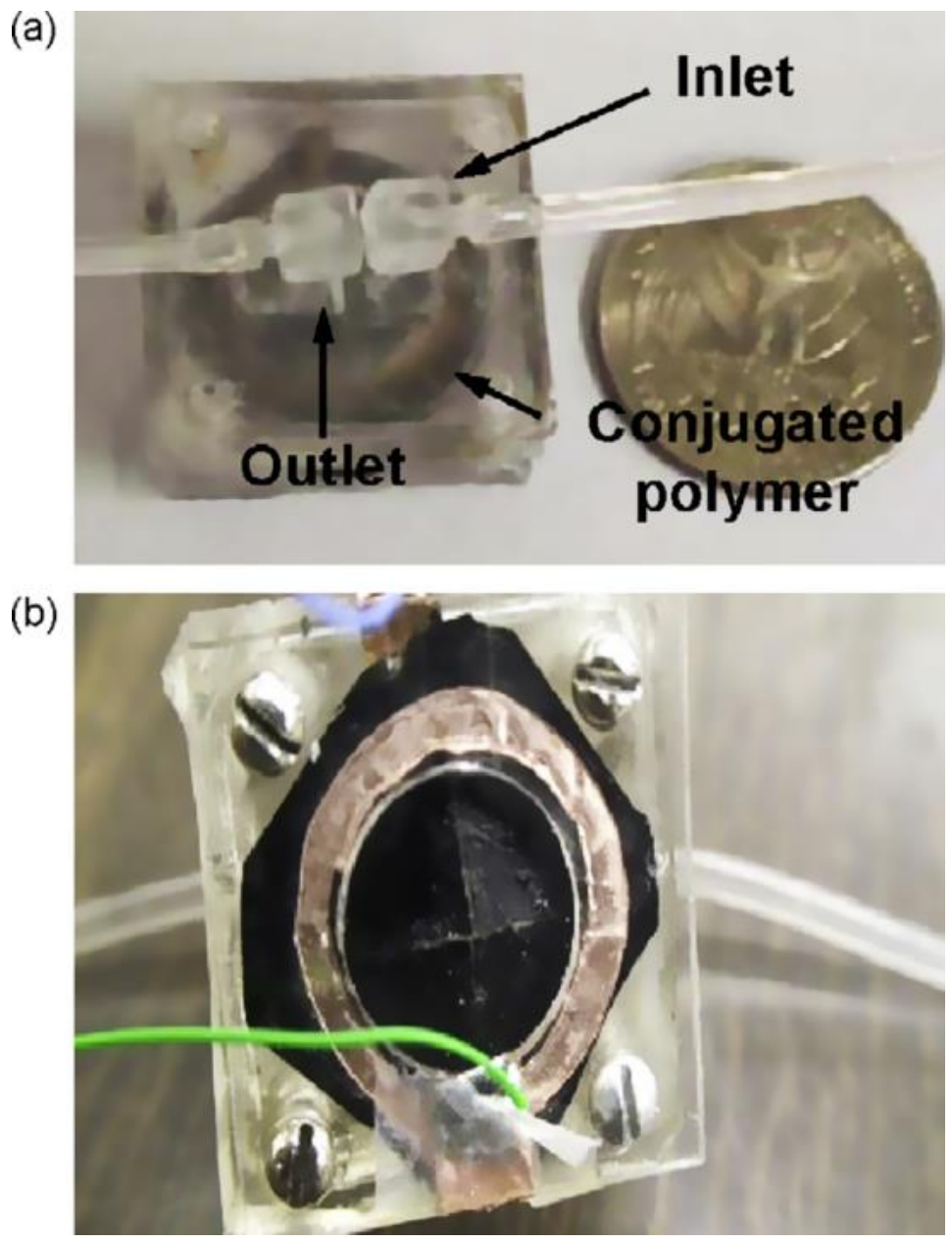

Fig. 58. Photographs of the assembled micropump presented in [95].(a) Top view .(b) bottom view.

### *CPA-based Micropump 6*

A new kind of valveless CPA-based micropump was presented by Naka et al. in 2010 [121] that has two belt-shape bimorphs PPy CPAs (Ppy has been electrochemically deposited on titanium sheets as the working electrode (WE)). These CPAs (**Fig.59**) make open/close actuation on a PDMS tube that can provide the required pressure for the pumping action.

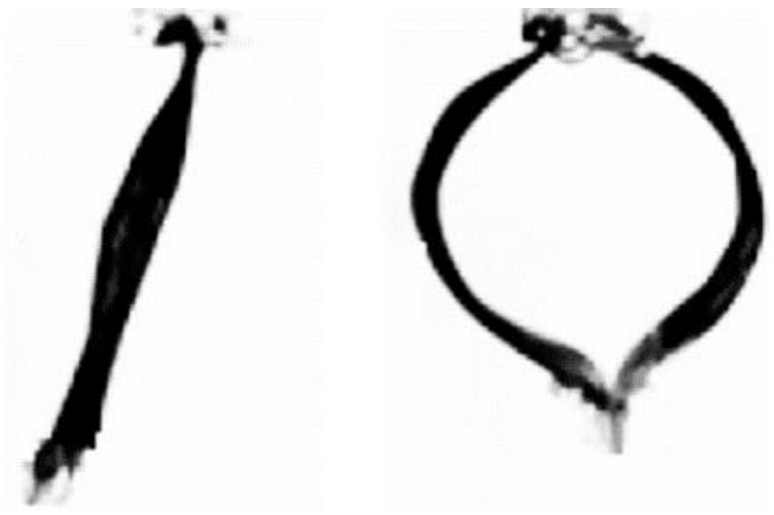

Fig. 59. A fabricated version of the belt-shaped bimorph PPy CPAs presented in [121]. Left: before applying voltage, right: after applying voltage.

As depicted in **Fig.60,** the pump is a kind of polyethylene capsule (Length: 60 mm and OD: 21 mm). The two mentioned CPAs had been placed inside the capsule surrounding a 20 µm thick PDMS tube (OD: 5 mm and ). A silver wire and a Pt sheet as reference electrode (RE) and counter electrode (CE) were also located outside the CPAs. The capsule is filled with the required electrolyte (a mixture of water and 0.5 mol/L LiTFSI solution for this work). As illustrated in **Fig.61,** a tank, containing the fluid that micropump is supposed to transport, is connected to the pump inlet. The first CPA near the tank is called actuator ***A*** and the second one is called actuator ***B***. The applied voltages to ***A***, and ***B*** actuator s are in the range of −1.2 to +1.0 V and −0.6 to +0.5 V, respectively with a frequency of 0.005 Hz and 180◦ phase difference. This type of the applied voltages to the CPAs ensure that they can make a continuous open/close deformation in PDMS tube which pumps the fluid of the tank even when the delivery head is upper than the tank (**Fig.61**). The effect of the viscosity of the pumped fluid was found to be small, and this micropump can pump viscose fluids even the fluids like silicon oil with the viscosity 400 times greater than that of water. The flow rate of this micropump also was measured to be in the range of 2.0 to 83.0 μl/min that is an appropriate range for the microfluidic applications.

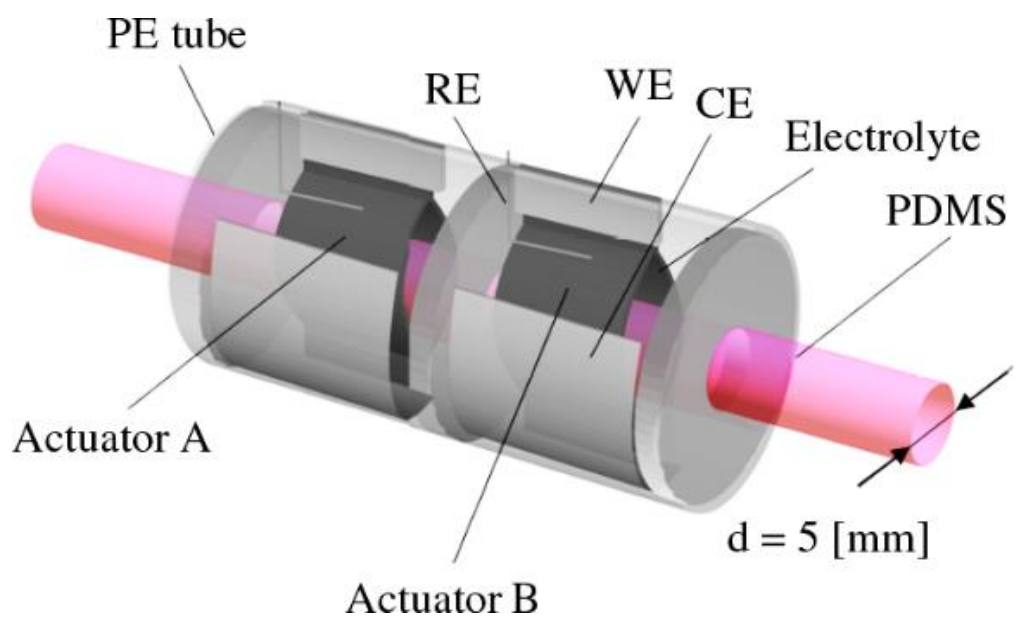

**Fig. 60. Details of the presented pump in [121].**

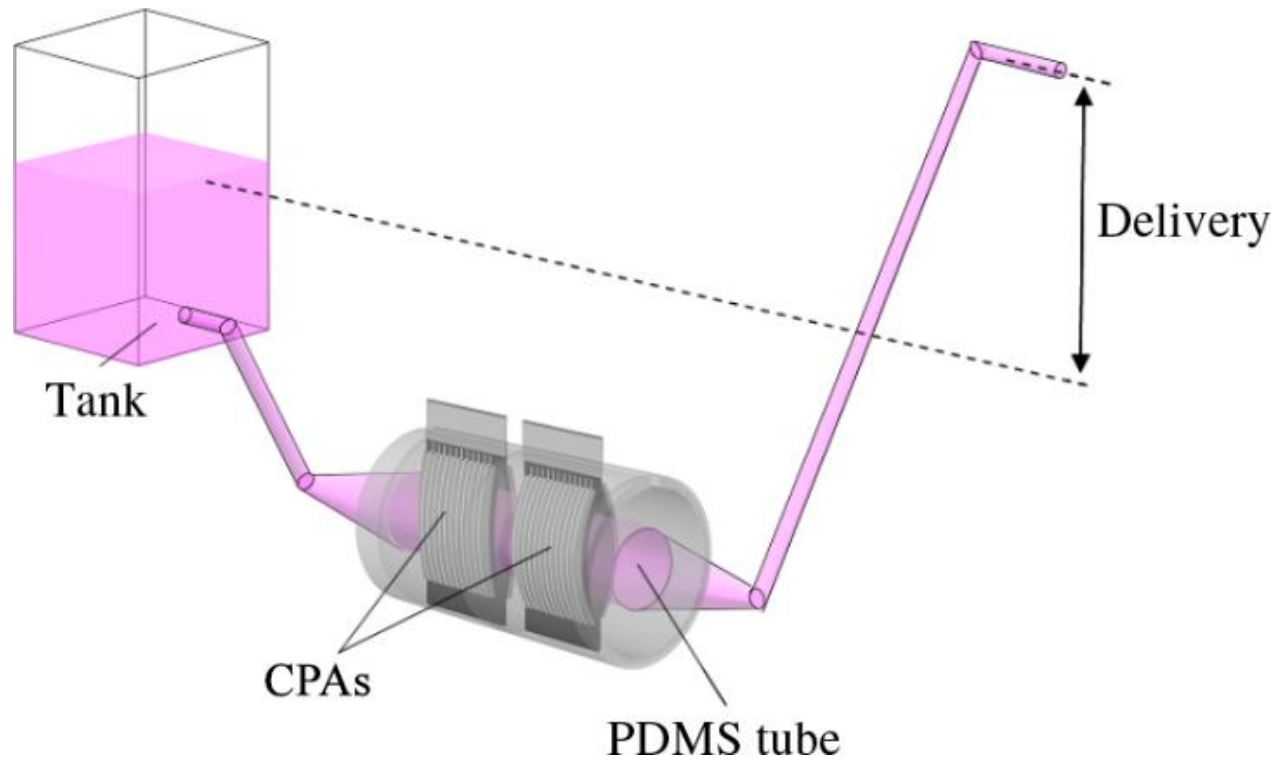

**Fig. 61. The presented pump in [121] under working.**

***CPA-based Micropump 7***

Another CPA-based micropump is a miniaturized 'syringe type' pump presented by Hiraoka et al. in 2012 [122]. This micropump has been fabricated by a stack of several conductive polymer (CP) layers interconnected with electrolyte layers. In order to integrate the micropump in the microfluidic and especially in the lab-on-chip devices, the stack is placed in a polycarbonate case. The proposed micropump can produce the required pumping pressure for the flow rate of 4.7 μl/min even by voltages lower than 2V. For a stack with eight single CPAs which are attached on top of each other, a maximum strain of 5% was also measured against atmospheric pressure. Since the diffusibility of the ions in conductive polymers is low (∼$10^{-7}$ $cm^2/s$) and on the other hand the working principle of CPAs is dependent on ions migration, increasing the thickness of the polymer layer (beyond∼100 μm) makes a serious constraint in having a large displacement of the CPA and also increases its response time. To solve these problems a proposed stacked CPAs is presented, the schematics of which is illustrated in **Fig. 62**. The CPAs are fabricated by several units; each unit consists of a ∼70 μm PPy-TFSI film as the active element of the CPA, a ∼33 μm capacitor layer soaked in the electrolyte (EMI-TFSI here), and ∼20 μm PPy dodecylbenzene sulfonic acid (PPy-DBS) layer as a counter electrode (CE). Here eight units have been stacked on top of each other, and the result is a ∼960 μm thick multilayer CPA with ∼560 μm active conductive polymer material. It is necessary to have several electrical connections between the stacked layers where for this purpose a silver wire was inserted into the tiny holes in the layers and fixed with silver glue. The multilayer CPA is kept in an 830 μm thick polycarbonate (PC) frame to which a polyimide (PI) membrane was attached at its bottom by 10 μm thick glue layer (**Fig. 62-b**). The PC frame was also sealed by attaching a thin glass on its top. Due to the tension *T* in the PI layer, a vertical force is produced which helps the stacked layers stay attached during actuation avoiding their separation (**Fig.62-a**). In **Fig. 62-b,c** the fabricated version of this micropump is shown that has been glued on a microfluidic system.

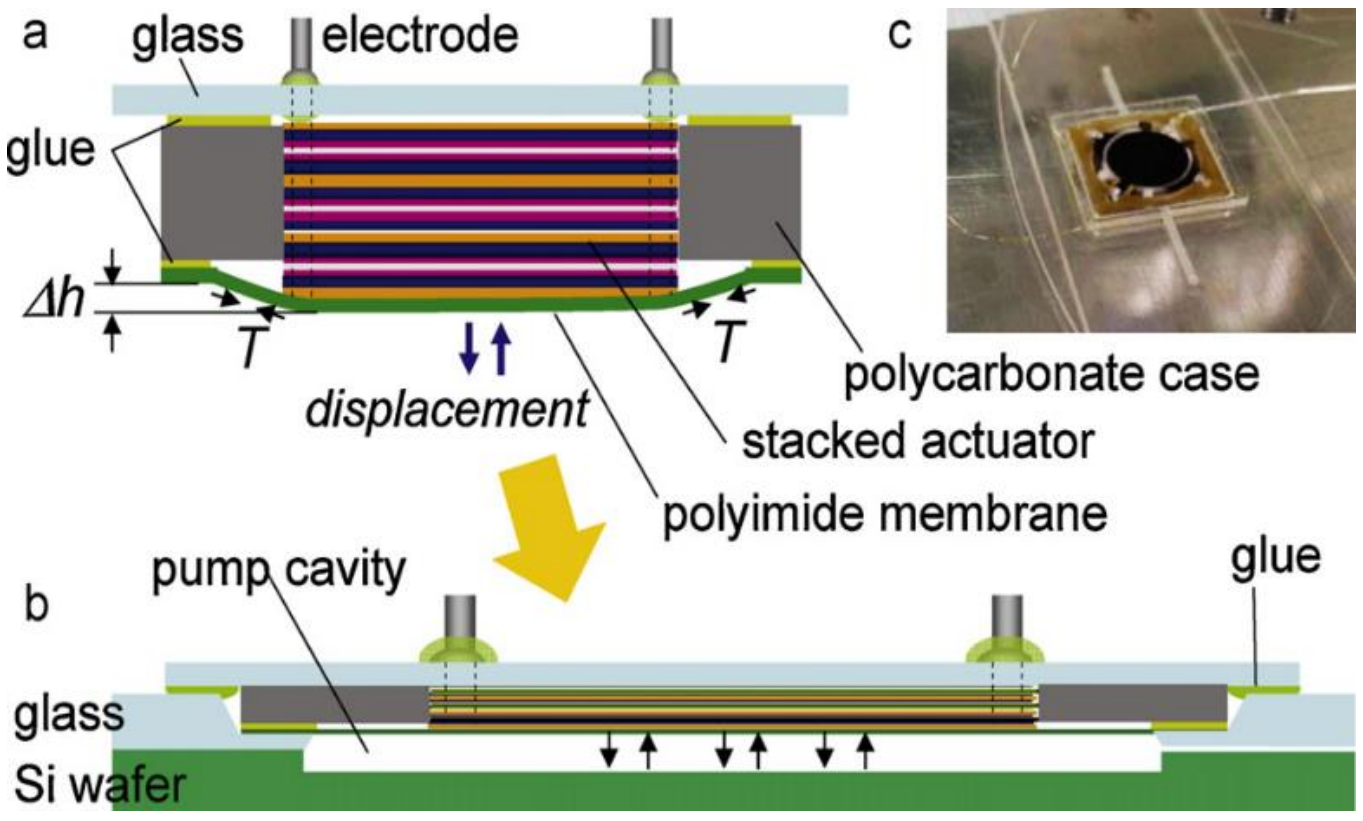

**Fig. 62. Proposed CPA-based micropump in [122]. (a) Magnification schematics of the stacked layers and other parts. (b) The schematics of the packaged stacked CPA attached to a microfluidic system. (c) Photograph of the fabricated version glued on a microfluidic system.**

***CPA-based Micropump 8***

In 2013 a Japanese company, EAMEX, developed the world's first semi-commercial micropump that does not require a driving source such as a motor, and it works based on CPA-based diaphragm [123]. **Fig. 63** shows the proposed CPA-based micropump presented by EAMEX. As shown here it has two PPy based bilayer CPA and there is electrolyte solution between them. When the input voltage is applied with the polarity of positive/negative, i.e. upper CPA diaphragm is biased positively and the lower negatively, the upper CPA will be expended to suck the fluid from the left top inlet, and the lower one will be contracted to expel the fluid to the right down outlet. In the opposite polarity, i.e. negative/positive polarity, the pumping procedure is inverted. The upper CPA will be contracted to expel the fluid to right top outlet, and the lower one will be expanded to suck the fluid from the left down inlet. There is a problem in the proposed micropump and that is the electrolyte solution between CPAs leaked to the pumping fluids. To solve this problem the seal coating of CPA is presented with a thin polymer film but this sealing cover is reported to reduce the output flow rate of the pump [124].

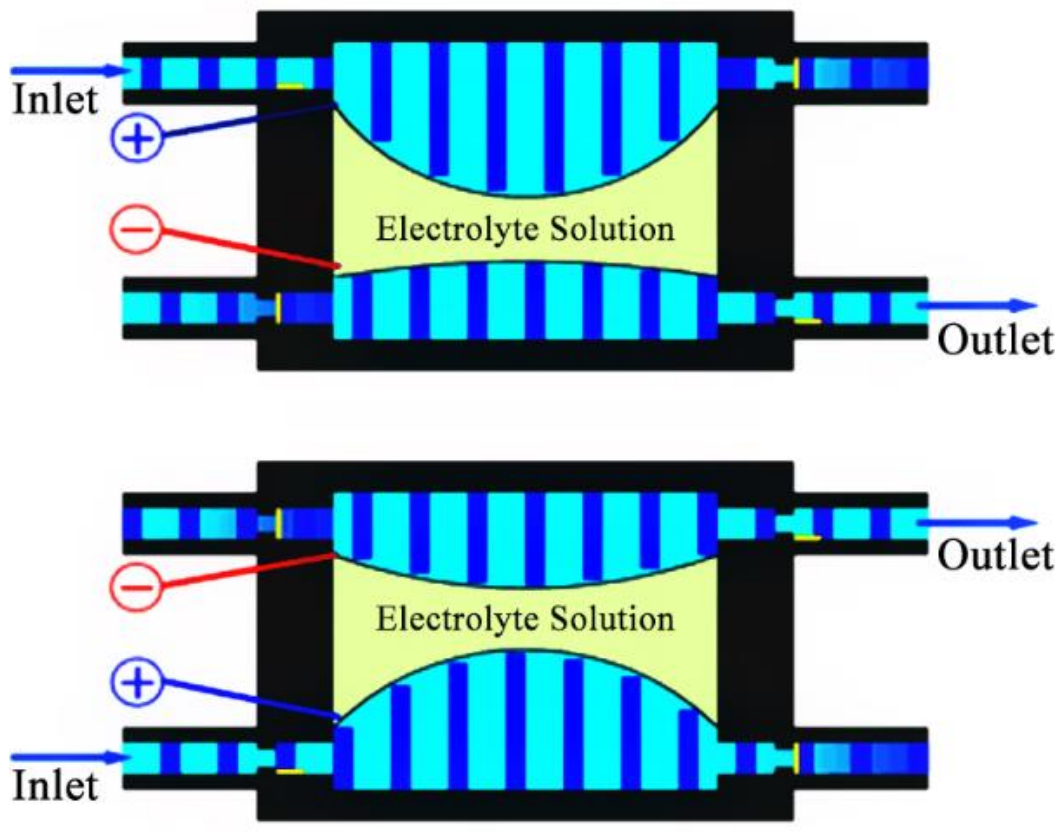

**Fig. 63. Schematics of proposed CPA-based micropump in for two different polarities[123].**

The proposed CPA-based micropump has been produced in several models and all of them work by less than 2V at 0.5 or 1 Hz frequencies where they can pump the fluids in the range of 2 to 2000 ml/min. Four main types of EAMEX micropumps have been shown in **Fig. 64**, and their main specifications also have been collected in **Table 1**. It is true that the proposed pumps are the semi-commercial version and they can work reliably, but except "type **a**" they do not have the proper specifications (especially the size) for microfluidics applications. And also the flow rate of them is not always a positive point for all microfluidic devices since most of the times we need micropumps with the flow rate in the range of µl/min, not ml/min. Hence, regardless of the variety of the promising EAMEX micropump products with great features, they are not necessarily the proper choice for the microfluidic usage.

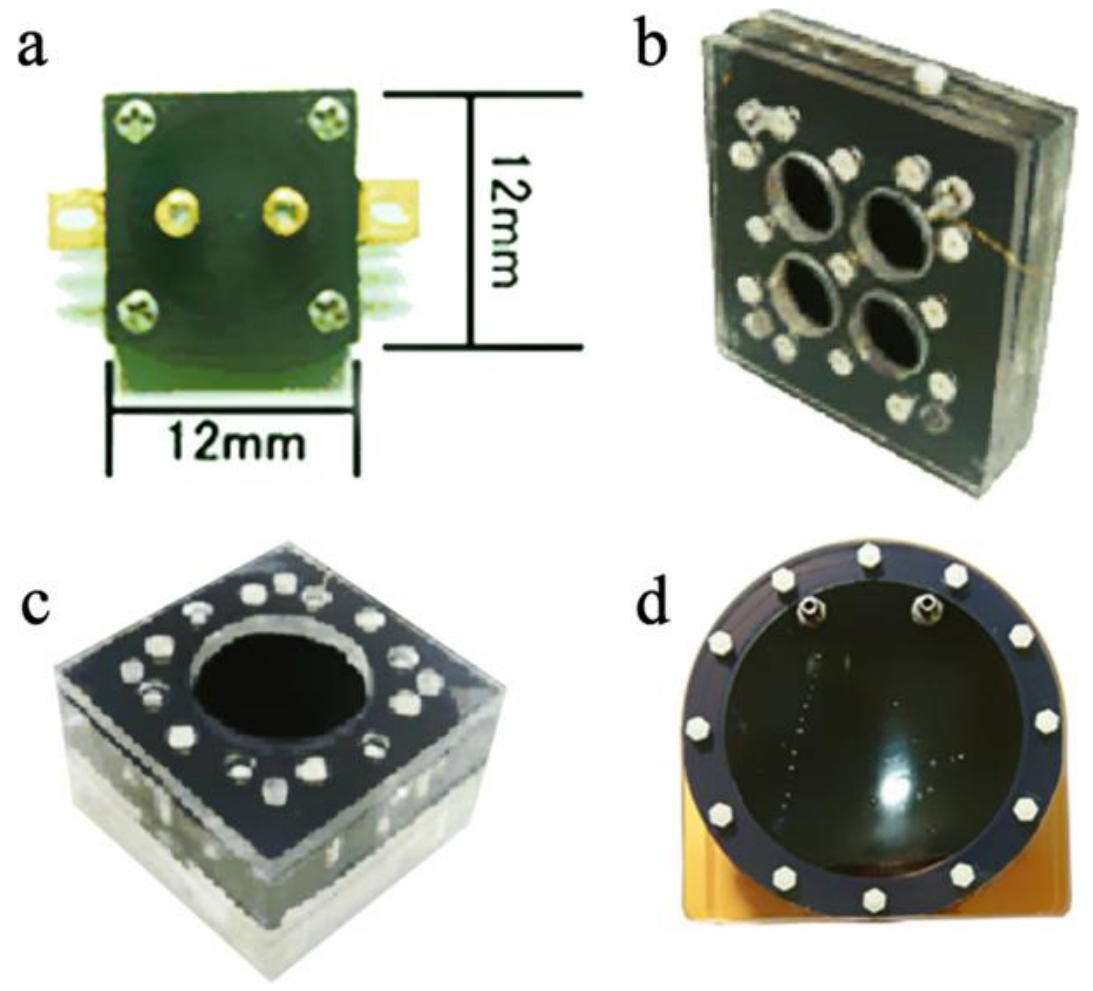

**Fig. 64. Four main types of EAMEX micropump products [123].**

**Table.1 : The specifications four main types of EAMEX micropump products[123].**

| T1-3ypes | a | b | c | d |
|---|---|---|---|---|
| Size | □ 12 mm | □ 27 × 40 mm | □ 30 × 60 mm | φ 106 (124.5) × 85 mm |
| Diaphragm diameter | φ 10 mm | φ 10 mm | φ24 mm | φ100 mm |
| Number of cells | 1 | 4 | 1 | 1 |
| Discharge pressure | 7 kPa | 55 kPa | 40 kPa | 13.6 kPa |
| Flow rate | 2 ml /min | 3 ml /min | 6 ml/min | 2000 ml/min |
| Drive frequency | 0.5 Hz | 1 Hz | 1 Hz | 1 Hz |
| Applied voltage | 1.4 V | 2 V | 2 V | 2 V |
| Current | 100 mA | 200 mA | 200 mA | 800 mA |

### *C. CPA-based microfluidic microvalves*

Like IPMCs, most of the microfluidic-based applications of CPAs are related to micropumps, but they also have great potential to be used as active elements of microfluidic microvalves. So far three works have published in the field of microfluidic microvalves and in this section they will be described. Though one of these works is an MSc thesis [125] that has not explained clearly its proposed CPA-based microvalves, and this part, we only focus on two other published works.

#### ***CPA-based Microvalve 1***

The first CPA-based microvalve was published by Berdichevsky and Lo in 2004 [126]. In this work, a PPy-based CPA has been integrated into a PDMS microchannel and by its expansion/contraction, it can close and open the microchannel (**Fig. 65**). As depicted in **Fig. 66** the whole system is fabricated using microfabrication method [126]. For PPy deposition using electrochemical plating, a sputtered layer of gold, and a solution of pyrrole monomers, and sodium dodecylbenzenesulfonate (NaDBS) dopant as the electrolyte have been used. This dopant allows the PPy does a redox reaction resulting in the expansion of CPA. More details about this reaction were already explained in *"CPA fabrication"* part. The proposed microvalve works as follows:

When an input voltage is applied between the working electrode (WE) and the counter electrode (CE), PPy(DBS) grown on a sputtered gold layer as WE and the gold deposited tracks shown in **Fig. 65** as CE, a redox reaction involving Na+ ions in the electrolyte is done, and the result is caused to CPA swelling and closing the valve. The mentioned redox reaction is represented in the following equation: [126] :

$$PPy^{+}\left(DBS^{-}\right)+Na^{+}+e^{-}\longleftrightarrow PPy^{0}\left(NaDBS\right)$$

The result of this reaction is an isotropic volume change and for achieving it in this microvalve only a low input voltage is required. Experimentally it has been shown that for having a proper valve operation, this voltage should be about 0 and -2.6 V for switching between it's opening and closing states.

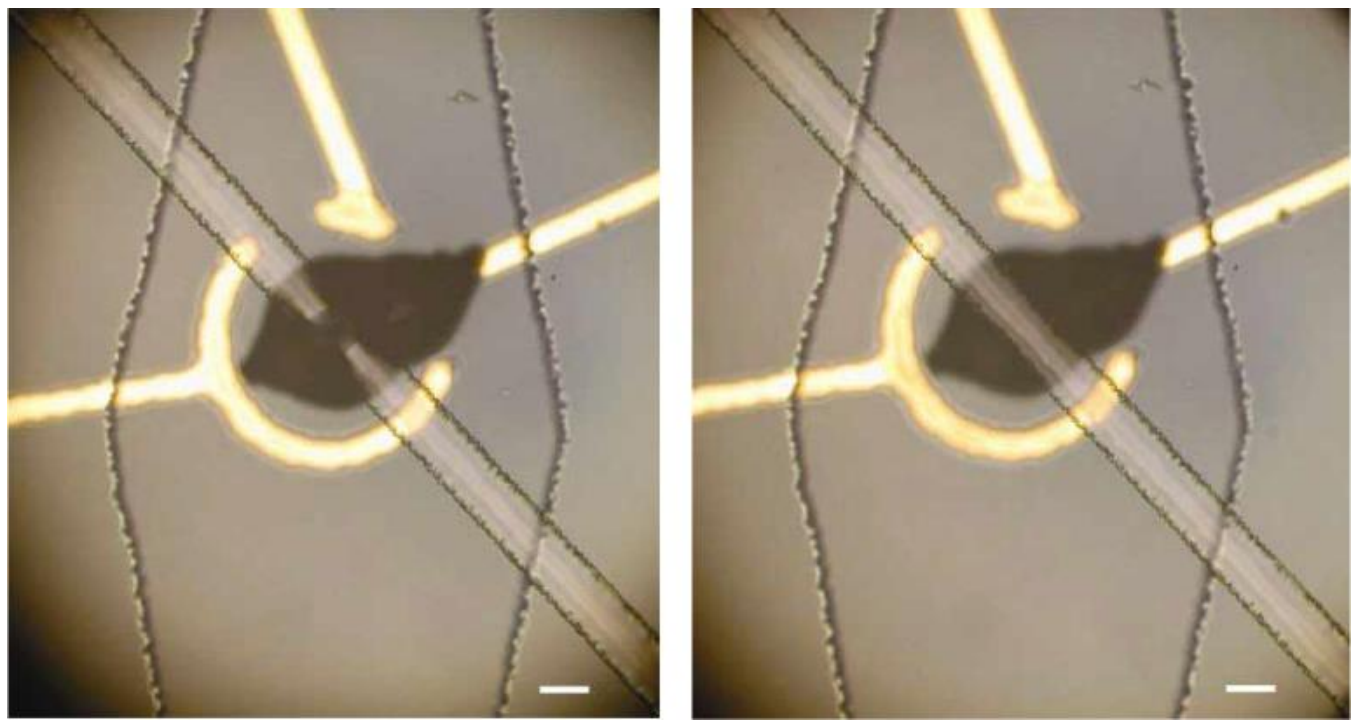

**Fig. 65. The working principle of the proposed CPA-based microvalve in [126], swelling the PPy (Dark area between Au lines) is the desired actuation for closing the valve. Left : closed valve, Right : open valve. The white scale bars are equal to 50 µm.**

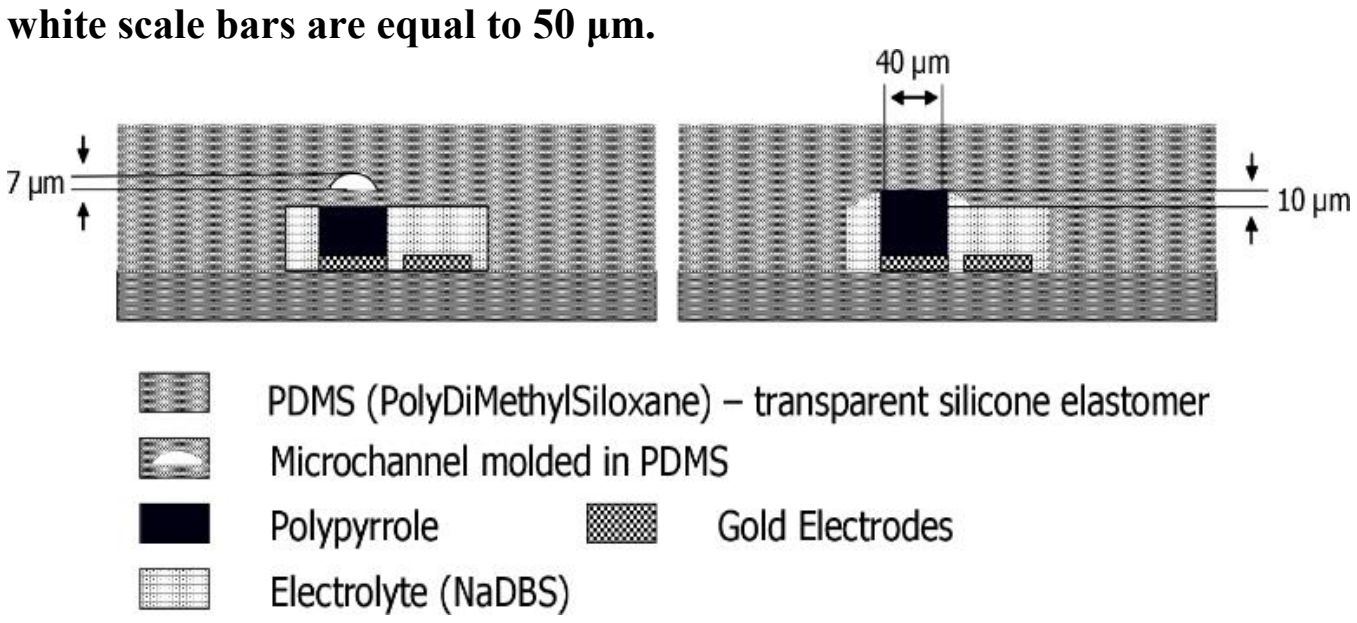

**Fig. 66. The schematic of working principle of the proposed CPA-based microvalve in [126] and it's materials, Left: open valve, Right: closed valve.**

### *CPA-based Microvalve 2*

The second CPA-based microvalve [127] was presented four years after the first one [126] but there is not any structural novelty in the more recent one and its authors have focused more on its fabrication problems like delamination at the gold/PPy interface. As illustrated in **Fig. 67**, its working principle is similar to that of the proposed microvalve in [126]. The optical micrograph of the fabricated version of this microvalve also has been shown in **Fig. 68,** and from the figures it is clear that this idea is very similar to the proposed idea in the [126] even in the selected geometry for the counter electrodes (CE). In addition, experimentally they found that in order to have a proper valve operation, a two-level input voltage (0 and -1.5 V) should be applied between the working electrode (WE) and the CE in order to switch between its opening and closing states.

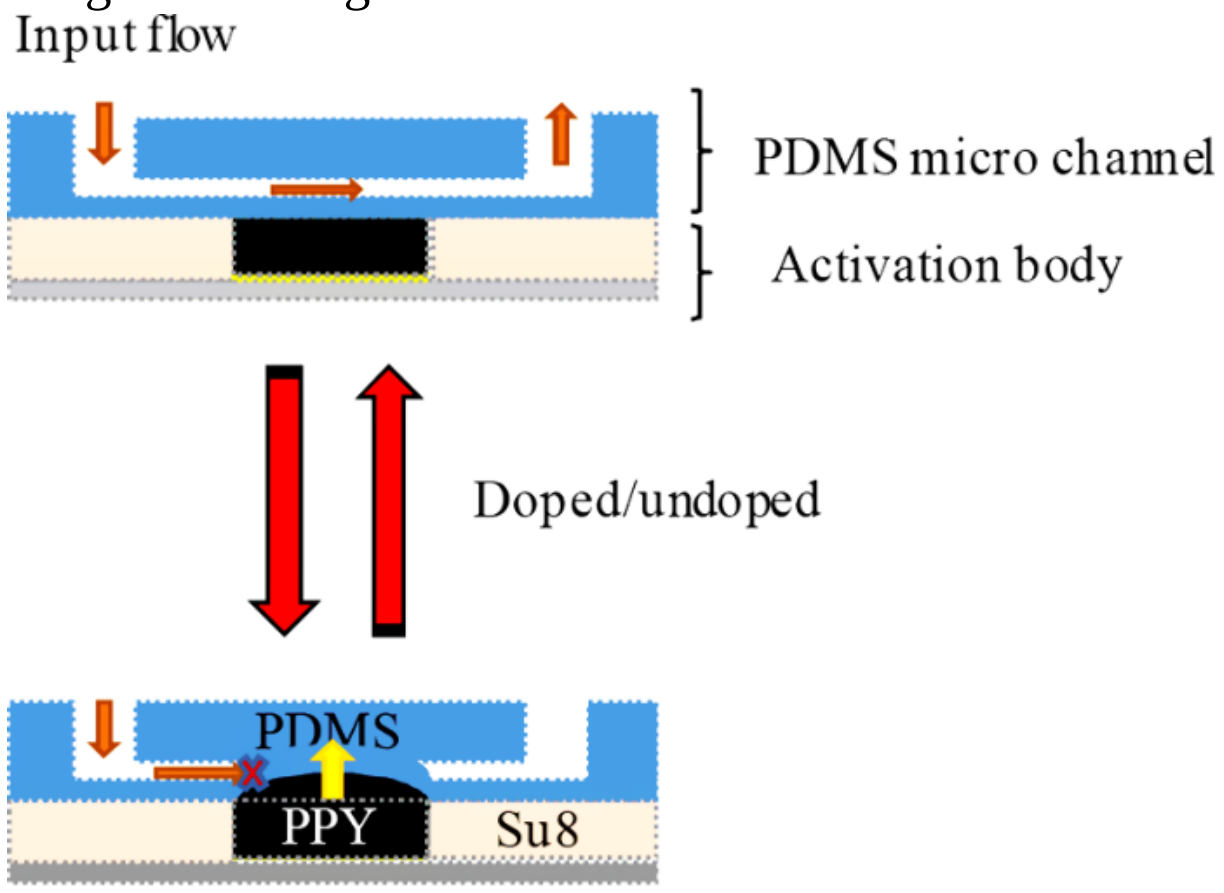

**Fig. 67. The schematic of working principle of the proposed CPA-based microvalve in [127] and it's materials, Top: open valve, Down: closed valve.**

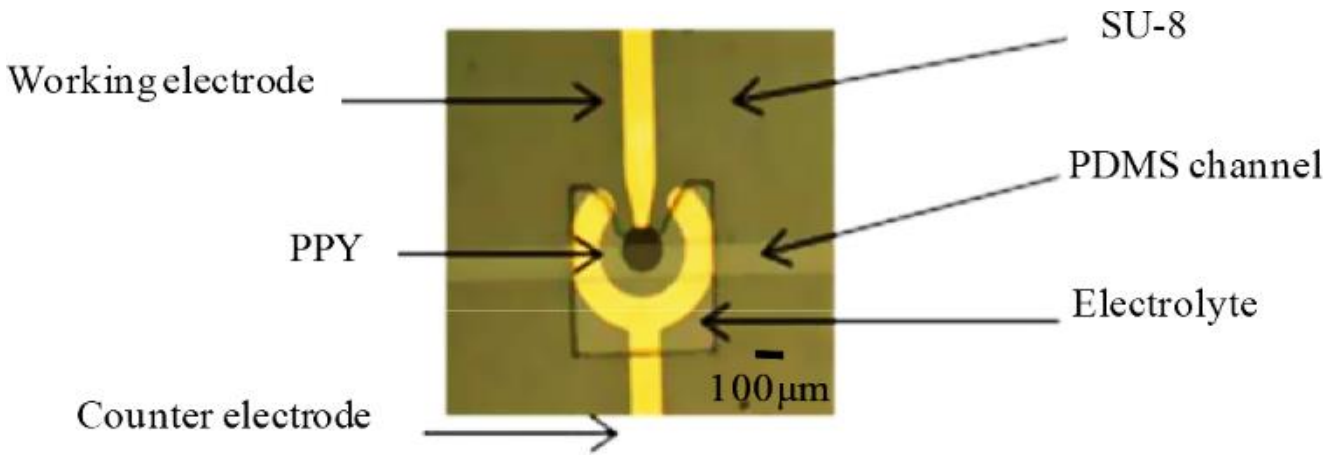

**Fig. 68. Optical micrograph at the top of the fabricated CPA-based microvalve in [127].**

## IV. Ionic Carbon Nanotube-Based Actuators

Due to attractive electrical, mechanical, optical, thermal, and chemical characteristics of nanocarbons, especially carbon nanotubes (CNTs) and graphene, they have been among the most interesting and useful materials during the last three decades, and we can find a variety of applications for this materials [128]. One of the most attractive applications of CNTs is CNT actuators and the first macroscopic actuator containing single-walled CNTs (SWNTs) was presented in 1999 by Baughman et al. [129]. They showed that the SWNT sheets have electrochemical actuation in response to applied voltage against the counter electrode in an electrolyte solution [130]. This work was a trigger point for this field, and now we can find many researchers all over the world that they pursue research on CNT actuators and CNT based artificial muscles [128]. So far several types of CNT actuators have been presented such as SWNT actuators [131], Multi-Walled CNT (MWNT) yarn-based actuators [132, 133], SWNT and MWNT paper-based actuators [134, 135], etc. All of the proposed CNT actuators need to have electrolyte solutions which is a constraint for their practical applications [130]. To solve this constraint a different form of CNT based *i*-EAP named

Bucky-Gel Actuator (BGA) was presented. This actuators can use the special features of CNTs while it does not need to have an electrolyte solution [130]. Hence, the focus of this part is about the microfluidic application of BGAs.

## *A. Principles of Bucky-Gel Actuators (BGAs)*

One of the most important problems in CNTs is their poor dispersibility and it is difficult to process them. But as discussed earlier, they have several significant features that makes them attractive for the researchers. Hence, in order to solve the dispersibility problem of the CNTs, various approaches in the literature have been proposed [130]. One of the most important approaches was presented by Fukushima et al., and they reported imidazolium-based ionic liquids (ILs) that could be a new type of dispersant of the CNTs[136, 137]. They showed that the mixing of imidazolium-based IL and SWNT suspension makes a kind of gel and this gel has both electrical and ionic conduction. They called it "Bucky-Gel" and the bucky gels  are easy to use in any shape. They also have many applications such as in fabrication of electrochemical biosensor for detection of organophosphate chemicals [138] or as a multifunctional material for an apparent four-electron ($4e^-$) reduction of oxygen ($O_2$) [139], etc.. Another application of bucky gels is the development of them as *i*-EAP actuators being called bucky gel actuators (BGAs) [130, 140-143].

As depicted in **Fig. 69**, a BGA is a trilayer composite like IPMC or trilayer CPA that has a membrane (electrolyte layer) sandwiched between two electrodes [140]. Similar to IPMCs and CPAs, they will be bent in response to a low applied voltage between their electrodes. Inversely, if we bend it, a low voltage is measurable between its electrodes across the membrane. Hence, the BGAs also can work as the sensors [144]. The electrodes are bucky gels in a polymer matrix, and the membrane is the same polymer matrix that has been mixed with an IL. ILs are the room temperature and non-volatile salts with high ionic conductivities and wide potential windows [130], and hence they are very effective for improving the electrolytes and the performance of the *i*-EAPs like the IPMCs [40, 145] and CPAs [146]. The working principle of the BGAs is explained by two theories, as charge injection and ion transfer [140]. The charge injection theory originates from the quantum and double-layer electrostatic effect [129, 140] [147]. The charge injection makes the dimensional changes in carbon-carbon bond length that these dimensional changes causing the BGA bending. But the ion transfer theory interprets differently, and expresses that the ion transfer between the layers is the main reason for the BGA actuation. Based on this theory and as shown in **Fig. 69,** the bending actuation of the BGA occurs when the positive and negative ions are accumulated separately in the electrodes, i.e one electrode for the positive ions, and another one for the negative ions. Like other *i*-EAPs, the BGAs also have some good features like inquiring low voltage, ease of fabrication, fast responsivity, air operability, high blocking force, and working at varying frequencies. All of these features besides the ability to be miniaturized for MEMS devices, makes the BGAs as the potential candidates for the active microfluidic systems.

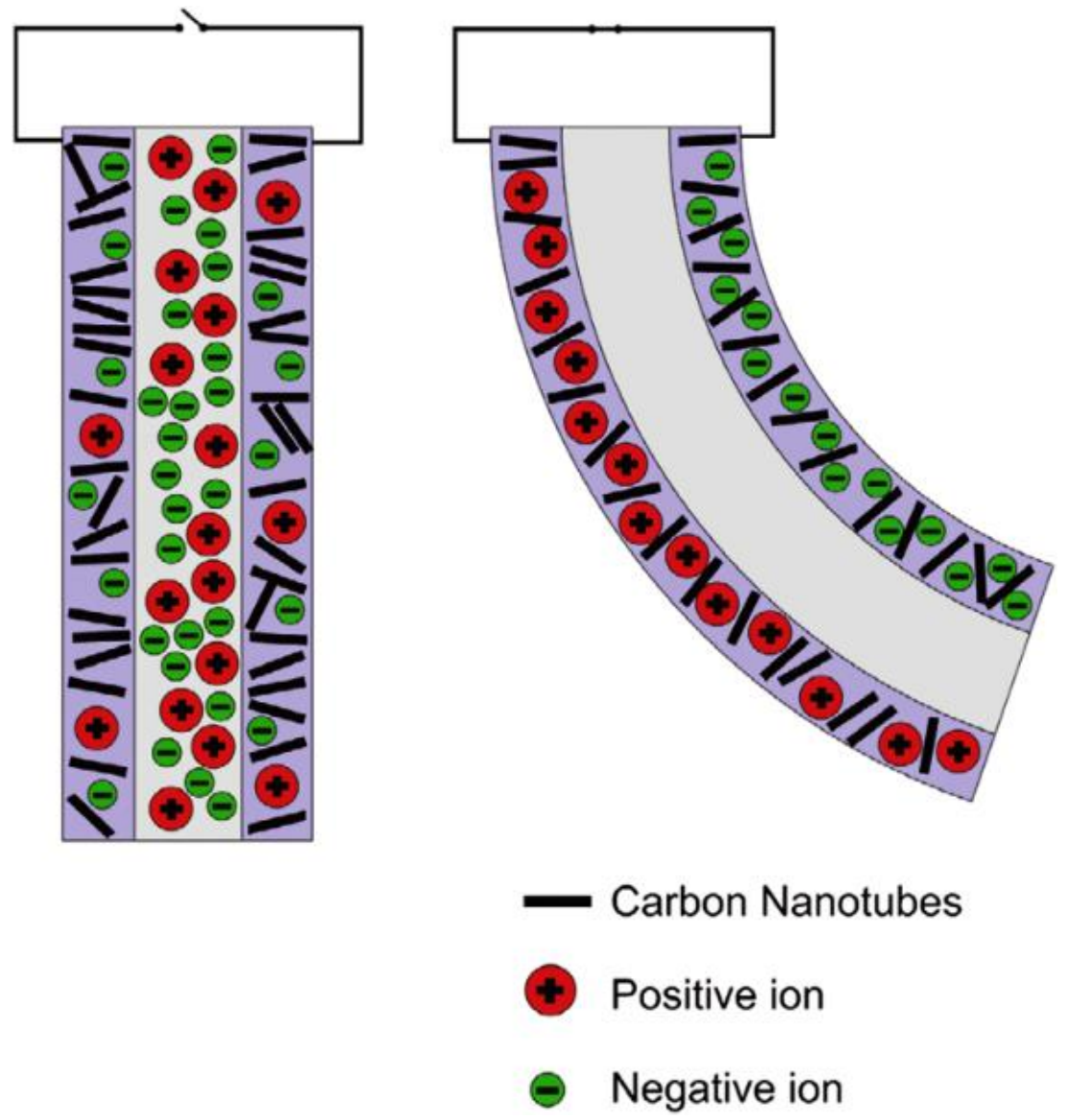

**Fig. 69. Schematics of BGA bending motion due to the theory of ion transfer[140].**

## *B. BGAs Fabrication*

The fabrication procedure of the BGAs is very easy but very time-consuming. There are only three steps in this procedure as follows; electrodes layer casting, the electrolyte layer casting, and hot press of them together as a composite where the two electrode layers have sandwiched the electrolyte layer in-between (**Fig. 70**). To cast the electrodes and electrolyte layers, four main elements are needed, as carbon source, IL, polymer matrix, and the solvent. For example, SWNT, $EMIMBF_4$ (1-ethyl-3 methylimidazolium tetrafluoroborate), PVDF and DMAC (dimethylacetamide) are some of the prevalent types of these elements. The electrode layers contained the nanocarbon (e.g., SWNT), IL (e.g., EMIMBF4), and polymer (e.g., PVDF). One of the common fabrication procedure of the electrode layers begins with the mixing of nanocarbons, IL, and the solvent (e.g., DMAC) with a ball-mill for a short time (e.g. 30 min). The resultant mixture of this step is a gelatinous black that the polymer and another amount of solvent should be added to it. The resulted mixture should be ball-milled again for another short time, and after that, it should be sonicated for a long time (e.g. 24 hr). Now the mixture is ready to cast, and it should be put in the silicon rubber molds and cured in the oven for half a day at 50 °C. Finally, for complete evaporation of the solvent, the casted sample should be kept in a vacuum oven at reduced pressure for a very long time (e.g. 100 hr). In order to fabricate the electrolyte layer, polymer and IL should be mixed (usually 1:1 ratio), and dissolved in the solvent and then cast like the casting procedure of the electrode layers [148].

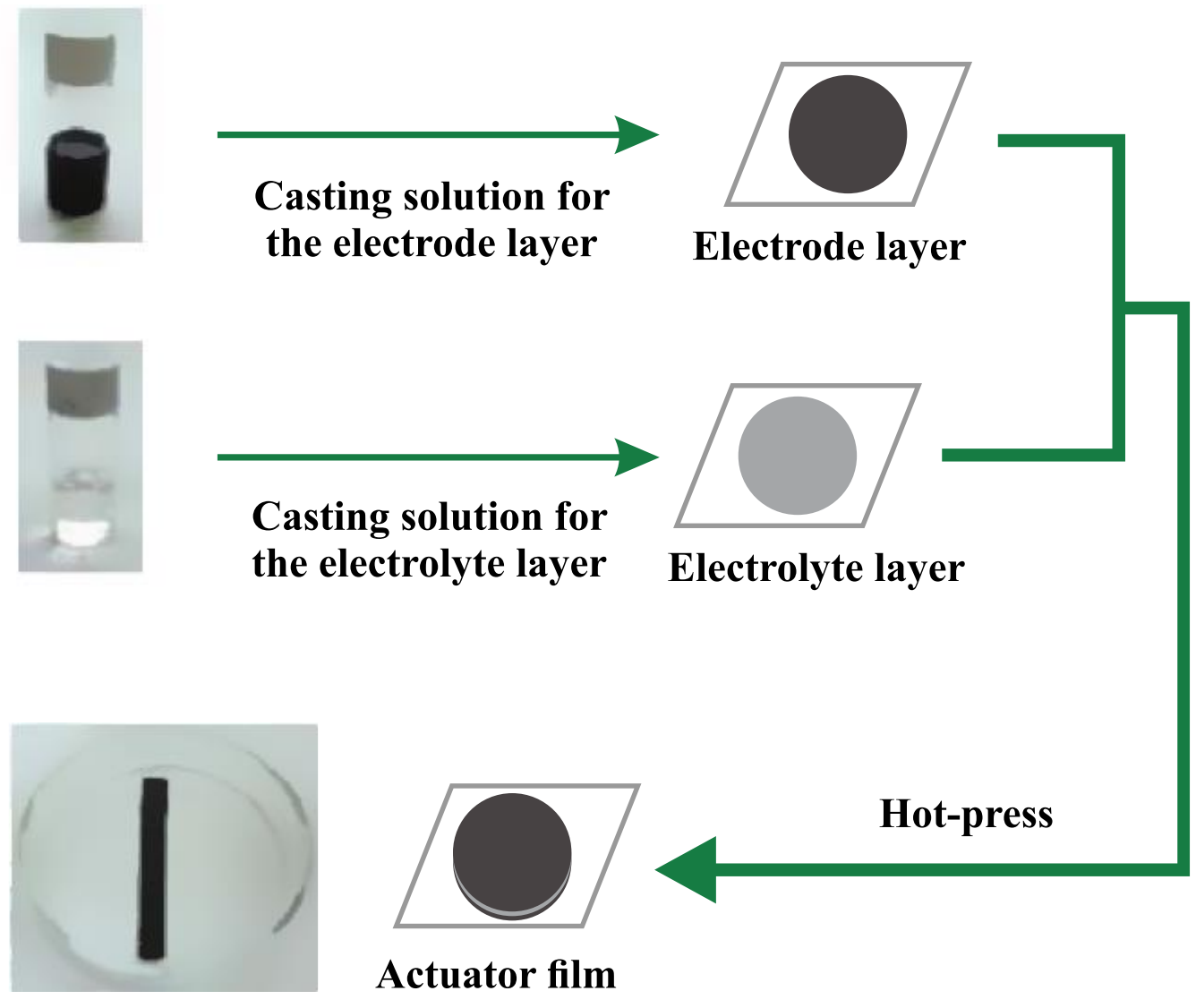

**Fig. 70. Fabrication procedure of BGAs [128].**

## *C. BGA-based microfluidic devices*

Despite the promising features of BGAs for microfluidic systems, incorporation of BGAs in microfluidic devices is still immature, and only two works have been ever published. The first work is about a BGA-based microvalve and the second one is a micropipette. In this part, these two works will be described.

### ***BGA-based Microvalve***

In the work of [148, 149], a BGA-based device was presented for flow regulation in microfluidic devices. It has a BGA microvalve that needs low input voltage, and it is fast in response and easy to fabricate while its power consumption is also low. The fabricated version of this device was experimentally tested and is shown that in this device the output flow rate has been reduced up to 93% based on the applied voltage and the frequency. The proposed BGA-based microvalve demonstrated that it has the adequate potential for the microfluidic flow regulation and it is useful regarding the point-of-care devices, drug delivery, and other microfluidics-based biomedical applications [148]. As depicted in **Fig. 71**, the proposed flow regulator device has two main passive parts, a base part and transparent PMMA cover, and BGA microvalve as an active part. In this device instead of a fixed channel, a tube-housing system has been considered. Hence by changing the tubes in different sizes, it can be feasible to test the device for different cross-sectional areas. To isolate the clamping part of the BGA from the flowing fluids, a smaller chamber also has been embedded in it where this clamping part was sealed with PDMS. The base part was printed by the 3D printer, and the PMMA cover was cut in the proper size and attached on the base part by four bolts. The working principle of this device is very straightforward. By controlling the input voltage of the BGA, we can control its bending and its controlled bending can open/close the fluid input, and consequently it can play the role of a microvalve. The key feature of this work is that for such applications it is necessary to have an actuator that has the proper blocking force to close the fluid path without leakage and it should be fast in response to control the open/close situation of the fluid path rapidly. In comparison to the IPMCs and the CPAs, the BGAs are more appropriate for these purposes because they have a higher blocking force and they can also work faster.

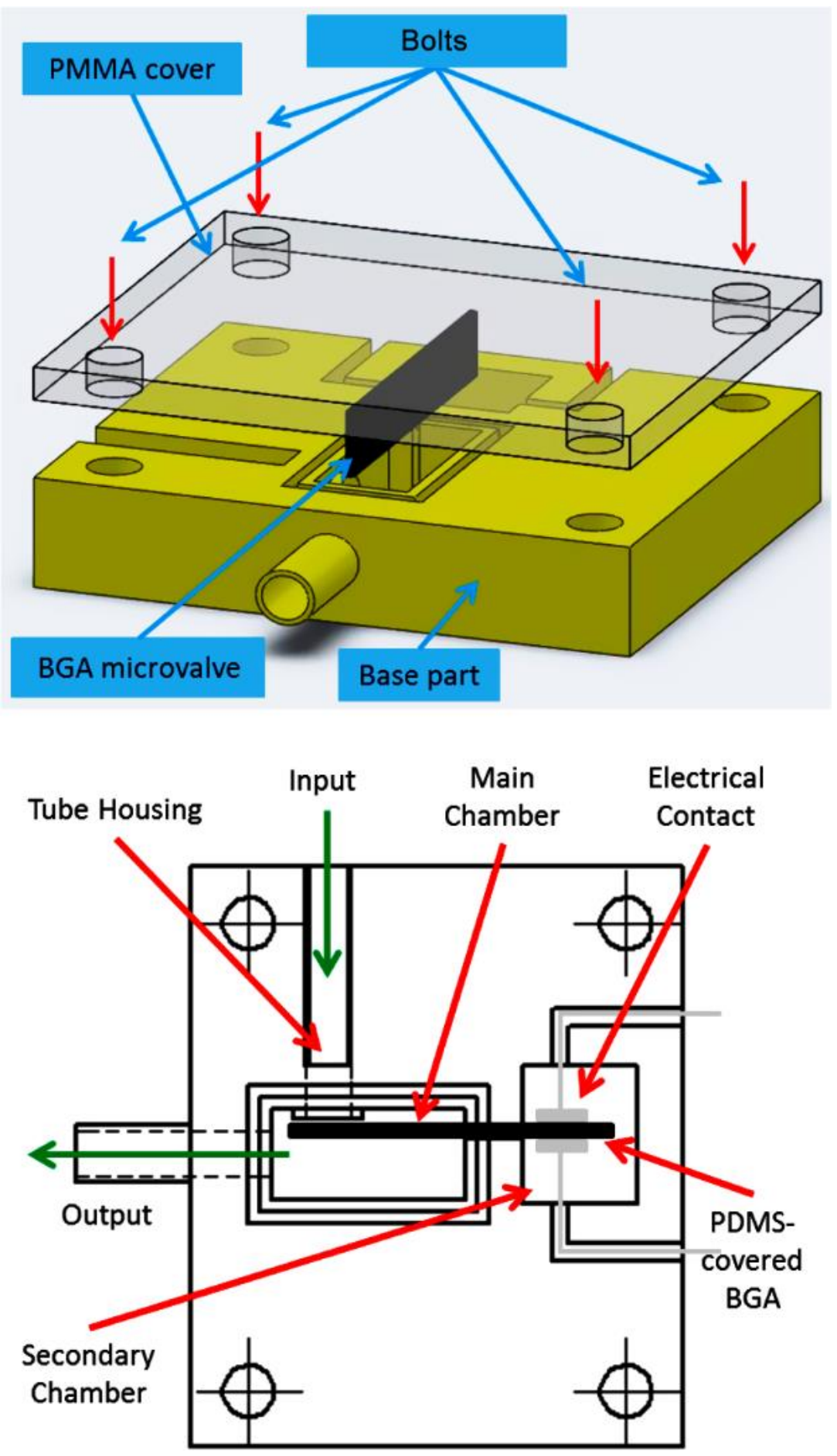

**Fig. 71. Proposed flow regulation device in [148]. Top: exploded view of the 3D schematics, Down: it's 2D plan.**

To test the performance of the fabricated device in **Fig.72**, the BGA microvalve was driven by a square wave voltage with three different amplitudes (5, 8 and 10V) at six different frequencies (250, 125, 100, 50, 25, and 10 mHz) and the flow rate results have been plotted in **Fig. 73**. It is seen that this device can regulate the input flow in the range of 0.5 ml/min to 2.5 ml/min. This variety of test helps to find proper amplitudes and frequency for the applied voltage to the BGA microvalve in order to have the most reliable (leakage less) and fastest flow regulation [148]. To fabricate the BGAs for microvalve, also SWNT, $EMIMBF_4$ (1-ethyl-3 methylimidazolium tetrafluoroborate), PVDF, and DMAC (dimethylacetamide) were used for the required carbon source, IL, polymer matrix, and the solvent, respectively [148].

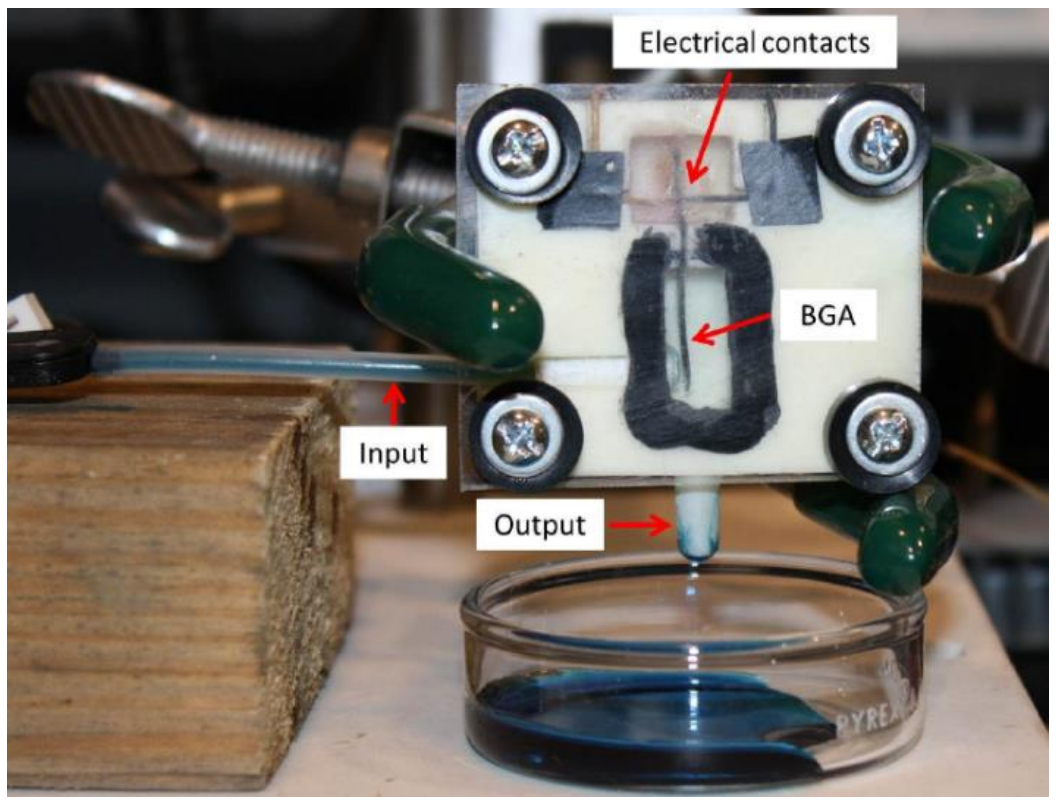

**Fig. 72. A fabricated version of the flow regulation device in [148].**

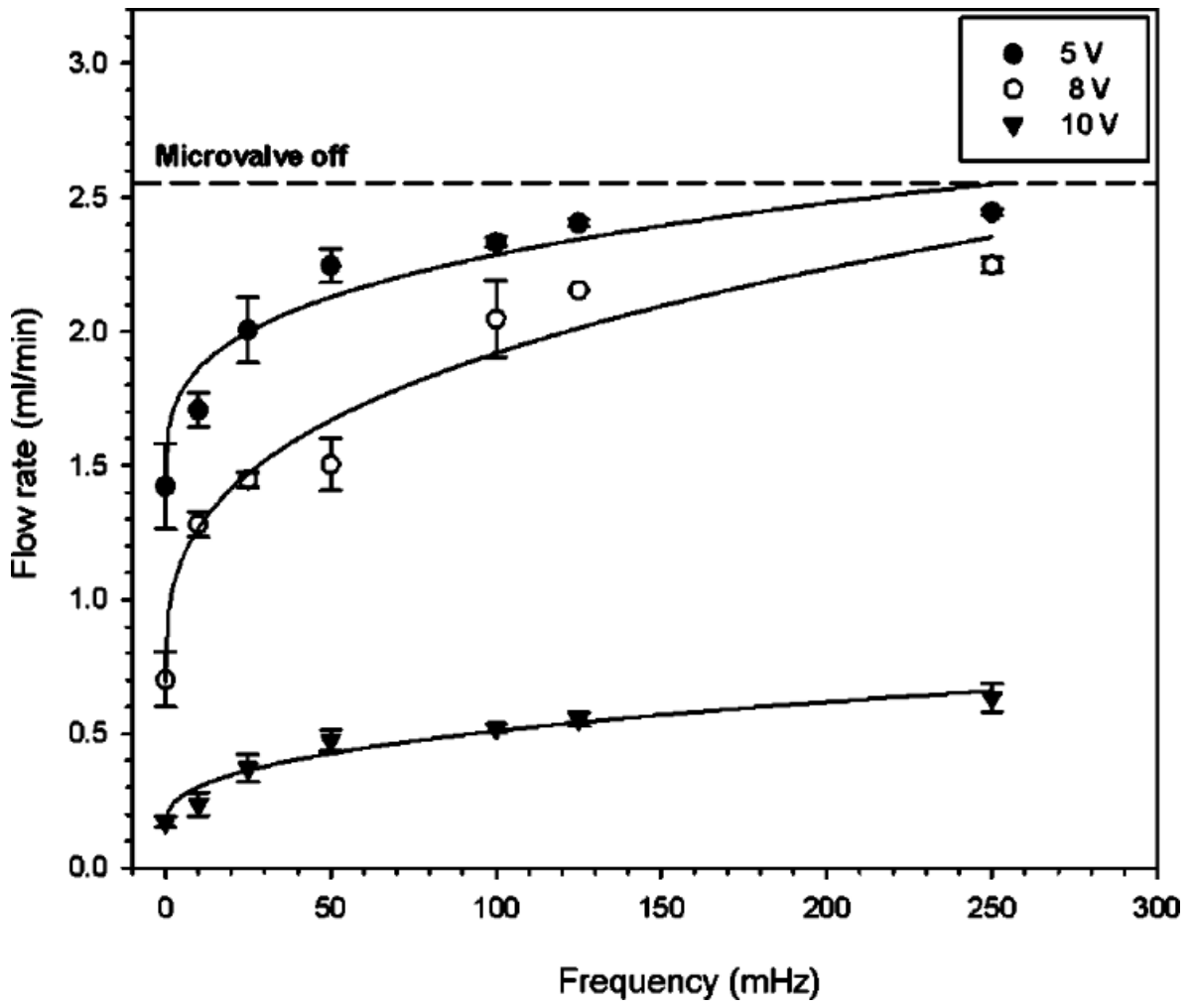

**Fig. 73. Measured flow rate of the BGA at different frequencies and three different voltages [148].**

### *BGA-based Micropipette*

In the work of [102, 150, 151], a BGA-based micropipette has been fabricated that is equipped with commercially available two and three way solenoid valves. This micropipette is fabricated and experimentally tested for suction, discharge, and dropping pure liquid water. The result shows that the proposed micropipette can work accurately and based on ISO 8655 standard for pipettes, it has the maximum permissible error [151]. As shown in **Fig. 74** this device has two PCB layers and a sheet of BGA as an active diaphragm. As shown here, in the bottom PCB layer, there are a flow channel and a circular chamber and the BGA sheet is put on it in an appropriate place to cover the chamber. Another layer of PCB is put on the BGA sheet which this sandwich is the proposed micropipette. The conductive sides of the PCBs are in touch with the BGA electrodes. Hence we can clamp the BGA by this PCB layers. Using some simple electronics and soldering the wires on the PCBs, we can apply an appropriate voltage to the BGA. In response to the applied voltage, the BGA will be bent and this bending changes the volume of the chamber (using a small volume of airflow). So this volume change makes a fluids movement into the flow channel that we can use it for sucking and releasing the fluids.

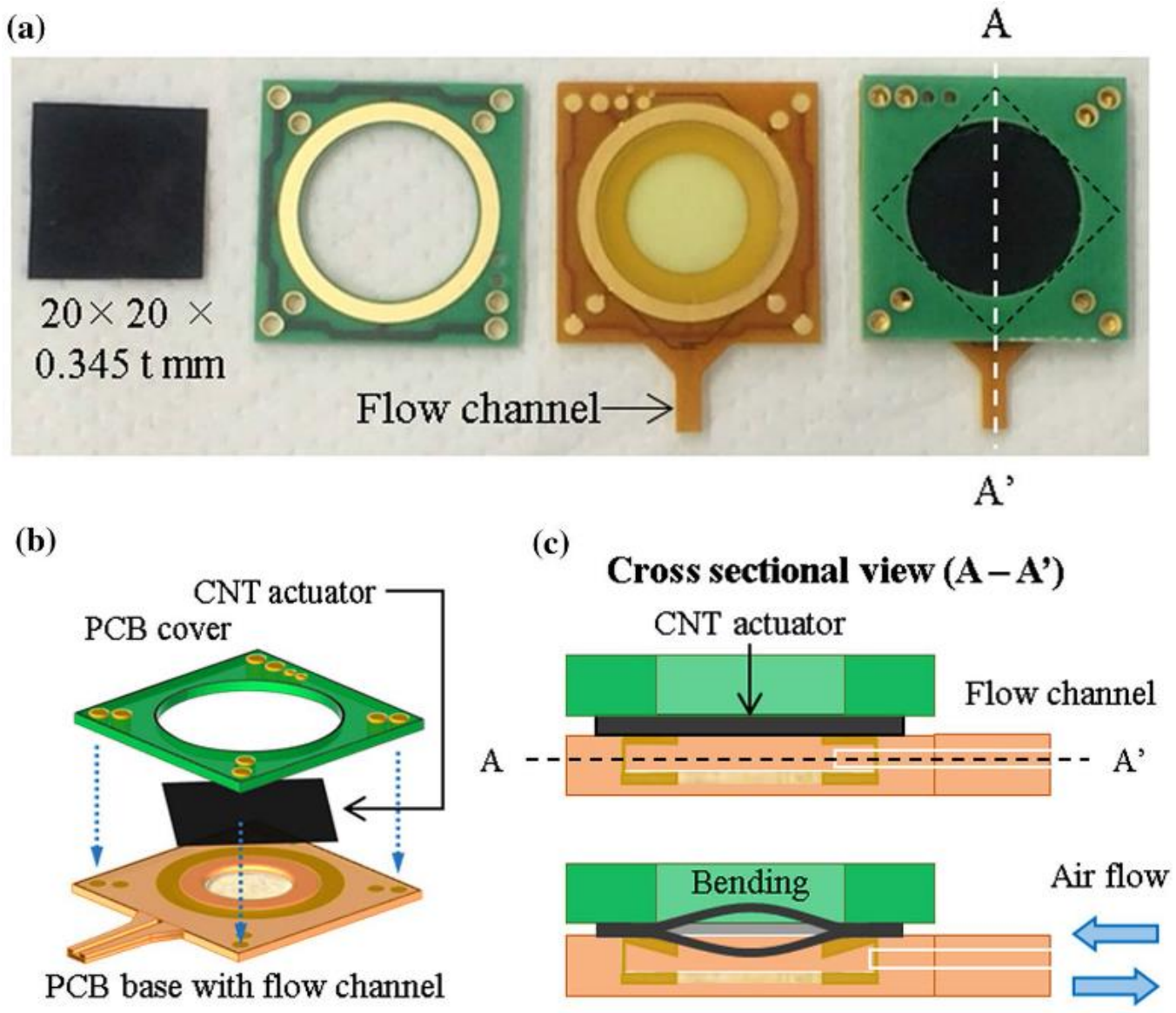

**Fig. 74. Proposed micropipette in [151]. (a) Photographs of the fabricated parts, (b) exploded view of 3D schematics, (c) cross-sectional view.**

In order to fabricate the BGA sheet for the actuator of the proposed micropipette, SWNT (Polyaniline was added as conductive additive), $EMIMBF_4$ (1-ethyl-3 methylimidazolium tetrafluoroborate), PVDF-HFP (polyvinylidene fluoride co-hexafluoroborate) and DMAC (dimethylacetamide) were used for the required carbon source, IL, polymer matrix, and the solvent, respectively [150]. And to investigate the pipetting performance, it was tested in response to a 2V input and in six voltage durations ($\Delta$t) of 1–6 s. The water displacement (mm) and the flow volume (µm) of the pipette were measured, and the results for sucking and discharging operations are plotted in **Fig. 75**. For example, after 6 sec it can suck 14.3 µl and discharge 17.4 µl, which are equivalent to 18.2 and 22.2 mm water displacement respectively.

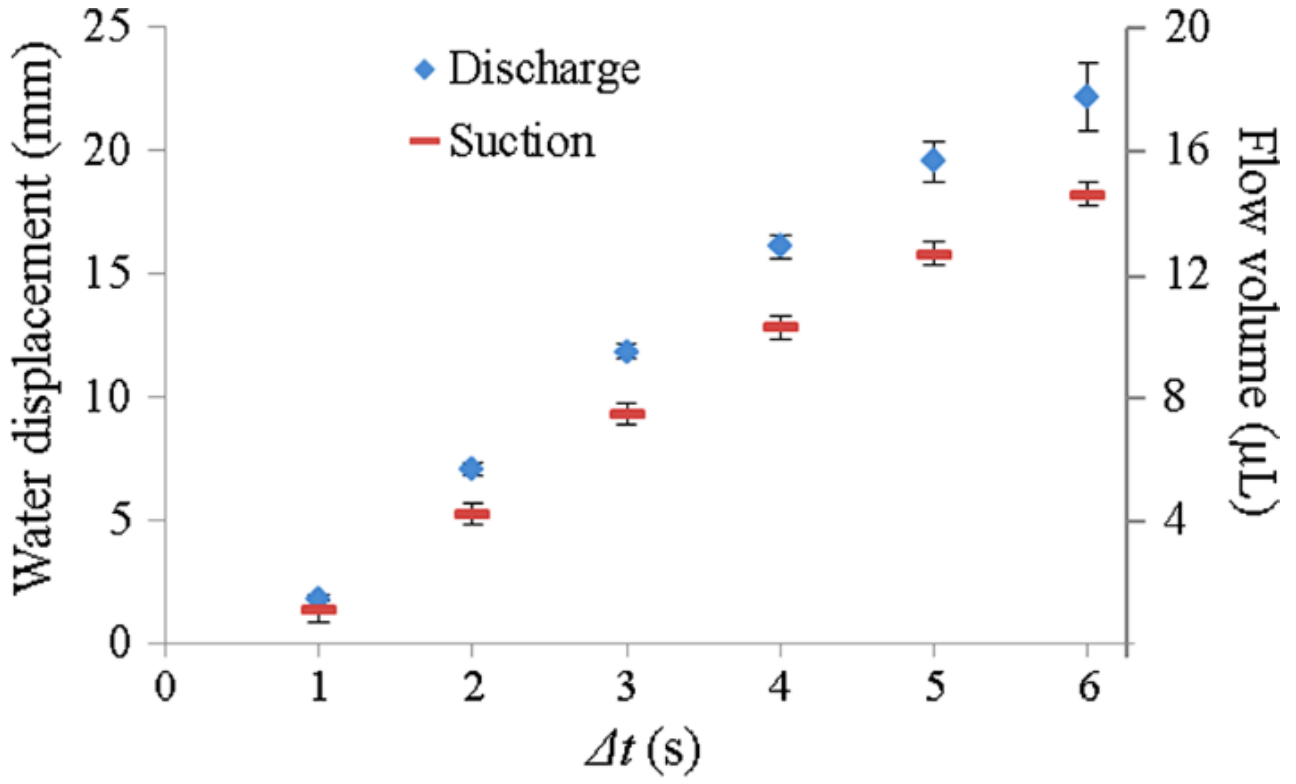

**Fig. 75. Average flow volume and water displacement of the proposed pipette in [151] as a function of the applied voltage duration (1-6 s).**

## V. Discussion

As discussed earlier, Ionic polymer-metal composites (IPMCs), conducting polymer actuators (CPAs), and Ionic carbon nanotube-based actuators (*i*-CNTAs) are three main types of the *i*-EAPs and we have reviewed all the reported microfluidic applications of them in the body of this paper. To compare all the discussed actuators, several technical data have been collected for each of the presented microfluidic devices in **Table.2** and also shown as several donut charts in **Fig.76**. As seen in **Table.2**, seven factors have been defined for each of the devices as; 1) the amplitude and frequency of the input applied voltage (Voltage/Frequency), 2) the main material used in the microfluidic devices (Device Materials), 3) Electrode material (active element material in CPAs and type of Nanocarbon in *i*-CNTAs), 4) membrane material, 5) Shape and structure of *i*-EAP, 6) type of the microfluidic devices (pump, valve and mixer), and 7) main measured indexes. As shown in the donut charts of **Fig.76**, we can find that for most of the *i*-EAP-based microfluidic devices, IPMCs and CPAs have been used with the usage weight of 55.56% and 37.04%, respectively, and only for 7.4% of them bucky gel actuators (BGAs) as *i*-CNTA were used. Almost all of the *i*-EAP-based microfluidic applications are related to micropumps, microvalves, and micromixers. As shown in **Fig.76**, more than 70% of them are micropumps and around 22% are related to microvalves and only less than 8% are micromixer. To justify these weights, we can say that in all of the micromixers and most of the microvalves, it is necessary to embed the actuator into the micro channels that it is technically very difficult. This is while most of the microfluidic micropumps are independent devices and are out of the micro channels. Nowadays microfabrication and MEMS technologies have been developed considerably and maybe we can find feasible methods to fabricate thinner and smaller *i*-EAPs and embed them into the micro channels.

Another important factor that we can find in the collected data in **Table.2**, is the used material in *i*-EAP-based microfluidic chips. As for the conventional microfluidic devices, the main material here is the same PDMS that is used in around 55% of all the devices. And 14% of the reviewed microfluidic devices use PMMA, and as shown in **Table.2** the rest are fabricated by verity of other materials such as Acrylic, Teflon, Polyethylene, etc. As stated earlier most of the reviewed devices use IPMCs and CPAs, which are further considered and discussed here. The first thing to be considered is the form of the used IPMCs and CPAs. As seen in the above two donut charts depicted in **Fig.76**, most of the IPMC actuators are in the form of a disk (45.45%) and the rest of the IPMCs have been shared equally in three forms of square sheets, cantilever beams, and petal-shaped actuators. Though in the CPAs, around 67% of them have been shared equally in their forms of cylindrical, square sheet, and cantilever beam actuators and the rest are related to other forms like petal-shape, belt-shape, etc. These results show that most of the researchers prefer to use disk shaped actuators in the IPMC based devices while for the CPAs based devices there is no remarkable preference to use as specific form of the actuator. Although IPMC researchers have presented more ideas in order to use IPMCs in the microfluidic devices, but the verity of their works are less than that of the CPA's one. For example in most of the IPMC-based micropumps, they try to use IPMC as diaphragm while in CPA-based micropumps we can find some other ideas like peristaltic pumps or valve based pumps using the CPA belts. The most common type of the IPMCs are the IPMCs that use Nafion membrane and Pt electrodes and as illustrated in the left donut charts of **Fig.76** around 80% of the IPMCs membrane and 60% of their electrodes have used pure Nafion and Pt, respectively. This is while this type of the IPMC has some practical problems. For example they are not durable enough and they have back relaxation problem, etc. Hence we can find that the most of the IPMC-based microfluidic devices haven't paid adequate attention to present practical applications and we believe there are many promising ideas to present modified version of the IPMCs in order to use them as feasible and reliable active elements in active microfluidic devices. Regarding the reviewed CPA structures in this work, we can find that more than 70% of them are bilayer and as shown in **Table.2**, 100% of their active element are Ppy while it has been not necessarily proven that these type of CPAs are the best ones and maybe some modification on CPAs can improve their performance. Finally we can see in **Table.2** that most of the mentioned *i*-EAPs work by low input voltage (less than 4 V) and low frequencies (around 1Hz). This low power requirement is the key point of *i*-EAPs that make them favorable candidates for the active microfluidic devices of POC systems. About the mentioned frequencies, it should be noted that the main resonance frequency of the developed *i*-EAPs is around 1 Hz and that's why they have chosen the operating frequency of their devices around this value. But in some applications like microfluidic micromixers we need to use *i*-EAPs in higher frequencies and maybe it is the main reason behind the limitation in the applications of the *i*-EAPs in the micromixers. Hence improving the frequency response of the *i*-EAPs can open the new window of their usage in microfluidic devices, especially as micromixers and fast micropumps and microvalves.

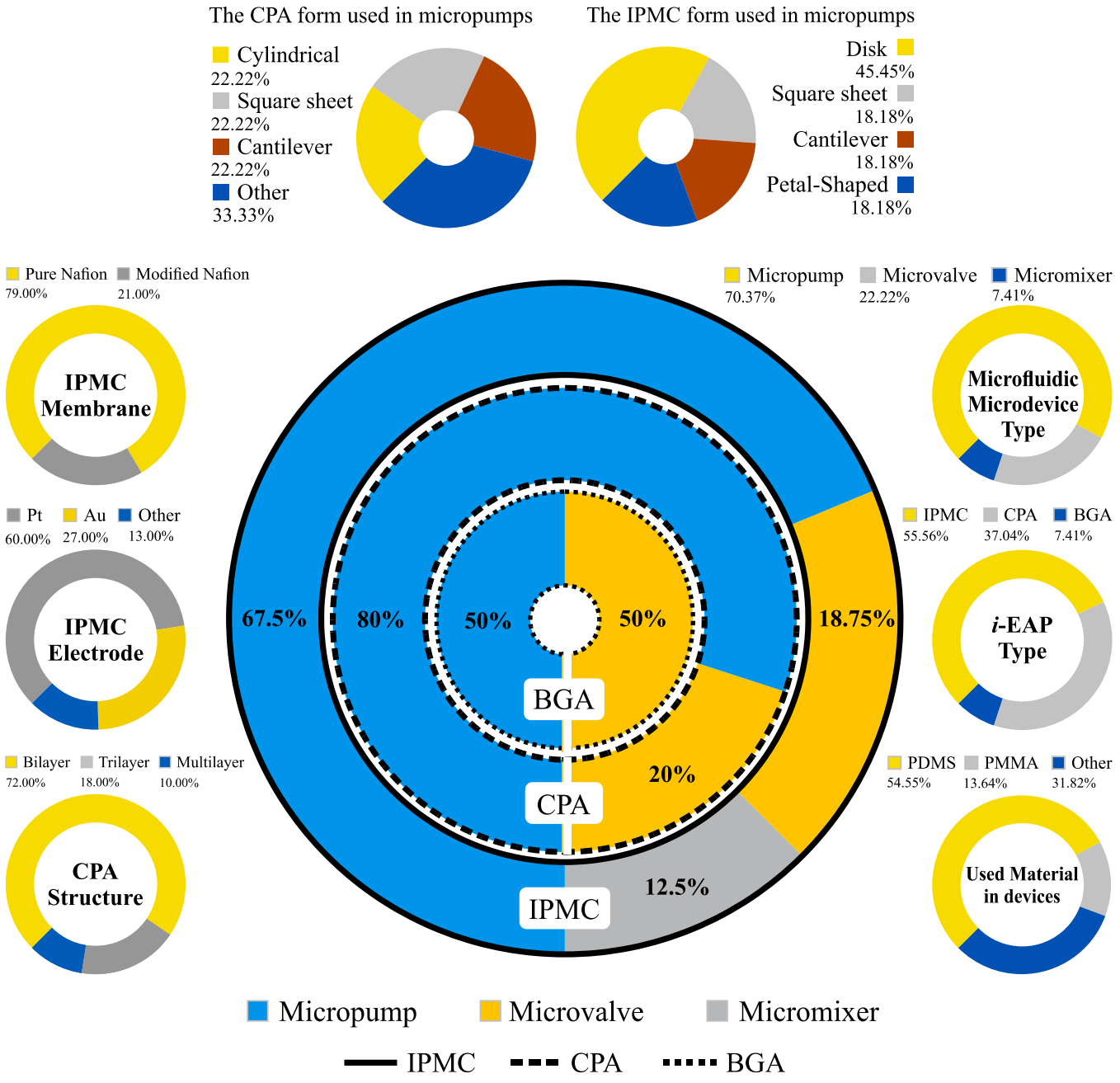

**Fig. 76. Several donut charts to provide some valuable statistics on all presented *i*-EAP-based microfluidic devices.**

**Table.2. Extracted features from all presented *i*-EAP-based microfluidic devices.**

| | Voltage / Frequency | Device Materials | Electrode Material | Membrane Material | Shape | Device Type | Main measured index | Ref |
|---|---|---|---|---|---|---|---|---|
| IPMC | 1.5 V / 0.1-15 Hz | Stainless materials | Pt | Nafion 117 | Disk/Cantilever Pump/Valve | Pump / Valve | FR : 3.5-37.8 µl/min | [87, 88] |
| | 8 $V_{P-P}$ / 0.5 Hz | PDMS | PPy | Nafion 112 | Square sheet | Pump | FR : 9.97 µl/min | [89] |
| | 2 V / 0.1 Hz | Unknown | Pt | Nafion | Disk | Pump | FR : 492 µl/min | [90] |
| | 3 V / 3 Hz | PDMS and PMMA | Nafion, layered silicate, and Ag nanopowder | Nafion/modified silica | Disk | Pump | FR :760 µl/min | [91, 92] |
| | Unknown / 0.1Hz | Acrylic, Teflon and Copper | Pt | Nafion | Disk | Pump | FR : 481.2 µl/min | [93] |
| | 5 V / 2 Hz | Unknown | Au | Unknown | Petal-shaped | Pump | FR: 202 µl/min | [94] |
| | 1 to 5 V / 0 to 8 Hz | PTFE | Pt | Nafion | Petal / Diced square-shaped | Pump | Diaphragm Displacement: 100-75 µm | [96] |
| | Unknown / 0.05 and 0.1Hz | Perspex and Latex | Pt | Nafion 117 | Cantilever | Pump | Diaphragm Displacement: ~300 µm | [97] |
| | -2 to 2 V / Unknown | Acrylic and stainless steel | Pt | Nafion 117 | Disk | Pump | FR : ~300 µl/min | [98] |
| | 2V / 0.1 to 0.5 Hz | PDMS and PCB | Au | Nafion | Cantilever | Pump | FR : ~780 µl/min | [99] |
| | 0.5 to 3 V / 1 Hz | Unknown | Pt | Nafion 117 | Petal-shaped | Pump | FR : 162 to 1611 ( µl/min) | [102] |
| | -5 to +5 V / DC | PDMS and Glass | Pt | Nafion 1110 | Cantilever | Valve | Valve Displacement: 80-150 µm | [31] |
| | -5 to +5 V / DC | PDMS and Glass | Pt | Nafion | Cantilever | Valve | Valve Displacement: 80-150 µm | [103] |
| | 1.5 V / 1 Hz | Unknown | Au | Improved Nafion with Ruthenium dioxide | Cantilever | Mixer | Mixing potential: 150% to over 375% | [104] |
| | 4.5 V (9 $V_{P-P}$) / 1 Hz | PDMS | Au (Gold leaf) | Improved Nafion 211 with PAH, gold nanoparticles and EMI-Tf ionic liquid | Cantilever | Mixer | Not reported any numerical index | [105] |
| | **Voltage / Frequency** | **Device Materials** | **Active elements Material** | **Separator Material** | **Shape / Structure** | **Device Type** | **Main measured index** | **Ref** |
| CPA | ±1V / switching | Multilayer with a plastic cover | PPy | PVDF | Cylindrical / Multilayer | Pump | FR: 2.5 µl/min | [115] |
| | ±1.5V / Unknown | PMMA | PPy | PVDF | Cantilever / Bilayer | Pump | FR :18-52 µl/min | [116-118] |
| | 1: 3V<br>2: 1V | PDMS | PPy | 1- Nafion<br>2- Polyurethane tube | 1- Cantilever / Trilayer<br>2- Cylindrical / Bilayer | Pump | FR 1 : 24 µl/min<br>FR 2 : Unknown | [119] |
| | −1.1 to 0.6V / Unknown | Vinyl, Teflon, Ti and Pt plates and Silver wire | PPy | Does not have | Bilayer | Pump | FR: 28 µl/min | [120] |
| | 4V / 0.5 Hz | PDMS | PPy | PVDF | Petal-shaped / Trilayer | Pump | FR :1260 µl/min | [95] |
| | Actuator A : −1.2 to +1.0 V / 0.05 Hz<br>Actuator B : −0.6 to +0.5 V / 0.05 Hz | Polyethylene and PDMS | PPy | Does not have | Belt shaped / Bimorph | Pump | FR : 2.0 to 83.0 µl/min | [121] |
| | -2 and 1 V switching / Square wave | Glass, Si, PI, PC and Silver wire | PPy | Does not have | Sheet / Multilayer (Stacked bilayer) | Pump | FR: 4.7 µl/min | [122] |
| | 1.4V / 0.5 Hz | Unknown | PPy | Does not have | Unknown / Bilayer | Pump | FR : 2 mL/min | [123] |
| | 0 and -2.6 V / Switching | PDMS | PPy | Does not have | Non-geometric / Bilayer | Valve | Properly operates | [126] |
| | 0 and -1.5 V / Switching | PDMS | PPy | Does not have | Circular / Bilayer | Valve | Properly operates | [127] |
| | **Voltage / Frequency** | **Device Materials** | **Nanocarbon Type** | **Polymer matrix / IL** | **Shape** | **Device Type** | **Main measured index** | **Ref** |
| BGA | (5, 8 and 10V) / (10-250 mHz) | PMMA and PDMS | SWNT | PVDF / EMIMBF4 | Cantilever | Valve | FR : >0.5 to 2.5 mL/min | [148] |
| | 2V / Unknown | PCB | SWNT | PVDF-HFP / EMIMBF4 | Square sheet | Pipette | Flow volume after 6 s<br>Suck: 14.3 µl<br>Discharge : 17.4 µl | [151] |

## VI. Conclusion

The most of active micropumps, microvalves, and micromixers are actuated by pneumatic, thermopneumatic, hydraulic, piezoelectric, electromagnetic, and electrostatic actuators. There are some remarkable constraint in these active devices, for example they are challenging to assemble and control, and need some special facilities and also they are expensive. These constraints are against some of the main targeted advantages of microfluidic devices such as POC development, use at home, availability, low-cost ability, etc. Against these actuators, *i*-EAPs are different and they are less in struggle with the mentioned constraints. Hence for the purpose of developing the active microfluidic chips without the need to special facilities, the *i*-EAP actuators have been presented in this review. *i*-EAP actuators have some promising features for the POC-based systems, home usable, ubiquitous, and low-cost active microfluidic devices. Since 1996 we can find many papers that have been presented for developing some ideas in order to use *i*-EAP actuators in the active microfluidic devices. But unfortunately they are not oriented or consistent and sometimes overlap each other in contents making it difficult to follow the directions of the research in this field. To address this problem, all the available considerable information about the developed *i*-EAP-based microfluidic devices to this date have been collected and discussed in this review for making it as a comprehensive reference for the researchers who want to use *i*-EAP actuators in active microfluidic devices. All in all, this review paper can show that *i*-EAPs and especially IPMCs and CPAs are the group of good candidates that can play the role of active elements in microfluidic devices. But the current version of the developed devices have significant constraints for their practical applications in the field of active microfluidics and it is necessary to apply some further modification on them.